\documentclass[a4paper,fleqn,usenatbib]{mnras}

\usepackage{newtxtext,newtxmath}

\usepackage[T1]{fontenc}
\usepackage{ae,aecompl}

\usepackage{graphicx}	
\usepackage{amsmath}	
\usepackage{amssymb}	

\newcommand\addtag{\refstepcounter{equation}\tag{\theequation}}

\title[S-PASS]{S-band Polarization All Sky Survey (S-PASS): survey description and maps}

\author[E. Carretti et al.]{E. Carretti,$^{1,2,3}$\thanks{Contact e-mail: \href{mailto:carretti@ira.inaf.it}{carretti@ira.inaf.it}}
M. Haverkorn,$^{4}$
L.~Staveley-Smith,$^{5}$
G.~Bernardi,$^{1,6,7}$
 \newauthor
B.M.~Gaensler,$^{8}$
M.J.~Kesteven,$^{3}$
S.~Poppi,$^{2}$
S.~Brown,$^{9}$
R.M.~Crocker,$^{10}$ 
 \newauthor
C.~Purcell,$^{11,12}$
D.H.F.M.~Schnitzler,$^{13}$
X.~Sun$^{14}$ 
\\
\\
$^{1}$INAF Istituto di Radioastronomia, Via Gobetti 101, 40129 Bologna, Italy\\
$^{2}$INAF Osservatorio Astronomico di Cagliari, Via della Scienza 5, 09047 Selargius (CA), Italy\\
$^{3}$CSIRO Astronomy and Space Science, PO Box 76, Epping, NSW 1710, Australia\\
$^{4}$Department of Astrophysics/IMAPP, Radboud University Nijmegen, P.O. Box 9010, 6500 GL Nijmegen, The Netherlands\\
$^{5}$International Centre for Radio Astronomy Research, University of Western Australia, Crawley, WA 6009, Australia\\
$^{6}$SKA SA, 3rd Floor, The Park, Park Road, Pinelands, 7405, South Africa\\
$^{7}$Department of Physics and Electronics, Rhodes University, PO Box 94, Grahamstown, 6140, South Africa\\
$^{8}$Dunlap Institute for Astronomy and Astrophysics, University of Toronto, 50 George St, Toronto, ON M5S 3H4, Canada\\
$^{9}$Department of Physics \& Astronomy, The University of Iowa, Iowa City, Iowa 52245, USA\\
$^{10}$Research School of Astronomy and Astrophysics, Australian National University, Canberra, Australia\\
$^{11}$Research Centre for Astronomy, Astrophysics, and Astrophotonics, Macquarie University, NSW 2109, Australia\\
$^{12}$Sydney Institute for Astronomy, School of Physics, The University of Sydney, NSW 2006 Australia\\
$^{13}$Bendenweg 51, 53121 Bonn, Germany\\
$^{14}$Department of Astronomy, Yunnan University, and Key Laboratory of Astroparticle Physics of Yunnan Province, Kunming, \\
           650091, People's Republic of China
}

\date{Accepted XXX. Received YYY; in original form ZZZ}

\pubyear{2019}

\begin{document}
\label{firstpage}
\pagerange{\pageref{firstpage}--\pageref{lastpage}}
\maketitle

\begin{abstract}
We present the S-Band Polarization All Sky Survey (S-PASS), a survey of  polarized radio emission over the southern sky at Dec~$< -1^\circ$ taken with the Parkes 
radio telescope at 2.3~GHz. The main aim was to observe at a frequency high enough to avoid strong depolarization at intermediate Galactic latitudes (still present at 
1.4~GHz) to study Galactic magnetism, but low enough to retain ample Signal-to-Noise ratio (S/N) at high latitudes for extragalactic and cosmological science.  
We developed a new scanning strategy based on long azimuth scans, and a corresponding map-making procedure to make recovery 
of the overall mean signal of Stokes $Q$ and $U$ possible, a long-standing problem with polarization observations. 
We describe the scanning strategy, map-making procedure, and validation tests. 
The overall mean signal is recovered with a precision better than 0.5\%. The maps have a mean sensitivity of 0.81~mK on beam--size scales 
and show clear polarized signals, typically to within a few degrees of the Galactic plane, with ample S/N everywhere (the typical signal 
in low emission regions is 13~mK, and 98.6\% of the pixels have S/N~$> 3$).
The largest depolarization areas are in the inner Galaxy,  associated with the Sagittarius Arm.
We have also computed a Rotation Measure map combining S-PASS with archival data from the WMAP and Planck experiments.
A Stokes $I$ map has been generated, with a sensitivity limited to the confusion level of 9~mK. 
\end{abstract}




\section{Introduction}

All-sky radio polarization surveys are extremely important for a number of scientific aims, ranging from the Galactic ISM to cosmology.
Radio polarization observations allow the study of the magnetic fields  of both the emitting source -- either compact or extended -- and the foreground medium.
Ultra relativistic electrons accelerated by magnetic fields emit synchrotron emission. 
The emission is intrinsically highly polarized (some 70\%), with polarization angle oriented by 
$90^\circ$ compared to the magnetic field orientation on the plane of the sky. 
The polarization fraction and its variation with frequency provide key information about a magnetised medium.  
Net polarization requires the field to be ordered to some degree; in a fully turbulent medium the  field is completely tangled and the resulting  emission unpolarized. 
A magnetic field with a projected direction on the plane of the sky which varies along the line of sight
can also depolarize the emission. 
This is a purely geometrical effect independent of the observing frequency.

The polarization fraction indicates the relative strength of  isotropic turbulent and ordered components, while its behaviour with the frequency provides information about the conditions of the source medium and its magnetic field (e.g. \citealt{Farnes14, Lamee16}).

Faraday Rotation (FR) rotates the observed polarization angle $\phi$ in proportion to wavelength squared $\lambda^2$  as the polarized emission passes through a magneto-ionic medium consisting of free electrons immersed in a magnetic field:
\begin{equation}
   \phi =  \phi_0 + {\rm RM}\, \lambda^2 \, ,
\end{equation}
where $\phi_0$ is the radiation intrinsic polarization angle (in the case of no FR or $\lambda = 0$) and the Rotation Measure RM 
measures the field component parallel to the line-of-sight $B_\parallel$ weighted by the free electron density $n_e$:
\begin{equation}
   {\rm RM \,[rad\,m^{-2}}] = 821 \int_{\rm source}^{\rm observer} n_e\,[{\rm cm}^{-3}] \,\, B_\parallel \,[\mu{\rm G}] \,\, dl\, [{\rm kpc}] \, .
\end{equation}
From these relations,  FR can be used to infer information on the magnetic field. 
In particular, the observed rotation of the polarization angle with frequency, a fit to multifrequency observations, or the application of the RM-synthesis algorithm \citep{Brentjens05} will give an estimate of the RM which, in turn, gives information on the magnetic field along the line--of--sight. 
FR can also have a destructive effect.
If the polarization angle variation  within the frequency channel, the telescope beam, or the emitting region is too large, FR causes signal cancellation (depolarization) (e.g. see \citealt{Burn66}).

The optimal observing frequency depends on the region, its complexity, and the amount of RM involved.
Low frequency observations are more sensitive to RM, but are more prone to depolarization. 
High frequencies are less affected by depolarization and therefore reveal much more polarized emission, even close to the Galactic Plane where  emission is complex. 
They also return angles closer to the intrinsic polarization angle, and therefore more directly inform us of the orientation of the magnetic field on the plane of the sky. 
Ideally,  observations over a broad frequency range are required.

For nearly three decades from the mid 1970s, the only available polarization surveys were the collection by \citet{Brouw76}, covering 5 frequencies from 408 to 1411~MHz, but sparsely and irregularly sampled, and with coarse resolution (a few degrees). 
In the mid 2000s  the first all sky survey at 1.4~GHz by the DRAO and Villa Elisa telescopes with 36~arcmin resolution appeared~\citep{Wolleben06, Testori08}, then followed at 23~GHz by WMAP with $\sim 1^\circ$ resolution~\citep{Page07, Bennett13}. 
Together these provided the astronomy community its first comprehensive view of the polarized sky.  
However, that view was incomplete. 
Depolarization is significant at low frequency. 
Maps made by \citet{Carretti05} from \citet{Brouw76} data show that at 408 MHz most of the sky is depolarized with the exception of the Galactic polar caps.  
With increasing  frequency, the depolarization starts to disappear and  signal appears at progressively lower latitudes. 
The 1.4~GHz maps of \citet{Wolleben06} and  \citet{Testori08} show that  polarized emission is still totally depolarized at $|b| < 30^\circ$ and with Faraday modulation visible up to $|b| = 50^\circ$~\citep{Carretti10},  with the exception of the Fan Region, in the outer Galaxy at longitude $l \sim 135^\circ$, where the disc RM is close to zero and Faraday effects smaller.
At the other end of the radio spectrum, WMAP high-frequency polarization observations resulted in the first 
all-sky image without FR. However, the WMAP image suffered from poor Signal-to-Noise (S/N) ratio, especially in the halo at mid and high latitudes, and are  insufficient for conducting high precision studies of cosmic magnetism  and polarized Cosmic Microwave Background (CMB) foregrounds.
Moreover, this image showed that 
a finer resolution was required to help beat depolarization in high RM regions and to reveal the full detail of the Galactic ISM and Galactic objects.
 
This paper presents the S-band Polarization All Sky Survey (S-PASS), a survey of the radio polarized emission at 2.3~GHz of the southern sky at Dec~$< -1^\circ$ conducted with the Parkes radio telescope at an angular resolution of 8.9~arcmin. 
 The frequency is a compromise: higher than 1.4 GHz allows us to beat  FR in the Galactic disc and to reveal the Galactic emission and structure in the disc and at the disc-halo transition,. However, the frequency is still low enough to retain ample S/N ratio  in the low emission areas at high Galactic latitude, which are required for high precision magnetism studies in the Galactic halo and for CMB foreground analysis.
The angular resolution of S-PASS is 4 times better than the DRAO and Villa Elisa maps (and 6 times WMAP's) delivering a much more detailed view of the sky.
Although S-PASS is designed and optimised for polarization measurements, we have also taken total intensity  data and produced a Stokes~$I$ sky map.

The science than can be addressed given the S-PASS characteristics is very diverse. 
Applications include:
\begin{itemize}
\item Galactic magnetism, which  can be studied through polarization angles and RMs. In particular, large scale Galactic magnetic field models can be optimised by fitting to these data (e.g., \citealt{Sun08,Jansson12}). 

\item The study of the polarized emission from the Galactic disc within $|b| < 30^\circ$, which can reveal structures in the disc and at the disc-halo transition, otherwise hidden by depolarization at lower frequencies.
Structures such as loops and lobes have much more contrast in polarization than in total intensity, making them easier to identify and analyse  (e.g., see \citealt{Vidal15, Carretti13a}). 
S-PASS reveals polarized emission down to a few degrees from the Galactic plane, even uncovering new structures in the Inner Galaxy~\citep{Carretti13a, Thomson18}. 

\item ISM turbulence , which can be identified and studied with the polarization gradient technique \citep{Gaensler11}. 
This has previously been applied to limited areas (e.g. \citealt{Gaensler11,Herron17}), but the area,  sensitivity, and resolution of S-PASS render it possible to construct an image of turbulence throughout the Milky Way  \citep{Iacobelli14,Robitaille17}.

\item The search for the B-Mode of CMB polarization is one of the most important topics in astrophysics today. 
Detection of this signal would constitute a discovery of the primeval cosmological gravitational wave background emitted at the time of Inflation  and would discriminate between the plethora of proposed Inflation models. 
B-Mode CMB polarization detection is hampered by Galactic foregrounds, of which synchrotron is one of the most important and is stronger than the cosmological signal  (e.g. \citealt{Carretti10,Krachmalnicoff16}).  
The high S/N, low FR, and high resolution of S-PASS are ideal to precisely characterise this foreground, estimate its impact on CMB experiments, and 
allow for the optimisation of CMB observing strategies.
Moreover, in combination with data at other frequencies  (e.g.\ C-BASS \citep{Irfan15} and QUIJOTE~\citep{Poidevin18}), S-PASS data allow for the construction of precise templates with which to  significantly reduce pollution by the Galactic foreground and thereby open up the possibility of detection of the gravitational wave background, even for pessimistic models.

\item The S-PASS angular resolution has made it possible to produce a catalogue of a few thousand polarized, compact sources \citep{Schnitzeler19}. 
The analysis of catalogues with hundreds of sources shows weak evidence for the evolution of magnetism for extragalactic radio sources (e.g. \citealt{Farnes14,Lamee16}). With an order of magnitude more sources,  an unambiguous detection is possible. 

\item The radio emitting Intra Cluster Medium (ICM) in galaxy clusters, including radio haloes and relics, is usually observed with compact interferometers with good sensitivity to extended emission and sufficient resolution to discriminate this emission from the blending of compact sources.
However,  the ICM of nearby clusters can stretch over angular scales which interferometers miss or to which they have poor sensitivity. 
The risk, therefore, is to underestimate such structures (because a significant part of their extended flux is missed) which, in turn,  leads to
incorrect inference of the total energetics, and the consequent physical misinterpretation of cluster phenomena. 
Single-dish telescopes have an unparalleled sensitivity to extended emission, retaining information on all angular scales. 
As previously shown~\citep{Brown11,Loi17,Vacca18}, observations with large, single-dish telescopes of a size that matches the minimum interferometer baseline are thus essential to obtain a full picture of large structures. 
S-PASS is in an excellent position to contribute to this effort with its high sensitivity and resolution~\citep{Carretti13b}.

\item  In addition to ICM diffuse emission, searches are underway for the synchrotron emission from the Cosmic Web. 
This would probe the flow of gas into filaments, and from filaments to clusters which are the nodes of the Cosmic Web. 
Moreover some 50\% of all cosmic baryons is expected to be in filaments \citep{Nicastro18} and its detection is essential to test the current scenario of structure formation.
Single-dish telescopes with large dishes are ideal for such a search, thanks to their sensitivity to extended emission and resolution. 
The transverse size of the filaments of the Local Cosmic Web is some 10~arcmin. S-PASS with its beam of 8.9~arcmin is perfectly suited for finding these.
Upper limits have been found so far with correlation techniques between radio emission and cosmic web tracers (e.g. \citealt{Vernstrom17,Brown17}) and a possible detection of diffuse synchrotron emission from a filament has recently been reported using the 64-m Sardinia Radio Telescope in combination with interferometric data to subtract blended compact sources \citep{Vacca18}. 
\end{itemize}

S-PASS is part of a `Golden Age' of single-dish polarization surveys. 
The radio polarization community is making a significant effort to map the sky in a broad frequency range and  fill the gaps in all-sky, diffuse polarized emission information. 
The Global-Magneto-Ionic-Medium-Survey (GMIMS) aims at mapping the entire sky from 300 to 1800~MHz with high frequency resolution \citep{Wolleben09}. 
Observations have been completed for the GMIMS-North-High-Band survey covering 1300-1800 MHz with the 26-m DRAO telescope~\citep{Wolleben10}, and the GMIMS-South-Low-Band survey covering 300-900 MHz \citep{Wolleben19} and the GMIMS-South-High-Band survey covering 1300-1800 MHz (also known as Southern Twenty-centimeter All-sky Polarization Survey -- STAPS, \citealt{Haverkorn15}), both with the Parkes radio telescope and using the same observing strategy developed by S-PASS. 
GMIMS and S-PASS complement each other: GMIMS is more focused on high RM sensitivity and the realisation of Faraday Tomography, while S-PASS is more focused on reducing depolarization and Faraday effects.  
C-BASS \citep{Irfan15} is an all--sky survey at 5~GHz, thus with a higher frequency than S-PASS, but with lower S/N ratio and coarser resolution.  
QUIJOTE~\citep{Poidevin18} aims at observing the northern sky at 10-20 GHz with resolution similar to C-BASS. 
The future combination of S-PASS, C-BASS, and QUIJOTE data holds out the promise of a high accuracy synchrotron emission map for high precision CMB foreground cleaning. 
Finally, Planck has delivered a polarization map at 30~GHz with S/N ratio similar to WMAP but finer angular resolution~\citep{Planck18}.

This paper presents the S-PASS survey, its major features, the observations, the map-making algorithm and validation tests, residual contamination, and the maps.
The scientific utilisation of the data is beyond the scope of this paper and has been described in separate papers, published and in progress. 
With this paper we describe and publicly release the data binned in one broad frequency band from all useful channels. 
A further paper will describe the multifrequency data cube.

Throughout the paper we use the IAU convention for polarization angles (PA), as normally used in astrophysics: PA is 0$^\circ$ for vectors pointing north and increases eastward. Note that this differs from the convention used in some experiments, like, e.g., WMAP and Planck, where PA increases westward. 

The paper is organised as follows: Section~\ref{Sec:obs} describes observations, calibration, and the main data features, Section~\ref{Sec:str} covers the observing strategy and its design,  Section~\ref{Sec:mapmaking} describes the map--making algorithm and its tests, Section~\ref{Sec:ground} covers the ground emission estimate and subtraction. 
In section~\ref{Sec:maps}  we show the maps obtained from S-PASS data together with an analysis of the signal distribution and main features. 
We also measure and display a RM map of the diffuse emission that combines S-PASS, WMAP, and Planck data. 
Section~\ref{Sec:science} gives a summary of the scientific results obtained so far using S-PASS data and described in separate papers, Section~\ref{Sec:data} lists the data set we will release, while Section~\ref{Sec:conc} presents our summary and conclusions.

\section{Observations and Calibrations}\label{Sec:obs}

Observations were conducted with the Parkes radio telescope, located in New South Wales, Australia.
This is a primary focus, 64-m diameter telescope. Observations were carried out in 8 sessions, either 17 or 18 nights each, 
from October 2007 to July 2009 approximately spaced 3 months each (October 2007, January 2008, April 2008, July 2008, October 2008, January 2009, April 2009, July 2009). Total observing time was approximately 1820~hrs. The main observational parameters are reported in Table~\ref{Tab:feat}.

Observations were conducted at night, sunset--to--sunrise, to prevent Solar contamination from sidelobes, which are significant in polarization at this frequency when sensitivities of some 1~mJy/beam are desired~\citep{Carretti05}.

\begin{table}
	\centering
  \caption{Main characteristics and parameters of S-PASS.}
  \begin{tabular}{lr}
    \hline
   Effective central frequency$^1$                    &  2303~MHz \\
   Useful bandwidth$^2$                                  & 168~MHz \\
   Frequency ranges                                                 & 2176-2216~MHz \\
                                                                        & 2272-2400~MHz \\
  FWHM                                      & $8.9$~arcmin \\
  Flux density--to--brightness Temperature gain    &  1.19~Jy/K  \\
  Channel bandwidth               & 0.5~MHz \\
  Binned sub-band bandwidth               & 8~MHz \\
  Sky coverage                   & Dec $ < -1^\circ$ \\
  Pixel size                                & 3.4~arcmin \\
  Sky projection				& HEALPix \\
  Total observing time                  & 1820~h \\
  Mean $Q$, $U$ beam-size pixel rms sensitivity  & 0.815~mK \\
  Stokes $I$ Confusion Limit                                 & 9~mK \\
  Flux density calibration accuracy   & 5\% \\
  Instrumental polarization               & < 0.05\% \\
  Mean source position rms error             & 33 arcsec \\
  Map systematic pointing error                             & 5.6 arcsec \\
  \hline
  $^1$~Mean frequency of used frequency channels.\\
  $^2$~After RFI channel flagging.\\
    \end{tabular}
    \label{Tab:feat}
\end{table}

The bandpass was centred at the frequency of 2300~MHz, 
with 256~MHz nominal bandwidth. The bandpass filter limits the usable range to 2176--2400 MHz.
The standard Parkes S-band receiver ({\it Galileo}) was used with 
a system temperature of $T_{\rm sys} \sim 20$~K. 
This is a circular polarization system delivering the Left--Handed and Right-Handed Circular Polarization $L$ and $R$, 
ideal for Stokes $Q$ and $U$ measurements with a single-dish telescope. 
 
The feed is installed on--axis and illuminates the reflector with an edge taper of 19~dB. 
The beam has a width of FWHM = 8.9' at 2.3 GHz and first sidelobe at -31~dB, 
the full--beam flux density--to--brightness temperature gain is $G$~=~1.19~Jy/K. 

The Digital Filter Bank Mark 3 (DFB3) backend was used, a digital spectro--polarimeter 
recording the two autocorrelation products ($RR^*$ and $LL^*$) and the complex cross--correlation product ($RL^*$) for full Stokes capability.  
A configuration with 256~MHz bandwidth and 512 frequency channels, 0.5~MHz each,  was used.
The backend was based on 8--bit samplers for a large dynamic range. 
The channelization is based on a polyphase filter 
technique to ensure an impressive isolation between frequency channels (more than 60~dB between adjacent channels) 
giving negligible cross-contamination by in-band radio frequency interference (RFI).
This represents a leap in capability compared to the 13~dB isolation of the old generation of Fourier--based correlators. 

Flux calibration was done using PKS~B1934-638.  
We assumed the flux density model by \citet{Reynolds94} that covers the range 0.4--9 GHz 
with an accuracy of 5\% \citep{Bernardi03}.
The source PKS~B0407-658 was used as secondary calibrator. 
Flux calibration was performed for each frequency channel, effectively delivering a flat calibrated bandpass.

Data were binned in 8~MHz bins and, after RFI flagging, 
21~bins were used
covering the ranges 
2176-2216 and 2272-2400~MHz, for an effective central frequency of
2303~MHz and bandwidth of 168~MHz. 

On-axis instrumental polarization calibration was done using the flux calibrator 
PKS~B1934-638\footnote{http://www.narrabri.atnf.csiro.au/calibrators/\\calibrator\_database\_viewcal?source=1934-638} 
and 
Ori~A~\citep{Gardner75}; these are assumed to be unpolarized down to 0.1\%.
We measured a typical instrumental polarization of 1\% in each frequency channel, rising to 2\% at the two very ends of the band.
Following this measurement, instrumental polarization was  subtracted using the standard technique. 
Once calibrated and corrected, the residual instrumental polarization on individual frequency channels was within $<\sim$0.2\% while 
averaged over the entire 2176-2400 MHz band, it was better than 0.05\%.
The fractional instrumental polarization is estimated as the measured $Q$ and $U$ response when observing the unpolarized calibrator 
divided by its Stokes $I$ flux. 
Then, for each segment of data, that fraction of its Stokes $I$ emission is subtracted from its measured Stokes~$Q$ and $U$.     

\begin{figure}
	\includegraphics[width=\columnwidth, viewport=140 570 500 820]{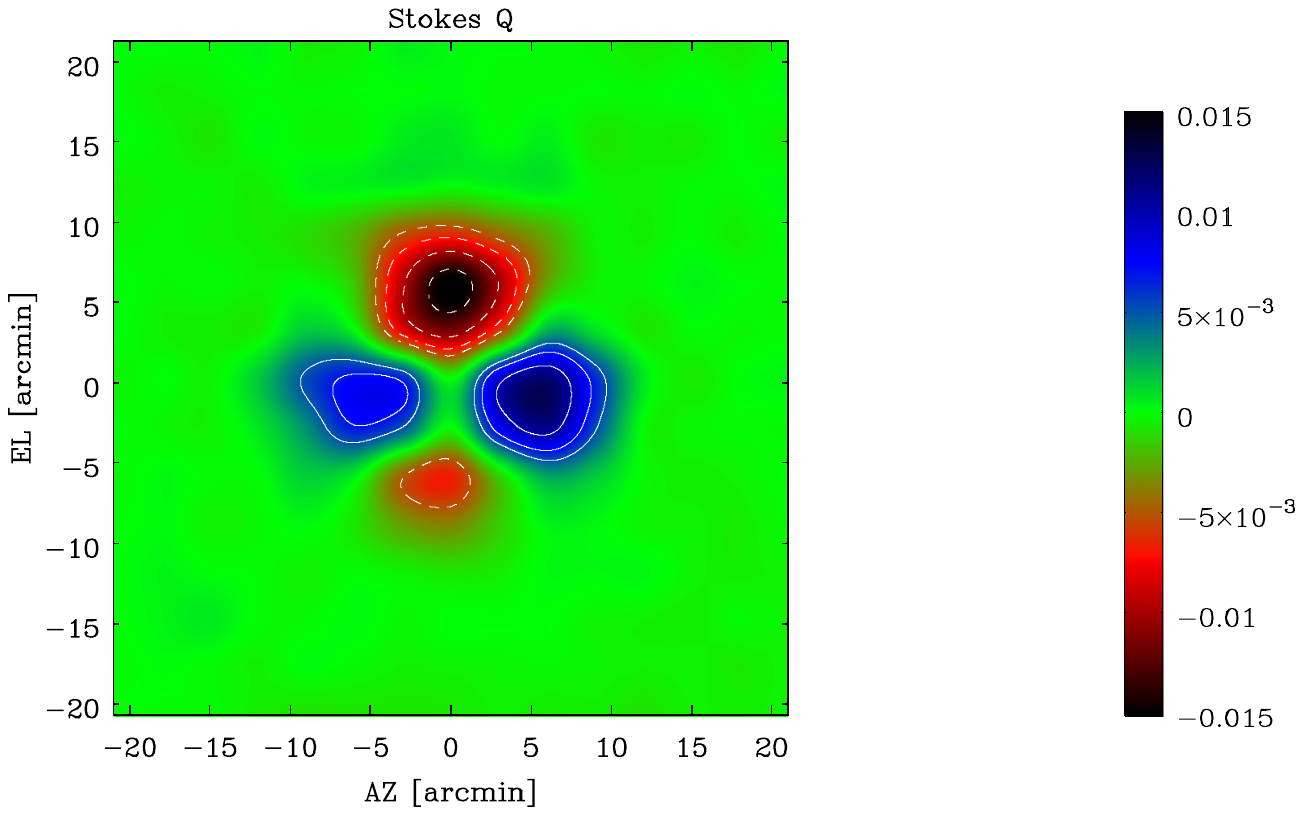}
	\includegraphics[width=\columnwidth, viewport=140 570 500 830]{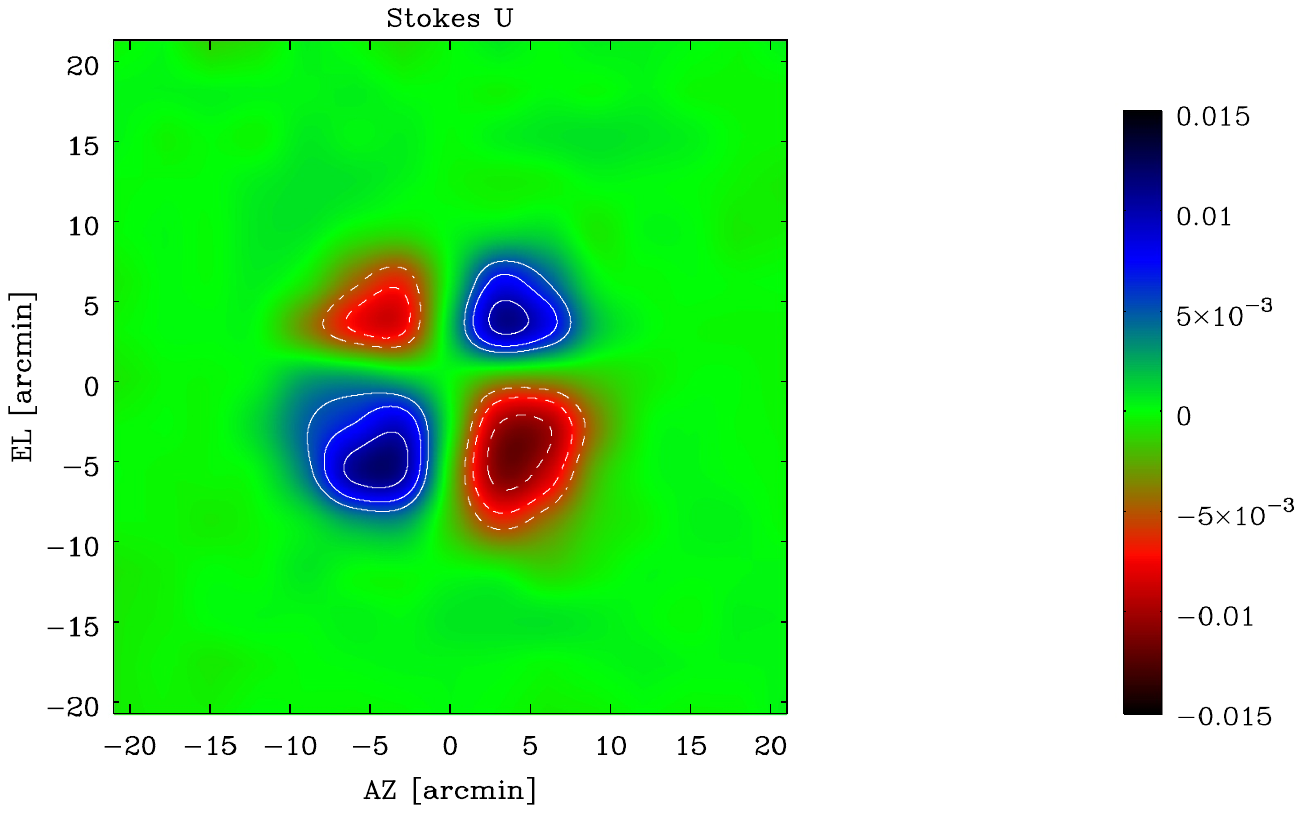}
    \caption{Instrumental polarization beam pattern for Stokes $Q$ (top) and $U$ (bottom) measured using the unpolarized source PKS~B1934-638. 
                  Unit is fractional polarization. Contour lines are for positive (solid) and negative (dashed) values,  start from $\pm 0.5$\% and scale with a $\sqrt{2}$ factor.}
    \label{Fig:offaxis_instpol}
\end{figure}

The off--axis instrumental polarization pattern after on-axis calibration is reported in Figure~\ref{Fig:offaxis_instpol}, measured using  PKS~B1934-638.
The impact of this on scientific results is marginal: it does not affect polarization measurements of compact sources. 
For diffuse emission, the pattern of opposite sign lobes leads to cancellation  on scales larger than twice the beamsize, with no significant residual effects  (e.g. see \citealt{Carretti04,Murgia16}).  
We estimate a residual instrumental polarization of 0.08\% and 0.06\% for Stokes $Q$ and $U$, respectively, when averaged 
over the instrumental polarization beam area, which represents the residual leakage of Stokes $I$ into $Q$ and $U$ in case of uniform emission on the beam scale.
No deconvolution of the instrumental polarization beam was done for this data release.

Polarization angle calibration was done using the sources PKS~B0043-424 and 3C~138, 
assuming polarization angles of PA$_{0043}$~=~140$\degr$ and PA$_{\rm 3C138}$~=~169$\degr$, respectively \citep{Broten88,Perley13}. 
The calibration was performed on each frequency bin (8 MHz width) 
to make in-band depolarization negligible. 
We found our data have a rotation of less than $3\degr$ over each 8 MHz bin before calibration, for a negligible depolarization smaller than 0.05~\%.
This reveals another obvious advantage of the modern spectropolarimetry: in-band phase equalisation is no longer strongly required
on the scale of the entire observing band --  only on the scale of the individual frequency channel width. 
This relaxes significantly the instrument design requirements. 

Pointing calibration was performed by the observatory staff at each session. 
Typical precision was better than 10~arcsec.
Measured on the final data set, the position error distribution of the compact sources identified in the maps has mean and dispersion of $4.7\pm24.7$~arcsec in R.A. and  $3.1\pm22.4$~arcsec in Dec. \citep{Meyers17}. The dispersion (33~arcsec combined) is the mean rms position error of the individual sources, which is limited by the source S/N. The mean (5.6~arcsec combined) is the systematic position error of the map.  Pointing errors are therefore insignificant compared to the beam size.  

The confusion limit was measured on our maps and found to be $\sigma_{I, (\rm CL)}  = 9$~mK \citep{Meyers17} at the beam resolution of 8.9~arcmin.

\section{Observing strategy}\label{Sec:str}

The practical construction of an all-sky  map of polarized continuum emission with a large telescope and a small beam size implies the realisation of several requirements including:
\begin{itemize}
 \item{} Recovering the emission at all scales, including the mean emission (offset) at the scale of the map size;
 \item{} Minimising the ground emission contamination;
 \item{} Scanning the sky at a fast rate and with small overheads to minimise 1/$f$ noise and reduce observing time.
\end{itemize} 

Recovering the mean emission is the most stringent and complicated requirement for large-scale polarization observations. 
Discrete object observations can make use of the background emission around the object as reference to correct and get the object emission. 
However, such reference emission does not exist for emission at very large scales (e.g. the ISM diffuse emission) and it is essential
to obtain a calibrated offset in order to correctly measure emission and polarization angle. 
An incorrect offset would result in incorrect polarization angles compromising all the potential science based on  the measurement of magnetic field direction and RM. 

Subdividing the sky in smaller, square areas (e.g. $10^\circ \times 10^\circ$ or $20^\circ \times 20^\circ$  patches) and mapping them with sets of short orthogonal scans is not an option. 
Basket--weaving in this manner is appropriate to recover signal up to the scale of the patch size, but information on larger scales is lost.

Ground emission is usually not an issue on short scans up to a couple of degrees, where a linear baseline subtraction is usually sufficient to remove it.
Moreover, the ground emission is more dependent on elevation (EL) than azimuth (AZ), one more reason to avoid standard orthogonal scans, where the scans would have an EL component. 
 
To address all these issues we used a scanning strategy based on fast, long azimuth scans. 
The scans were conducted at the elevation of the south celestial pole at Parkes (EL = $33.0^\circ$), and were made long enough to cover all declinations 
from Dec.$\sim -90^\circ$ to Dec.$\sim 0^\circ$.  
Scans were conducted  in a back--and--forth manner at a speed of 15 deg min$^{-1}$ with the telescope recording the position at full precision. 
That is a very fast rate for the Parkes telescope,  being 62.5\% of its AZ slewing speed (24 deg min$^{-1}$), and required a special drive system setup.

The Parkes telescope has an altazimuth mount and tracks the sky through the Master Equatorial (ME), a small equatorial mount system that easily tracks objects in Celestial coordinates. 
The ME is equipped with a laser whose beam illuminates a mirror on the telescope structure which is reflected back to laser light sensors on the ME.
The telescope drives are computer operated so that the laser beam is reflected back on the mid--point of those sensors. 
When this happens the telescope is locked to the ME and the sky position is known at the precision of the ME. 
When the lock is lost, the telescope position is unknown and the observation aborted.
This setup does not allow scans through the south Celestial pole that is a singular point for the ME equatorial mount.

\begin{figure}
	\includegraphics[width=\columnwidth]{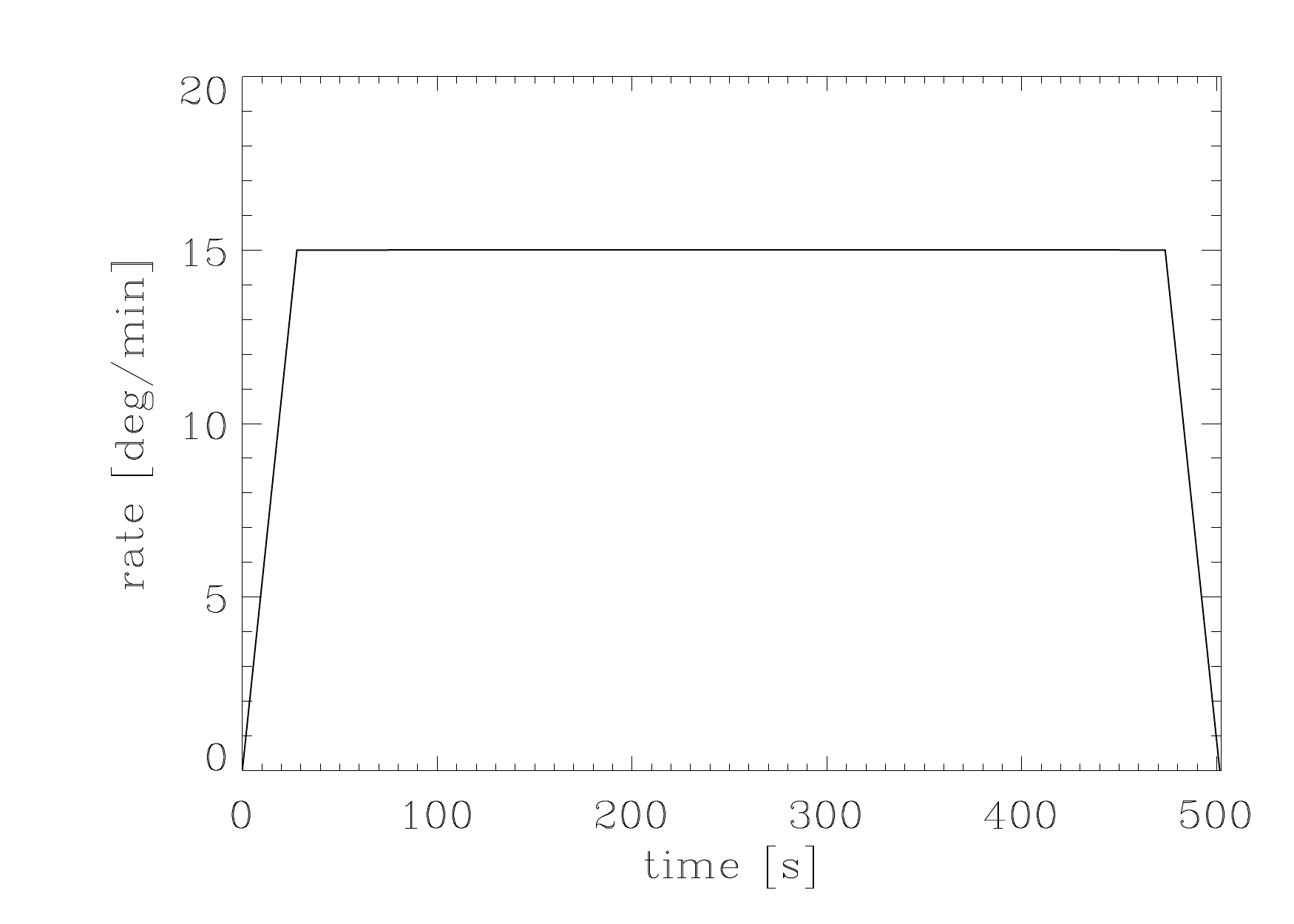}
	\includegraphics[width=\columnwidth]{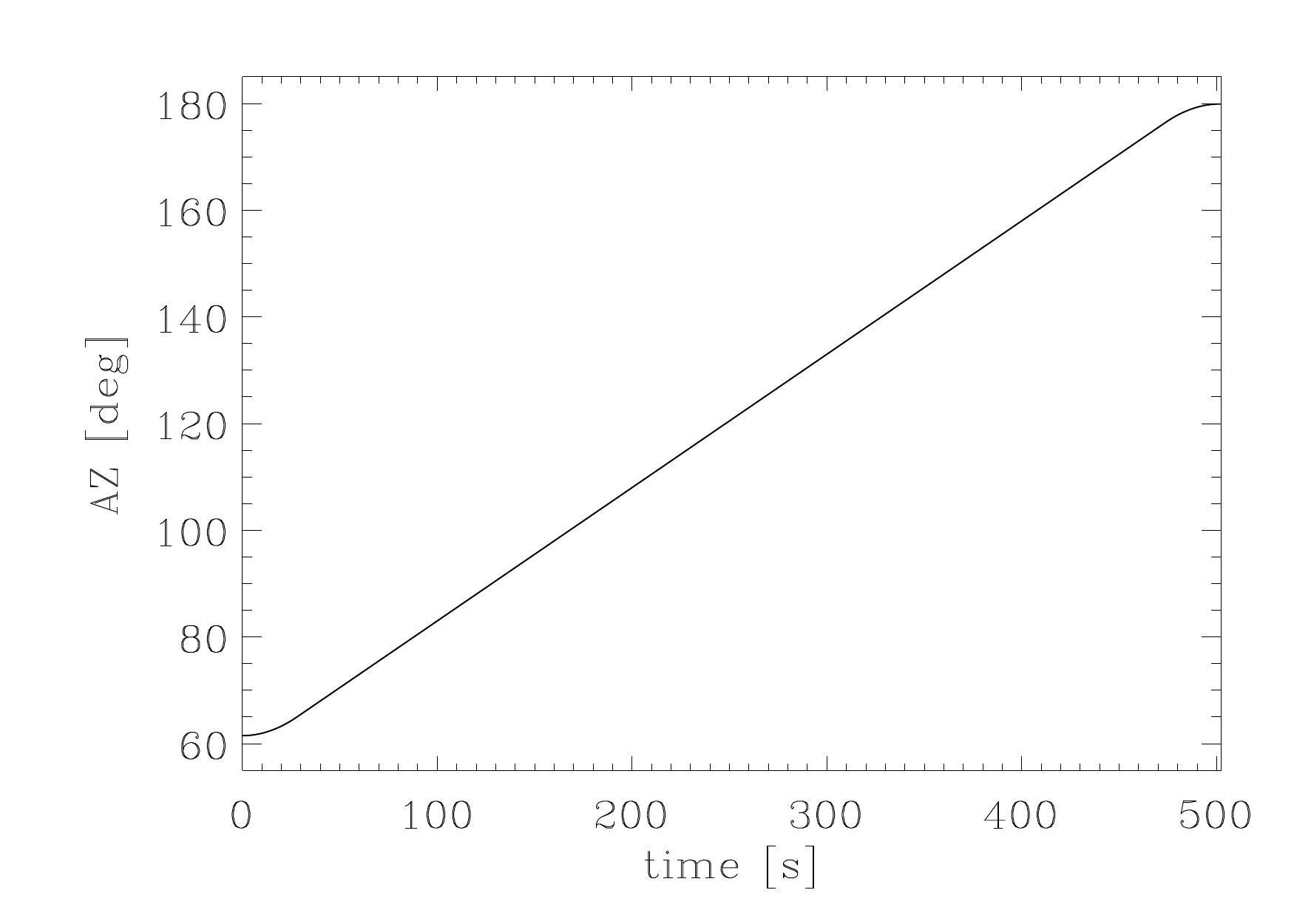}
    \caption{Scan rate (top) and azimuth behaviour (bottom) versus time for a forward eastward scan. The phase at cruise speed of 15$^\circ$/min is preceded and followed by a 28-s ramp up and down at constant acceleration.}
    \label{Fig:scanrate}
\end{figure}

Normally, scan rates of up to some 3--4 deg min$^{-1}$ are possible at the Parkes telescope. 
Faster rates tend to end up with the telescope's position tracking system losing lock soon after the scan starts, leading to observation abort. 
The fundamental problem here is the rate during the acceleration is too fast and the telescope cannot keep up. 
The drive system was thus modified  to have a gentle, constant acceleration ramp up until the nominal cruise speed was reached. 
The acceleration rate was 0.536 deg min$^{-1}$ s$^{-1}$, with the nominal speed being reached after 28~s. 
Tests showed that 15~deg~min$^{-1}$ was the fastest robust rate, without occasional lock loss. 
Indeed, no break--lock episode occurred during the entire project. 
The backend also acquired data during ramp up. 
The same was set for the end of the scan, with a gentle, constant deceleration ramp down preserving full position information. 
Figure~\ref{Fig:scanrate} shows the scan rate versus AZ for a full scan. 
The system was not able to acquire data while ramping down.

The AZ scan range was [61.5$^\circ$,  180.0$^\circ$] eastward, including the two acceleration and deceleration ramps, covering the Dec range [-90.0$^\circ$, +2.2$^\circ$]. The area around the south pole was acquired during backward scans that start from the AZ~=~180.0$^\circ$ end.
The westward scan range was limited by the telescope's south wrap AZ limit at 294.4$^\circ$ spanning  the range [180.0$^\circ$, 294$^\circ$], covering the Dec range  [-90.0$^\circ$, -0.6$^\circ$].

\begin{figure*}
	\includegraphics[scale=0.8, bb=80 30 450 280]{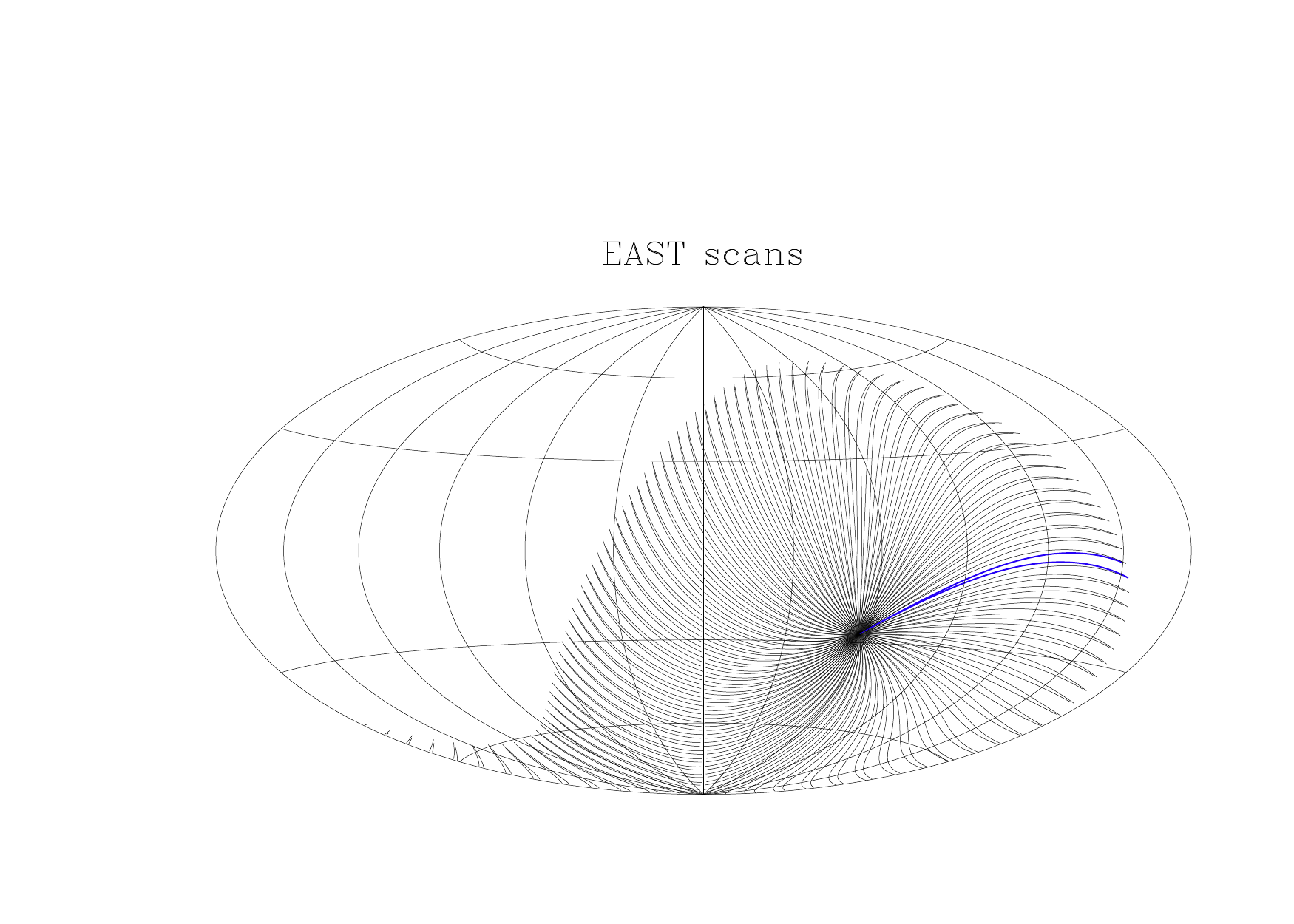}
	\includegraphics[scale=0.8, bb=80 30 450 280]{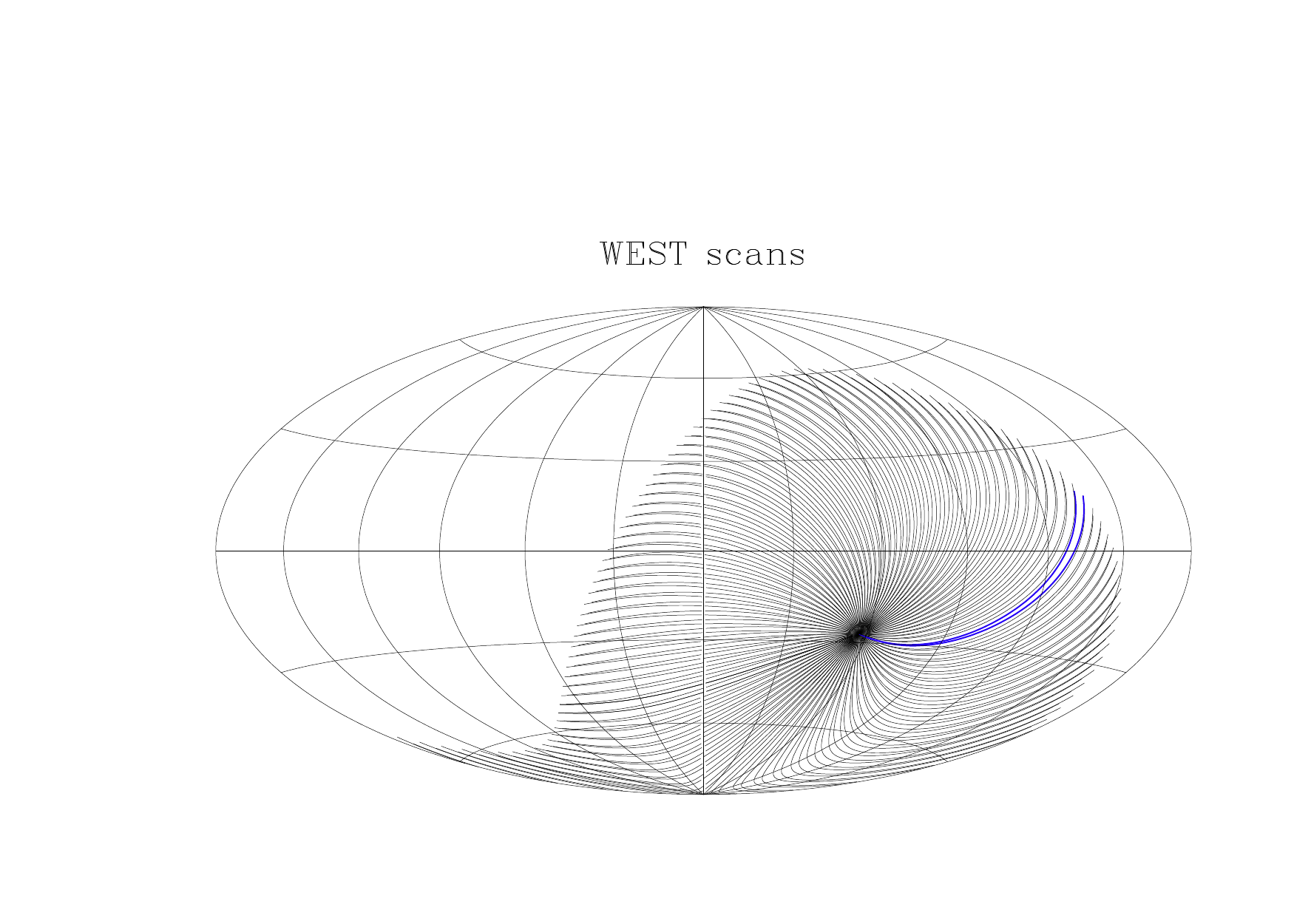}
	\includegraphics[scale=0.8, bb=80 30 450 280]{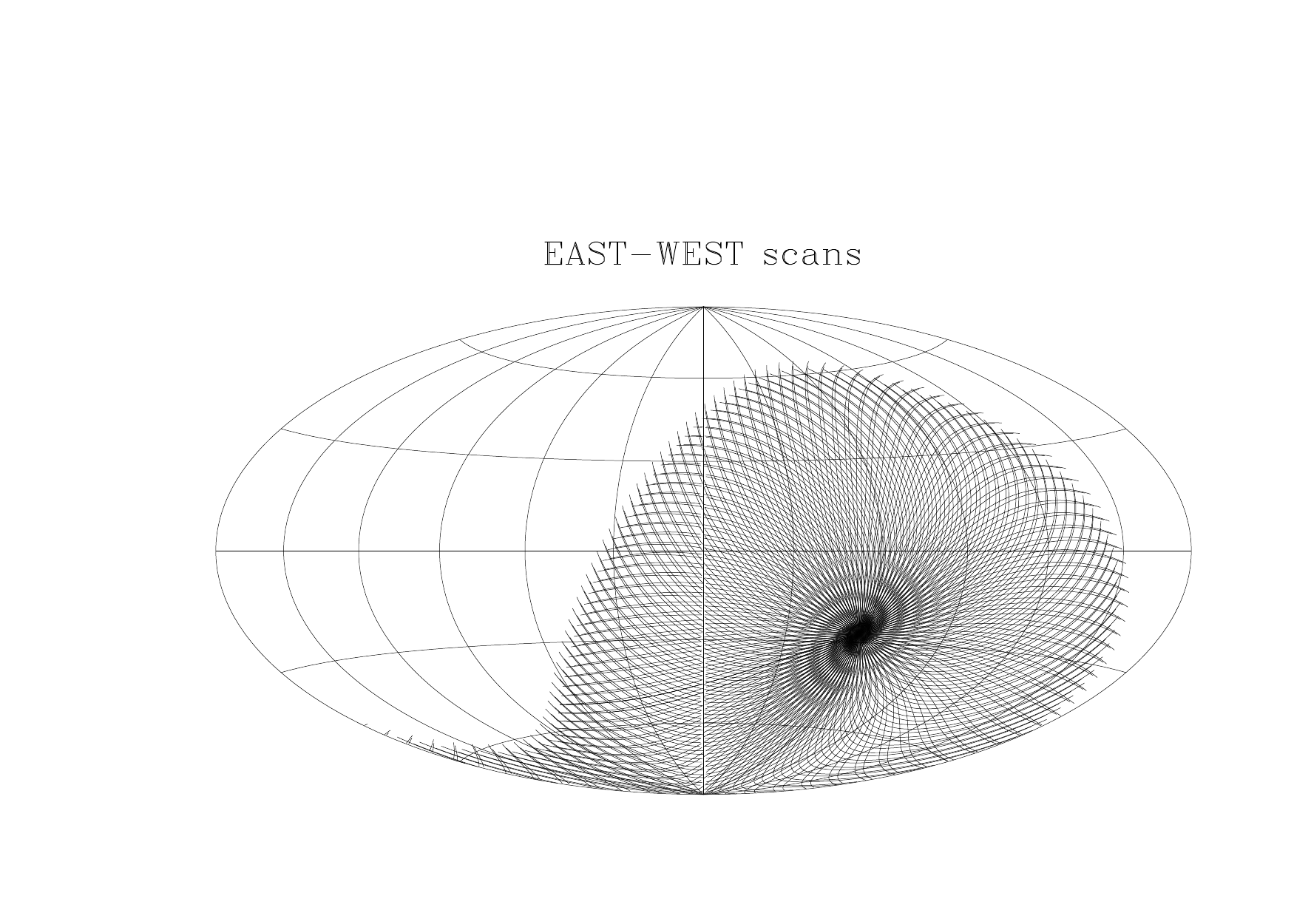}
    \caption{{\bf Top:} Full sequence of east scans as taken with an uninterrupted observation of 24~h sidereal time. Positions are shown in Galactic coordinates in Aitoff projection with longitude 0 at the centre, north up, and west to the right. Note that the loop is closed so, were the observation to continue for another 24~h, the entire sequence would repeat exactly the sequence. A single back--and--forth scan pair is highlighted in blue. {\bf Mid:} As for top panel except for west scans. {\bf Bottom:} The two full sequences of east and west scans are plotted together to show how they cross each other. }
    \label{Fig:scans}
\end{figure*}

Earth rotation was used to cover the entire 24-h R.A. range. 
The sky rotates during each scan, so that each back-and-forth pair does not repeat the same track in the sky advancing in R.A.. 
Each night the sky coverage was a zig--zag in the sky (Figure~\ref{Fig:scans}).  
The scans were conducted both eastward and westward when the sky rises and sets to have scans along two different directions and realise a basket--weave pattern (see Figure~\ref{Fig:scans}). 

With AZ scans at a fixed EL, there is no capability to arbitrarily scan a specific position in the sky at any time, but only once a day.
Thus, to realise a regular grid in the sky, scans need to have well defined geometry, duration, and speed, 
and start at the desired Local Sidereal Time (LST).  That makes each scan repeatable and scans can be spaced by the desired amount,
even though each scan can be executed only once a day.

Each night a sequence of back--and--forth scans was performed. 
To fill the sky regularly with scan sequences taken on different nights, an appropriate waiting time was left at the end of each scan, so that an integer number of  back--and--forth scans fitted into 24h of LST and the sequence of scans was a closed loop (Figure~\ref{Fig:scans}). 
Each night a different sequence was observed with exactly the same geometry except an offset in RA, that is, an offset in the LST between start times. 
Scan separation depends on declination, smallest at the south pole (all scans get there), largest at Dec~$= -18.4^\circ$ (Figure~\ref{Fig:scans}). 

\begin{figure}
	\includegraphics[width=\columnwidth, viewport=100 420 600 800]{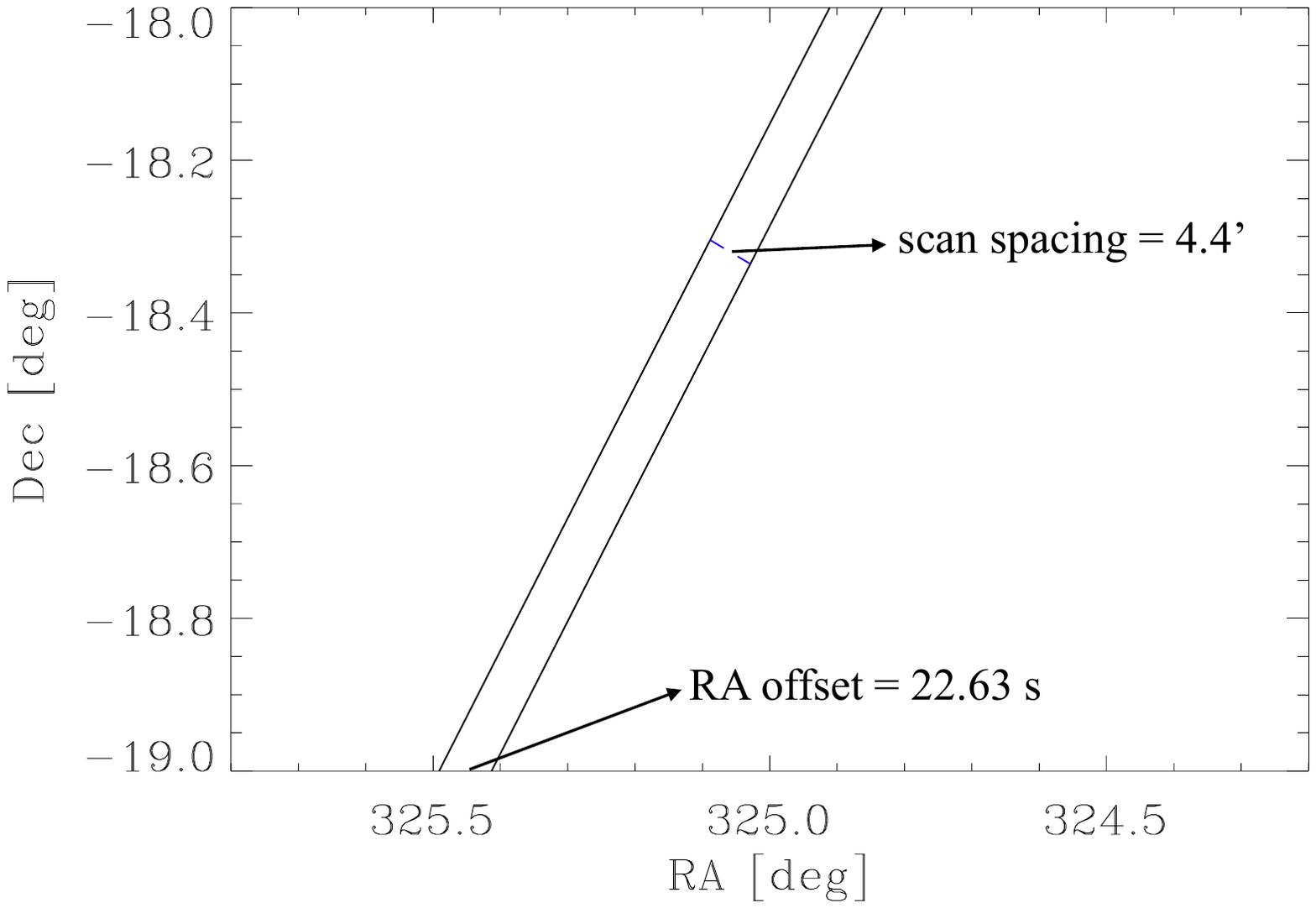}
    \caption{Portion of two east scans of two different zig-zag sequences spaced by just one RA/LST offset unit $d_{\rm RA_{east}} = 21.63$~s, realising
                  a spacing of 4.38' at the Dec. where the spacing is largest (Dec $\sim -18.4^\circ$).}
    \label{Fig:spacing}
\end{figure}

For east scans, an RA/LST offset of $d_{\rm RA_{east}} = 21.6270$~s was used, for a maximum scan spacing of 4.38 arcmin at Dec~$= -18.4^\circ$ (Figure~\ref{Fig:spacing}). 
The maximum spacing was chosen to be smaller than half a beam size, ensuring Nyquist sampling and making all sequences equally spaced. 
Along with the closed loop of each zig-zag sequence, this gave a regular sky coverage with all scans equally spaced. 
In detail,  47 zig-zag sequences of 85 back-and-forth scan pairs each were observed, for a total of 3995 scans in each heading.

Since east scans alone are sufficient to full sample the sky, west scans  were performed mainly to ensure scan crossing and basket-weaving for efficient map-making. 
To save observing time in this direction we sampled the sky with a spacing of only one scan per beam. 
RA/LST offset between successive scans was set to  
$d_{\rm RA_{west}} = 42.2081$~s, for a scan spacing of 8.67~arcmin, resulting in 24 zig--zag sequences of 89 back-and-forth scan pairs, for 2136 scans in each heading.
Note that the number of west scans was slightly more than half of the number of east scans. 
This was in order to realise the regular pattern  described above.  

Given such a scanning strategy, Solar System objects will be observed.  
Therefore, all data closer than 60~arcmin from the Moon and 10~arcmin from Jupiter and Mars were excluded.  
Considering the intensity of the first sidelobe, we estimated a worst case contamination of some 1~mK in Stokes I, negligible compared to the confusion limit, and more than one order of magnitude smaller in polarization, which is negligible compared to the polarization sensitivity.

\section{Map--making}\label{Sec:mapmaking}

\subsection{Method}

Basket--weaving techniques are applied to build maps and combine scans taken along crossed scans. 
Here we use the algorithm of \citet{Emerson88}. 
Basket-weaving, however, can recover emission only up to the size scale  of the map. 
The baseline of each scan is usually estimated and removed, and the average signal on the map area is lost.

\begin{figure}
	\includegraphics[width=\columnwidth]{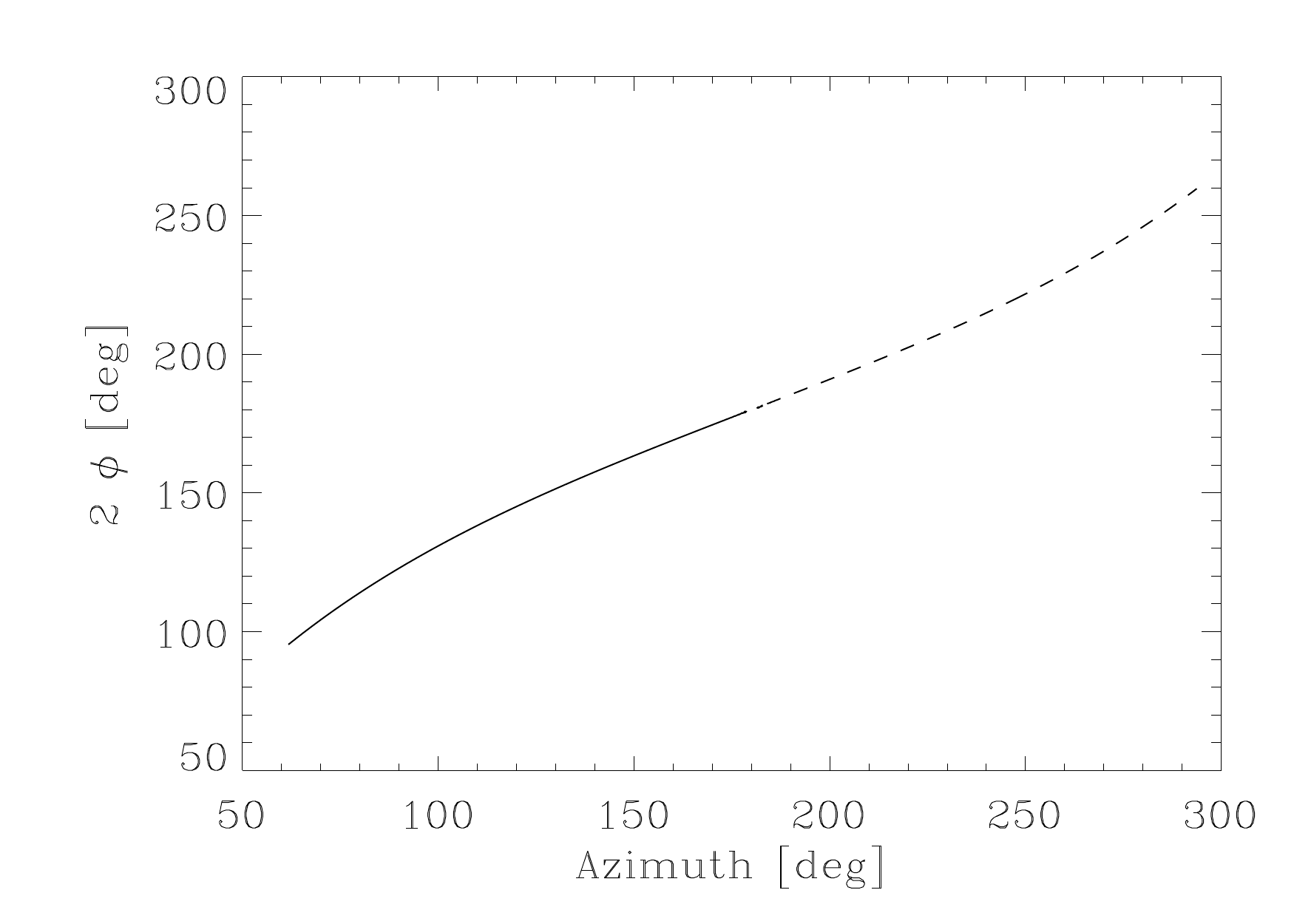}
	\includegraphics[width=\columnwidth]{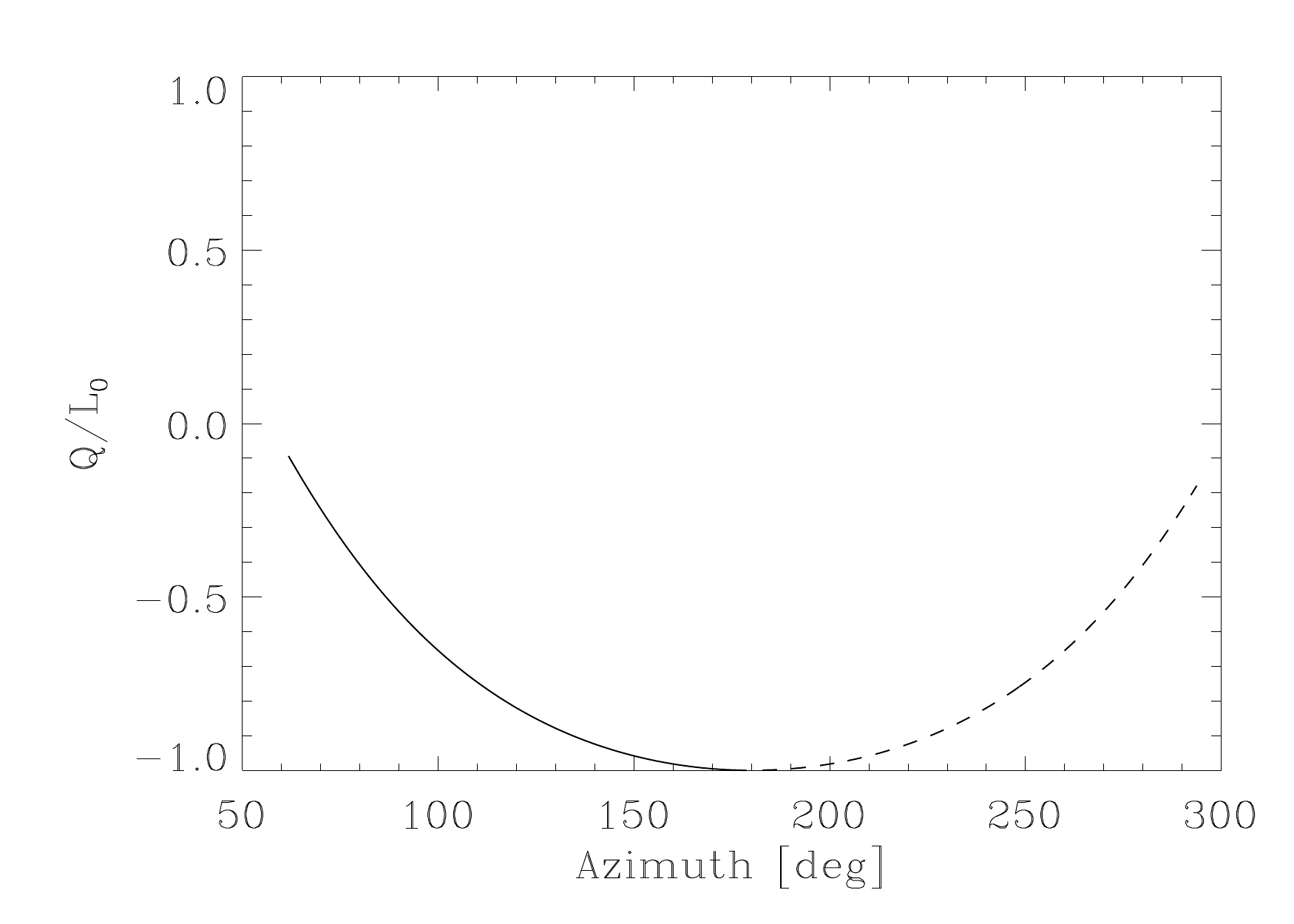}
	\includegraphics[width=\columnwidth]{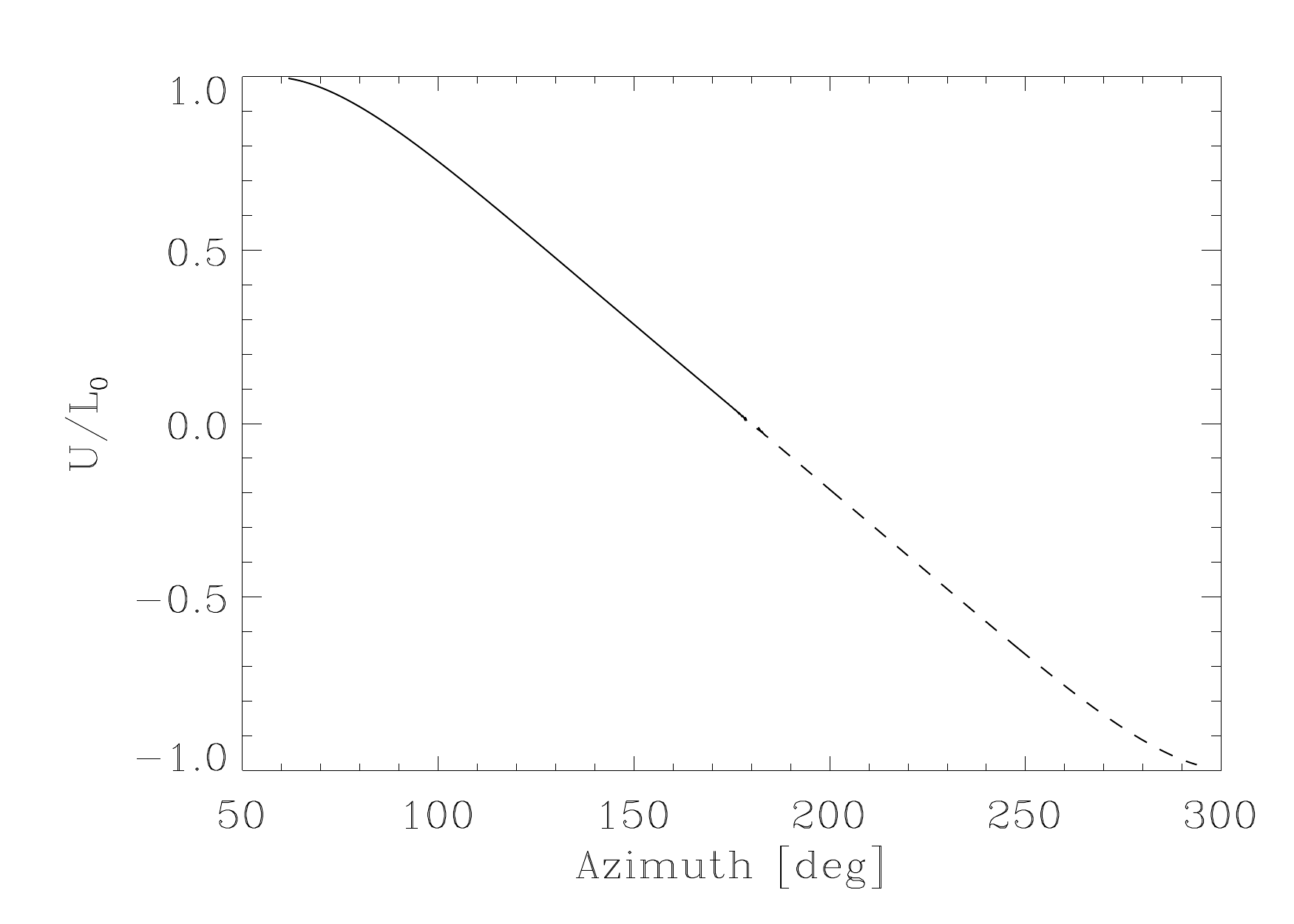}
    \caption{{\it Top}: range covered in the east (solid) and west (dashed)   scans of S-PASS of $2 \times$ the parallactic angle ($2 \phi$), which is the angle the 
                  two Stokes parameters 
                  $Q$ and $U$ are modulated with. {\it Mid}: Example of Stokes~$Q$ behaviour along a east (solid) and west scan (dashed) modulated by the variation 
                  of $\phi$ of the top panel. Stokes $Q$ is normalised to the polarized emission amplitude $L_0$.  The case of emission polarization angle 
                  $\theta_0 = 0^\circ$ is shown. {\it Bottom}: as for mid panel 
                   except for Stokes~$U$.}
    \label{PA:Fig}
\end{figure}

In the context of S-PASS this applies only to the Stokes~$I$ signal (which  is not modulated during a scan). 
The polarized components, Stokes $Q$ and $U$, are different: 
For these, the  variation of the parallactic angle $\phi$ in a long azimuth scan modulates a constant polarized signal by a sinusoidal function in the instrument reference frame as:
\begin{eqnarray}
  Q &=& L_0 \cos{[2  (\theta_0 + \phi)]}\, , \\
  U &= &L_0 \sin{[2 (\theta_0 + \phi)]}\, ,
\end{eqnarray}
 where $L_0$ and $\theta_0$ are the amplitude and polarization angle of a constant polarized signal. 
 Figure~\ref{PA:Fig} shows the variation of the modulation angle $2\phi$ in S-PASS east and west scans, as well as an 
 example of how $Q$ and $U$ behave.  The large range spanned by $2\phi$ -- some $90^\circ$ for each east or west scan, or $180^\circ$ combined -- introduces a large modulation of Stokes $Q$ and $U$ in the instrument reference frame, and a baseline subtraction does not cancel a constant sky signal. 
Because short scans would produce little modulation,  long azimuth scans are essential for this technique to be effective.

To recover the full constant signal of each scan, an unknown constant  offset (in the  instrument reference frame) is added to each scan, for either Stokes $Q$ and $U$.
This is also called destriping, because the lack of the correct offset on each scans would led to striped maps if made just binning the data on same pixels.
Scans cross each other in several points where each are required to have the same sky signal. 
The solution is found with a maximum likelihood procedure. 
The best offset parameter set is that which minimises the sum of all the squared differences between scans at their crossing points. 
The least square criterion is justified given the Gaussian distribution of the errors on Stokes $Q$ and $U$. 
The system of equations to solve are reported in Appendix~\ref{Sec:appA}. 

The system to solve is computationally challenging when all scans are considered: given 12,000 scans,  24,000 unknown offsets must be solved for, 
requiring some two months' CPU time for each frequency channel.
However, given the procedure is mainly useful in recovering the large-scale emission, it is not, in general, necessary to work  at full resolution.
Indeed,  binning the data in a smaller number of wider scans and solving for the consequently coarser resolution map produces an adequate estimate of the large-scale emission.

After offset estimation, the \citet{Emerson88} basket-weaving technique is applied. This is performed in Fourier space and requires two maps taken along two crossing directions. After solving for the offset at coarse resolution, two maps with east and west scans only are generated and combined together. The two maps are generated in equatorial coordinates using a cylindrical (Carr\'ee) projection, which gives the rectangularly shaped images required by the technique. 
         The technique is usually employed using straight scans, whose Fourier transforms are straight lines, allowing simple  and effective Fourier filtering. However, our scans are not straight lines. Their Fourier space images are quite complex and spread over almost the entire Fourier space, making it harder to set up dedicated filters. East and west scans were therefore approximated with linear fits to set the Emerson and Gr\"ave Fourier filters.
The coarse resolution maps so obtained were then converted to Galactic coordinates.
\begin{figure}
	\includegraphics[width=\columnwidth, viewport=20 350 580 790]{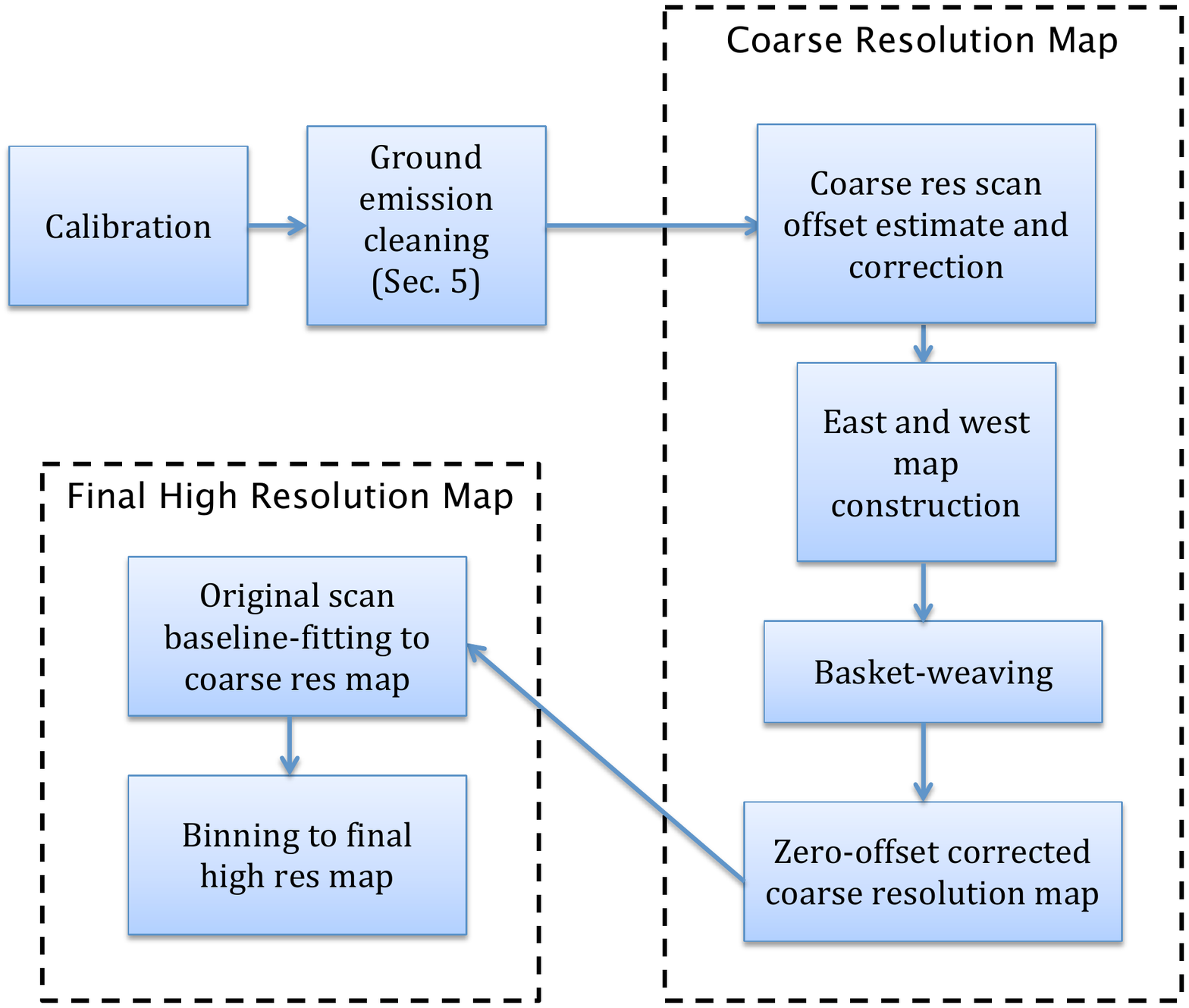}
   \caption{ Block diagram of the map-making procedure.}
\label{Fig:cartoon}
\end{figure}

The final, full resolution map is then generated using the coarse resolution, absolutely calibrated map from which the correct baseline for each scan can be determined. 
The baseline is obtained using a running median of the difference between each scan and the absolutely calibrated map.
Once corrected, the data from the original scans are binned in the final map at full resolution.
The polarization vectors were all parallel transported to the centres of the nearest map pixel in order to avoid false depolarization due to the variation of the reference direction across each pixel. The effect increases with decreasing distance to the pole. We apply this to all pixels, regardless of their distance from the pole.
The HEALPix pixelation~\citep{Gorski05} in Galactic coordinates was used with parameter $nside = 1024$, corresponding to pixels of $\theta_{\rm px} = 3.4$~arcmin, which ensures adequate sampling of the beam. HEALPix pixels are equal area and isolatitude. 
\\
\\
In summary, the steps of the procedure to obtain the final maps at full resolution are (see also the diagram of Figure~\ref{Fig:cartoon}):
\begin{enumerate}
  \item{} Scans are binned in $0.5^\circ$ wide scans. Data are also binned along a scan to $0.5^\circ$ pixels;
  \item{} The equation system determining the optimal set of offset values is solved (Appendix~\ref{Sec:appA}) for these coarse resolution scans;
  \item{} Offset--corrected scans are used to generate two maps generated from data taken in different scan directions, i.e.\ east and west;
  \item{} East and west maps are combined with the technique of~\citet{Emerson88}  which 
             gives the final zero--offset calibrated map at coarse resolution; 
  \item{} The baseline for each original scan is corrected by matching to the coarse resolution zero-offset calibrated map;
  \item{} The final maps at full resolution are obtained from the offset-corrected scans binned using a HEALPix projection with a pixel size of 3.4~arcmin ($nside = 1024$).   
\end{enumerate}

Stokes $I$ maps are obtained in a similar manner, with the equations to estimate the scan offsets defined in Appendix~\ref{Sec:appB}.
However, the map-making procedure and scanning strategy were optimised for $Q$ and $U$ only.
Stokes $I$ is not modulated, so the output map has the overall mean value undetermined.

Instead, this is recovered using archival, absolutely calibrated observations of the south celestial pole (SCP) taken at 2.0~GHz by \citet{Bersanelli94} with a horn radiometer of FWHM$_{\rm h} = 22^\circ$ at the South Pole Station pointing to the zenith. These authors measured a Galactic component of 
\begin{equation}
T^{\rm SCP, 22^\circ}_{2.0} = 325\pm 100\, {\rm mK}.
\end{equation}
Scaling with a brightness temperature spectral index of $\beta = -2.7$ and assuming a spectral index uncertainty of $\sigma_\beta = 0.2$ (1-$\sigma$) we get an estimate of the emission at 2.3~GHz of  
\begin{equation}
T^{\rm SCP, 22^\circ}_{2.3} = 220\pm 70\, {\rm mK},
\label{eq:offset_I}
\end{equation}
including measurement and spectral index errors. 

The Stokes~$I$ map obtained with the map--making procedure is smoothed to a FWHM$_{\rm h} = 22^\circ$ and the difference between the value of Equation \eqref{eq:offset_I} and that of the smoothed map at Dec~=~$-90^\circ$ is then added to the unsmoothed map to offset it appropriately.
This delivers a Stokes~$I$ map absolutely calibrated for the Galactic emission. (Note that CMB emission has not been included.) Proceeding this way the error of 70~mK applies to the mean level only. For all other scales the error is smaller and dominated by the statistical noise or the confusion limit (9~mK).

\subsection{Tests: Q and U}\label{Sec:testQU}

\begin{figure*}
	\includegraphics[angle=90, width=\columnwidth]{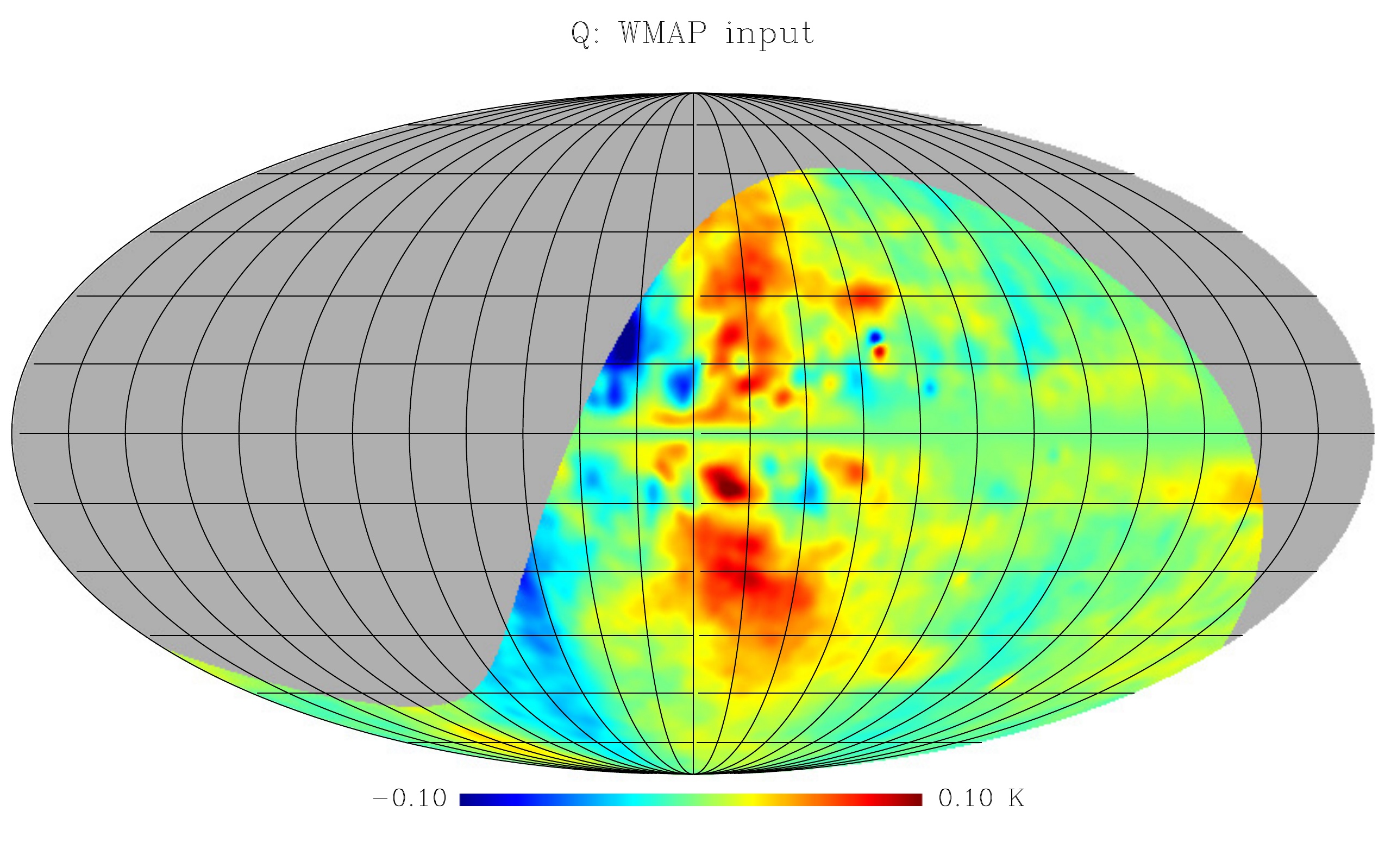}
	\includegraphics[angle=90, width=\columnwidth]{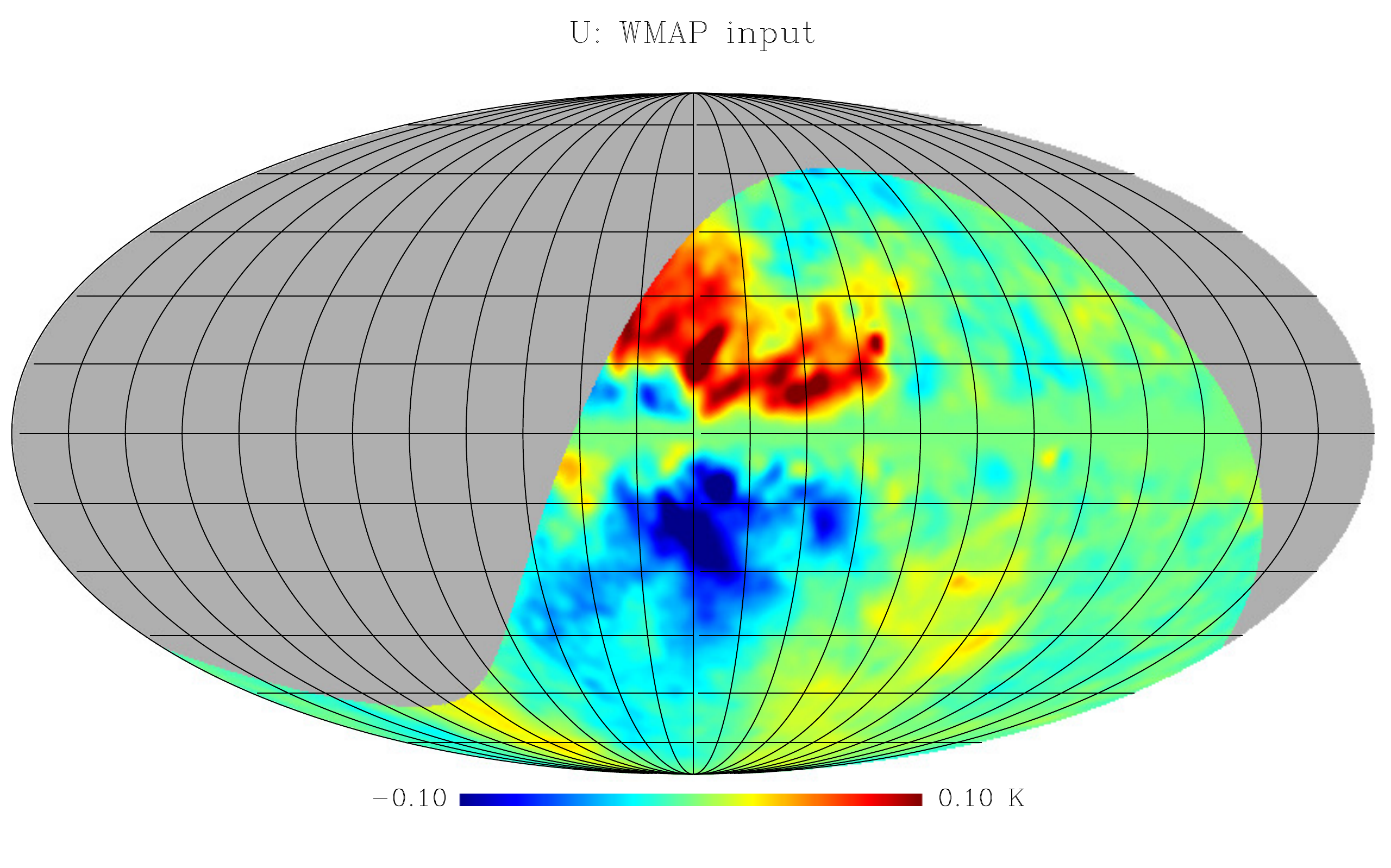}
	\includegraphics[angle=90, width=\columnwidth]{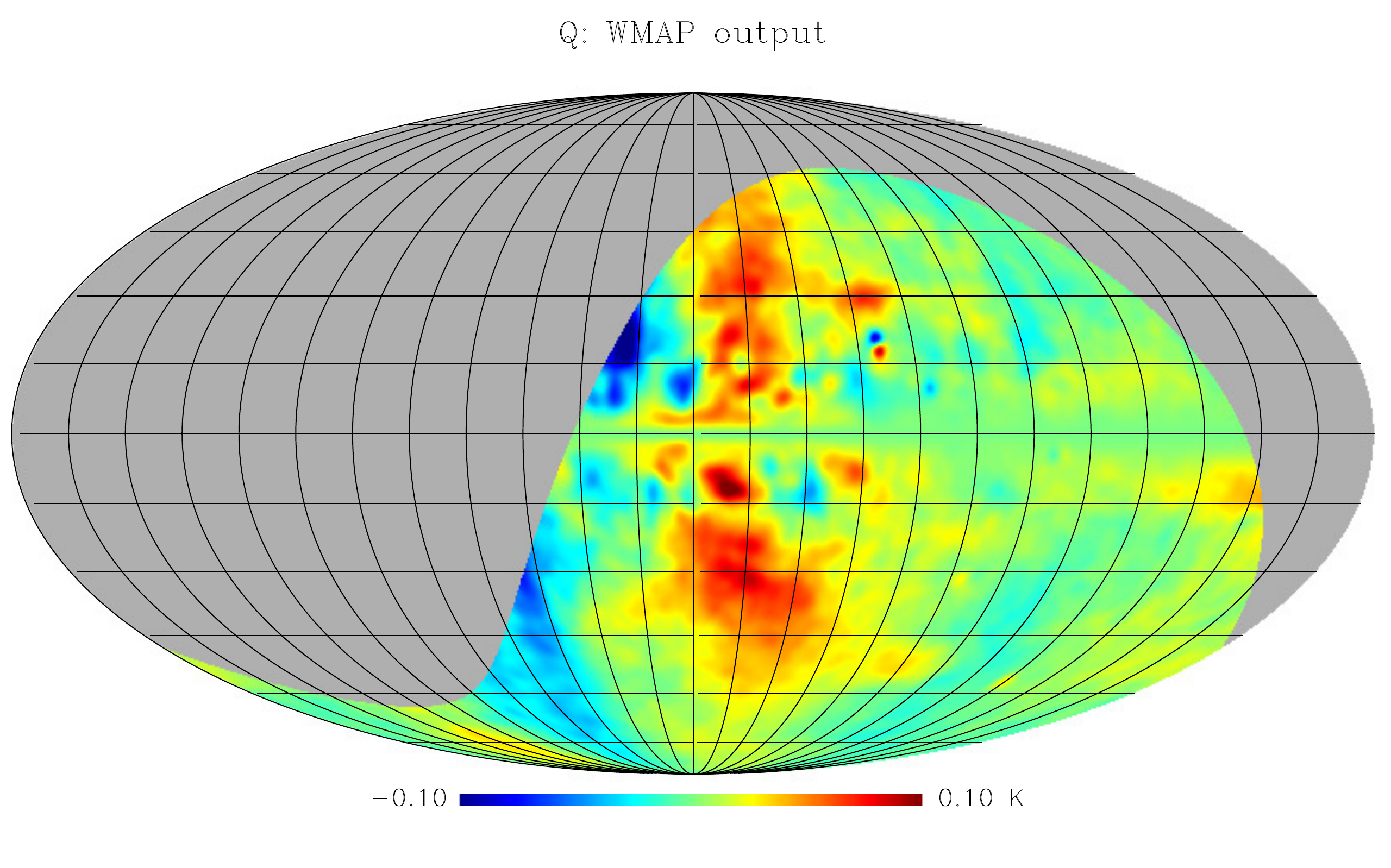}
	\includegraphics[angle=90, width=\columnwidth]{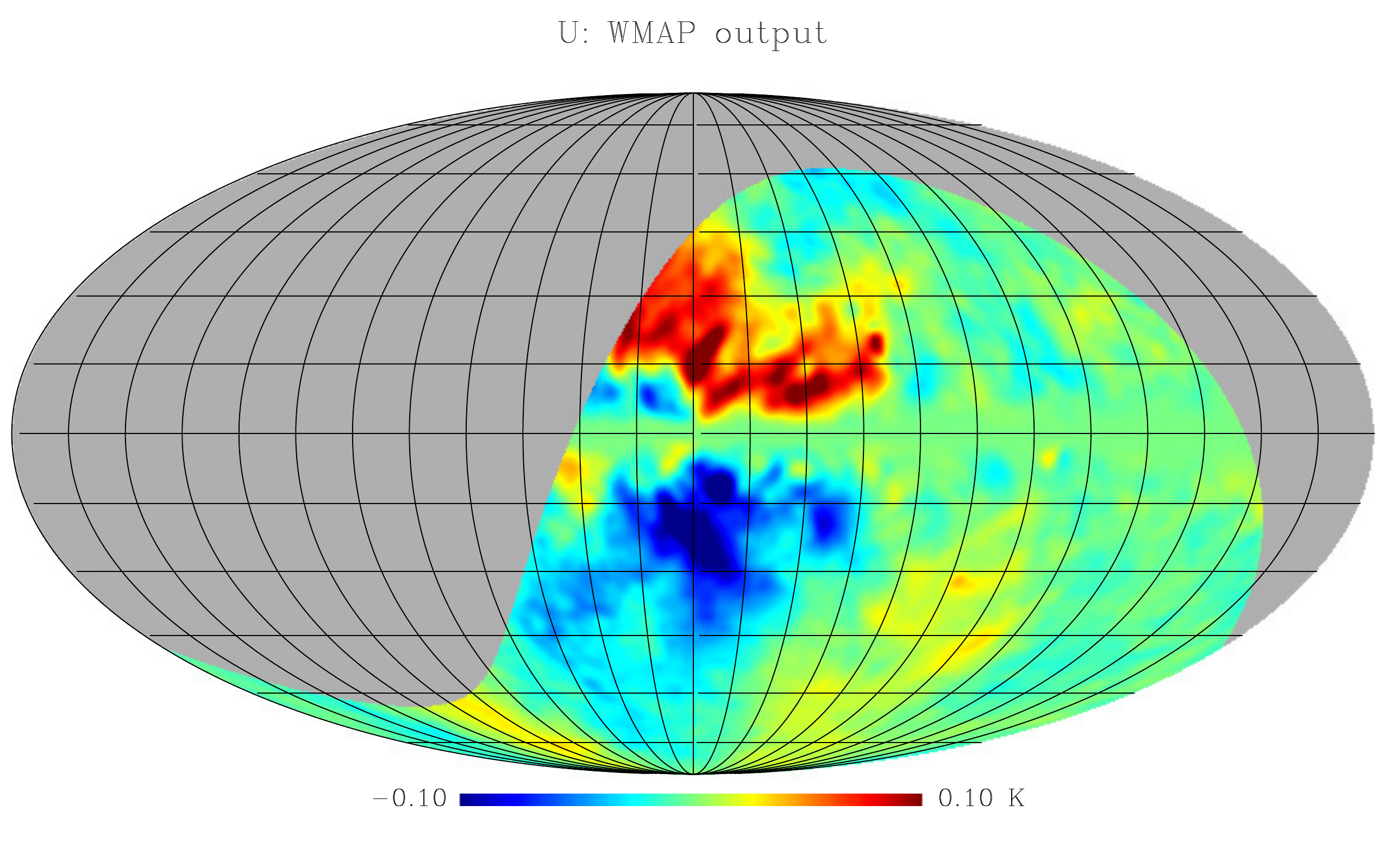}
    \caption{{\it Top}: Stokes $Q$ (left) and $U$ map (right) used as input maps for the simulation to 
	                       test the capability of the map-making procedure to recover the real sky. The images are in Galactic coordinates centred at the Galactic Centre, with 
	                       gridlines spaced by 15$^\circ$. Note that the IAU convention of polarization angle is used here, 
	                       and the Stokes $U$ map shown here has the sign changed 
	                       compared to the original WMAP data set. 
	          {\it Bottom}: As for the top panels, but for the output maps resulting from the map-making procedure applied to the simulated observations extracted 
	                        from the input maps.} 
    \label{Fig:testwmapQU}
\end{figure*}

To test the map-making procedure we performed a simulation with a realistic input $Q$ and $U$ maps.
Observed scans were extracted from the input maps and Gaussian noise added to each sample with the same 
rms as observed (assuming the noise measured for the entire useful band). 
The baseline was then subtracted from each scan 
to mimic the process in a normal data reduction and then the entire map-making procedure was applied. 
The baseline removing procedure excludes strong compact sources and areas with strong diffuse emission to avoid strong deviations of individual scans. 
Baseline removal was performed by an automatic procedure that runs the baseline fitting once, flags outliers, and then repeats the fit. 
This was repeated a few times with decreasing outlier threshold to ensure that only the strongest sources  were flagged in first iteration, with the threshold being progressively refined to flag weaker compact sources.

The output maps were then compared to the input ones to assess how well it was reconstructed.
 We used the same HEALPix pixelation as for the S-PASS maps.

For input maps for the simulation, we used the WMAP polarization maps at 23~GHz \citep{Bennett13}, which contain emission that is mostly Galactic synchrotron. 
For a realistic signal level, the 23~GHz amplitude was scaled to 2.3~GHz with a brightness 
temperature spectral index of $\beta_{\rm scal} = -3.2$, the typical slope at high Galactic latitudes \citep{Carretti10}. 
Mid and low Galactic latitudes have flatter indexes, ensuring the map presents a worst-case scenario.

Faraday depolarization is insignificant even on the  Galactic plane at 23~GHz, so the WMAP sky reveals strong disc emission in the inner Galaxy.
As we will see later, the emission at 2.3~GHz is strongly depolarized there, making the signal much weaker. 
To account for this, we apodised the 
WMAP map in the Galactic plane with a Hanning filter of 15$^\circ$ width. 
This effectively weakens the signal within a few degrees around the Galactic Plane, mimicking a more realistic measurement.
The resulting maps are shown in Figure~\ref{Fig:testwmapQU}, top panels.

The mean signal in the input maps is $\overline{Q}_{\rm WMAP} = 8.8$~mK and  $\overline{U}_{\rm WMAP} = -3.2$~mK for Stokes $Q$ and $U$, respectively. 
These are the mean signals (or offsets) that the map-making procedure must be able to reconstruct to recover the absolutely calibrated signal. 
Since $Q$ and $U$ are signed quantities, the mean measures the offset, but does not measure the typical intensity of the signal. 
This can be measured by the rms values of  $Q_{\rm rms, WMAP} = 24$~mK and $U_{\rm rms, WMAP} = 29$~mK, or  by the mean polarized intensity $\overline{L}_{\rm WMAP} = 28$~mK. 

Simulated observational data were generated as described above and then the map-making procedure was applied. The output maps (Figure~\ref{Fig:testwmapQU}, bottom panels) are nearly identical to the input ones.
We find the rms of the difference between the two maps is 2.26~mK on pixel scales (3.4~arcmin) and 0.81~mK on the beamsize scales (FWHM= 8.9'), consistent with the expectation from instrumental noise alone. 
The error from the map-making procedure is thus not adding significantly to the error budget. 
\begin{figure*}
	\includegraphics[angle=90, width=\columnwidth]{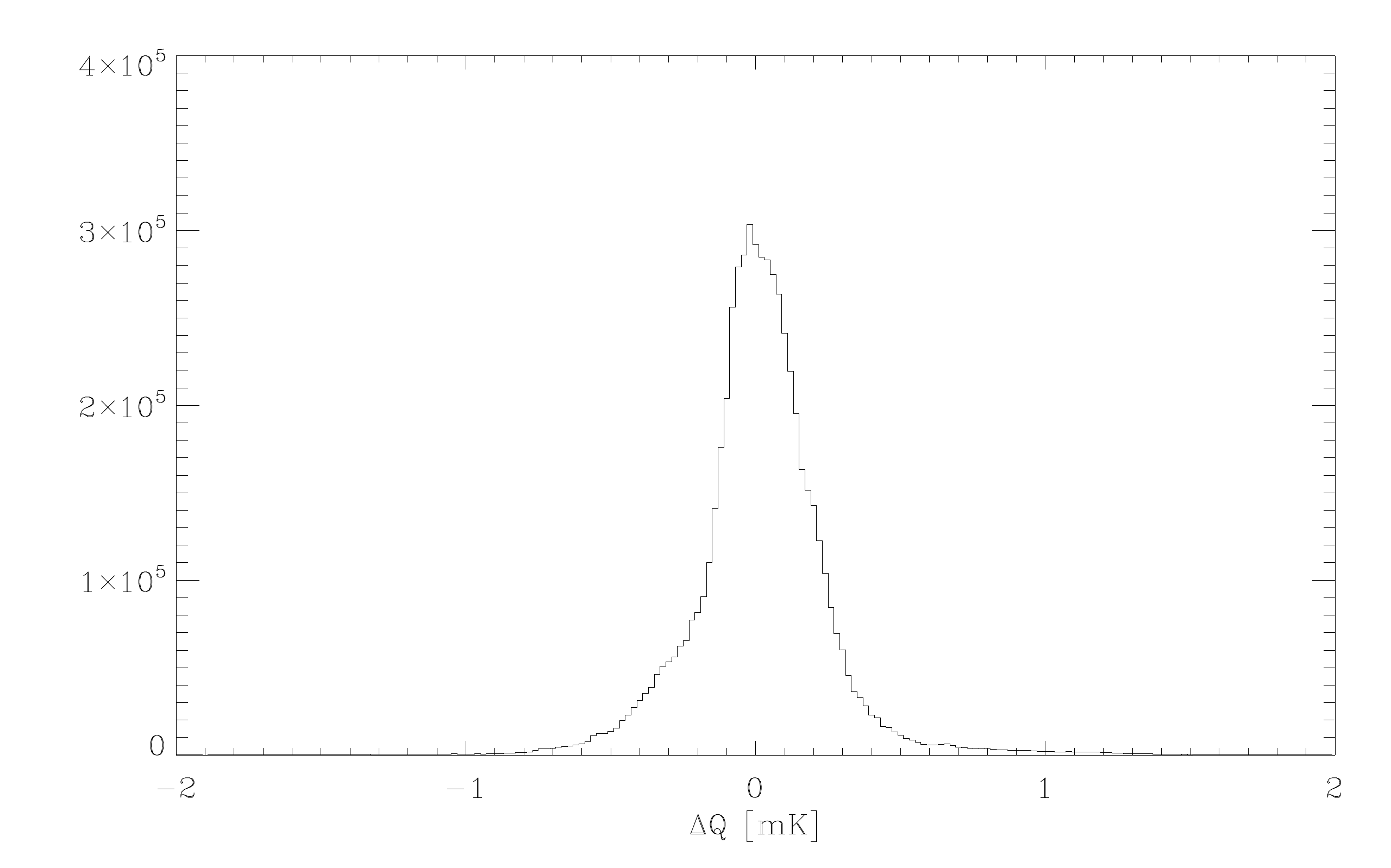}
	\includegraphics[angle=90, width=\columnwidth]{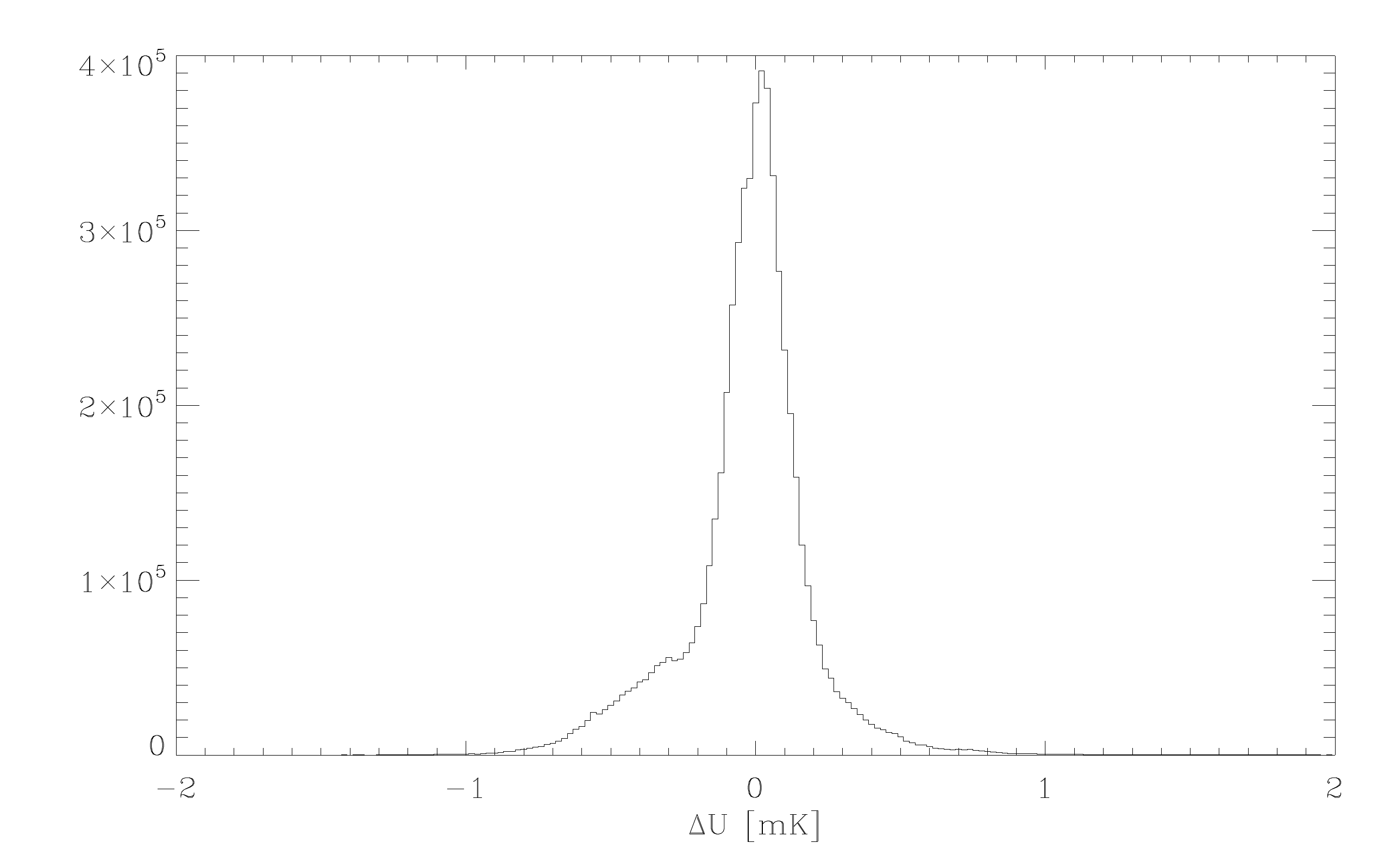}
    \caption{Distribution of the difference between input and reconstructed maps in the case with no instrumental noise from which one may estimate the scale of errors arising solely from the  map-making procedure;  Stokes $Q$ (left) and $U$ (right).}
    \label{Fig:testwmapHistNoNoiseQU}
\end{figure*}
\begin{figure*}
	\includegraphics[angle=90, width=\columnwidth]{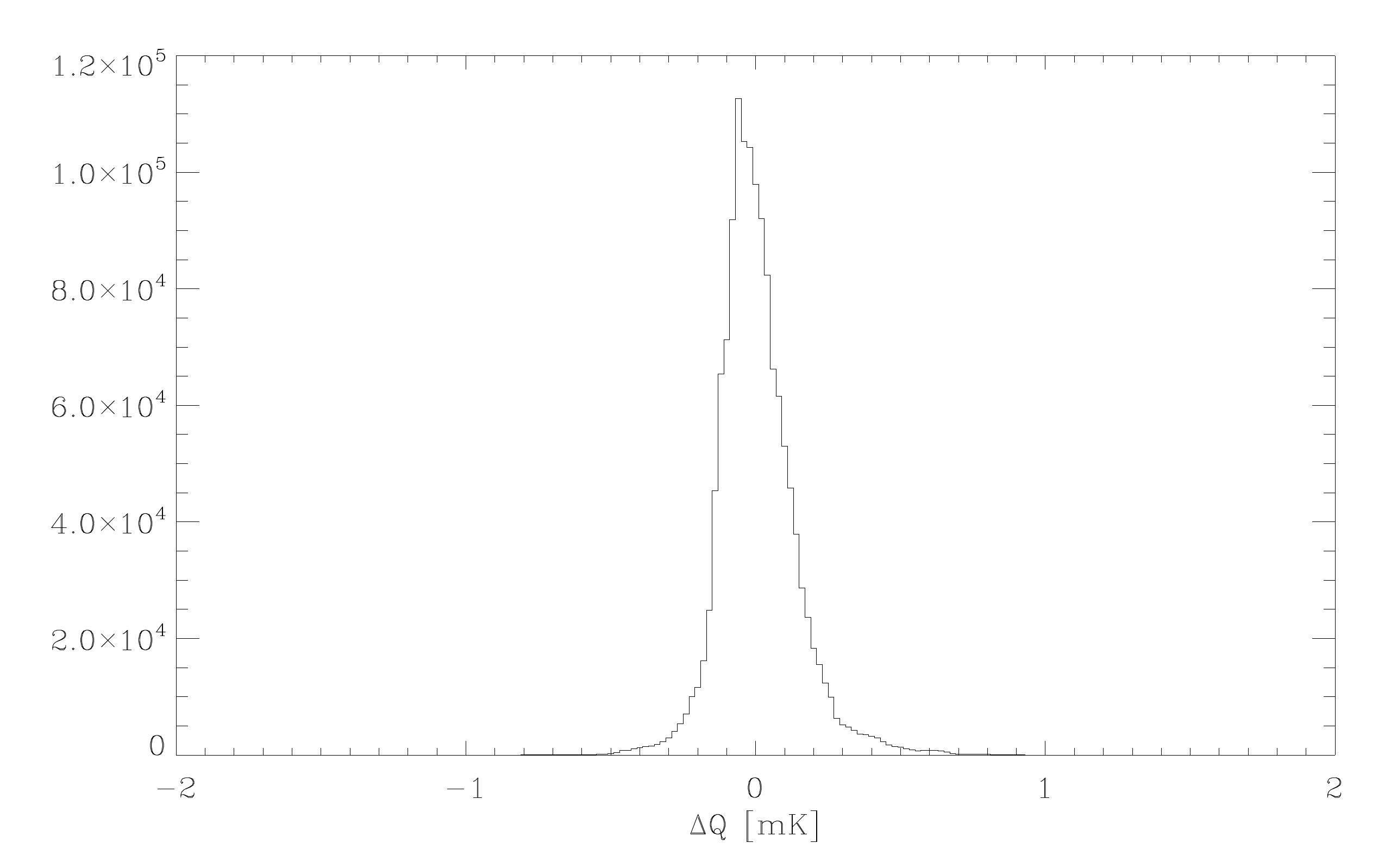}
	\includegraphics[angle=90, width=\columnwidth]{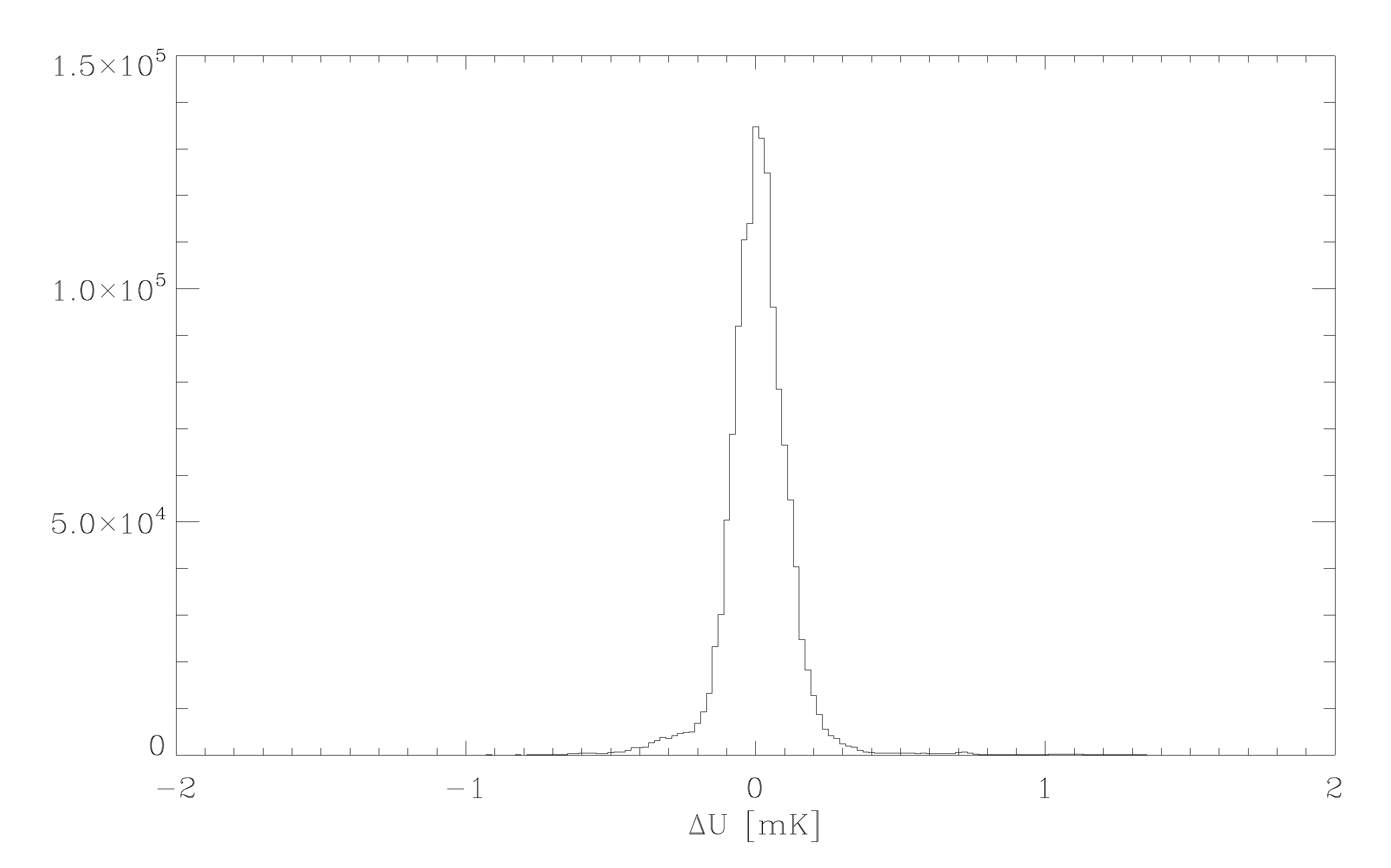}
    \caption{As for Figure~\ref{Fig:testwmapHistNoNoiseQU} except only for areas with polarized emission $L < 10$~mK. This restriction allows for a more accurate
	estimate of the performance of the map-making procedure  in low emission regions.}
    \label{Fig:testwmapHistNoNoise10QU}
\end{figure*}

The mean values of the difference maps measure how well the mean values of the input maps are recovered (i.e., the zero-offset calibration). 
We find they 
are $\overline{\Delta Q}_{\rm WMAP} = 0.021$~mK and $\overline{\Delta U}_{\rm WMAP} = 0.017$~mK, consistent with zero within 2-$\sigma$ (the error on the mean with our instrumental noise is 0.010~mK).  
The  mean emission (offset) is then recovered with a precision better than 1\% (0.24\% and 0.53\% for Stokes $Q$ and $U$, respectively). 
The map-making procedure thus recovers the input maps correctly.
Indeed, even   the mean emission  is recovered with high precision.  

To estimate the error contributed by the map-making procedure alone, a simulation with no noise was also made. 
The output maps are essentially identical. 
The histograms of the differences between output and input maps are shown in Figure~\ref{Fig:testwmapHistNoNoiseQU}. 
The distribution is symmetric at the centre, but shows some asymmetry in the wings. 
The 16\% and 84\% cumulative distribution limits are ($-0.18, +0.17$)~mK and ($-0.21, +0.10$)~mK  for $Q$ and $U$, respectively. 
The uncertainty in the map-making procedure is thus much smaller than both the typical signal in the maps (by 3 orders of magnitude) and the pixel noise (by more than a factor of 10), further confirming that its contribution to the error budget is negligible.

The mean values of the differences are  0.023~mK and 0.012~mK for $Q$ and $U$, similar to the case with instrumental noise included. 
Fractional errors are 0.26\% and 0.38\%.

For  very faint emission regions, Figure~\ref{Fig:testwmapHistNoNoise10QU} shows the distribution of the differences only for pixels with  $L < 10$~mK. 
This distribution is more symmetric and approximately half as wide as the general case, with  16\% and 84\% cumulative distribution limits of ($-0.09, +0.09$)~mK and ($-0.12, +0.10$)~mK  for $Q$ and $U$, respectively. 

All the above indicate excellent performance. 
In the observational noise case, the error distribution is consistent with the instrumental noise, with negligible contribution from the map-making procedure. 
The no--noise case suggests that the analysis error is approximately 0.1~mK in the low emission regions where $L < 10$~mK, 23~times weaker than the pixel sensitivity and 8~times better than the sensitivity in beam-sized areas.  
In higher emission areas the error is a bit larger in absolute terms at some 0.2~mK, but in relative terms it is negligible with  signal-to-noise ratios S/N~$> 50$.

\subsection{Tests: Stokes $I$}
\begin{figure}
	\includegraphics[angle=90, width=\columnwidth]{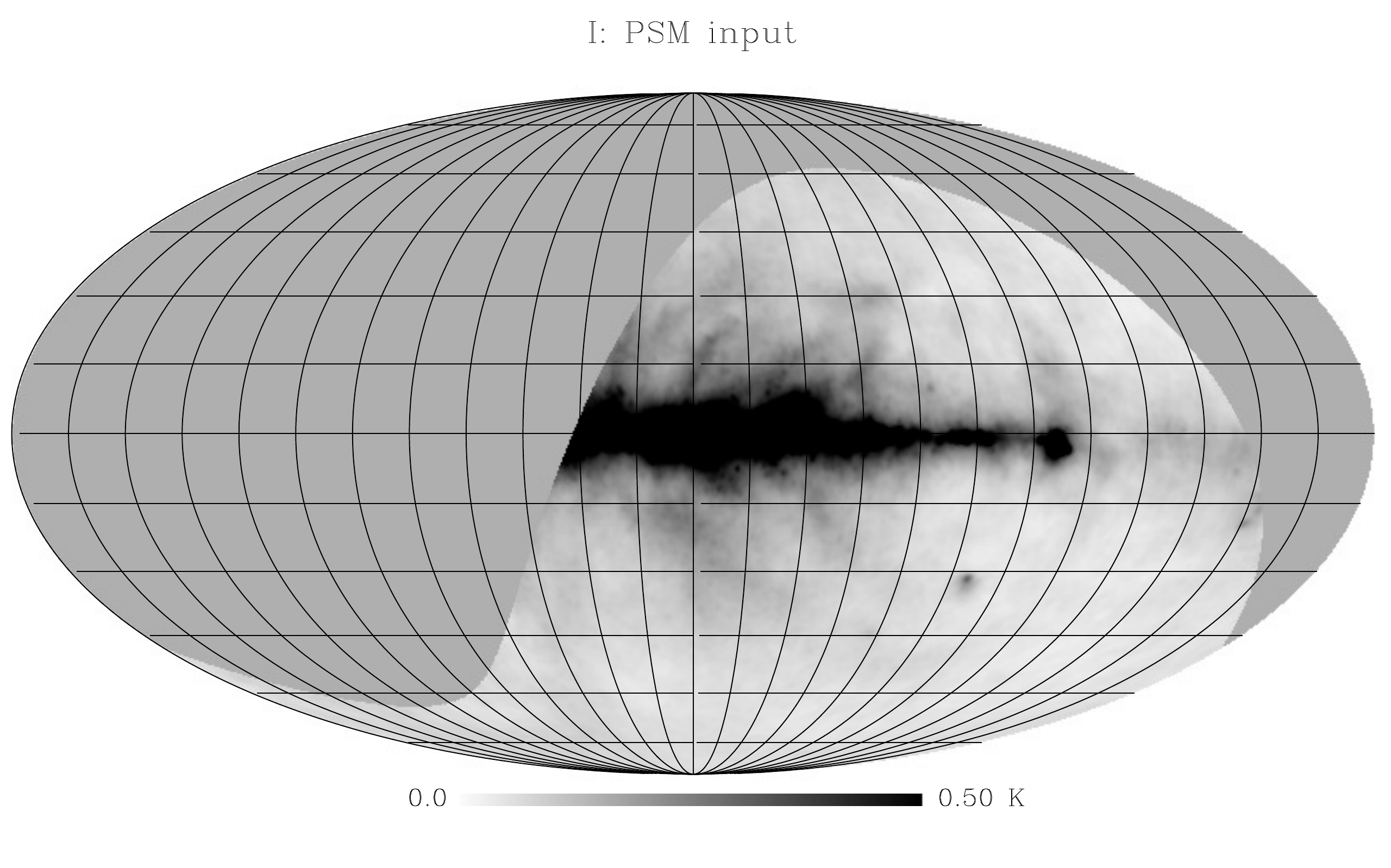}
	\includegraphics[angle=90, width=\columnwidth]{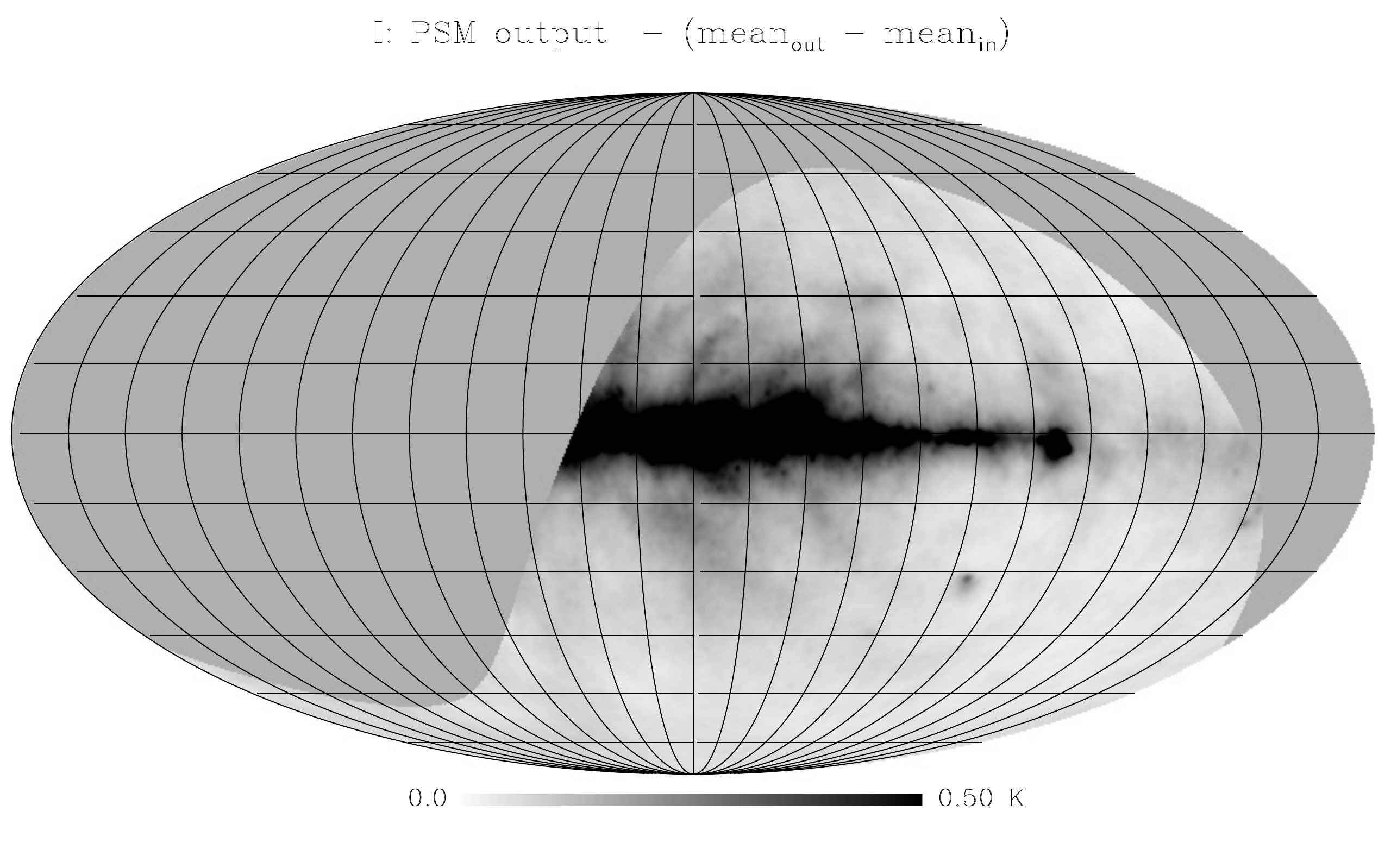}
    \caption{{\it Top}: Stokes $I$ input map used in the simulation to 
	                       test the map-making procedure. The images are in Galactic coordinates centred at the Galactic Centre, 
	                       grid lines are  spaced by 15$^\circ$.  
	          {\it Bottom}: As for top panel, but for an output map that is the result of the map-making procedure applied to the simulated observations extracted 
	                        from the input map. The mean value of the input map is lost, so the map shown here is offset by the same amount to better show how 
	                        other features were well recovered.}
    \label{Fig:testwmapI}
\end{figure}

The Stokes~$I$ map-making procedure was tested in a similar manner, with the major difference being that the input map is more complicated because the total intensity emission is a combination of several components including synchrotron, free-free, and CMB.
We used the Planck Sky Model \citep{Delabrouille13} computed at 2.3~GHz that uses several data sets from radio to millimetre wavelengths to appropriately scale in frequency the relevant components. 
The input map we used is shown in Figure~\ref{Fig:testwmapI}. 
The mean signal is $\overline{I}_{\rm PSM} = 155$~mK, and the weakest emission  ${I}_{\rm PSM, min} = 24$~mK.

Simulated  data were generated as described in Section~\ref{Sec:testQU}, then the map-making procedure was applied. Some of the mean emission is lost as expected, and the output map is offset by $\Delta I = -99.6$~mK. 
Setting this aside, Figure~\ref{Fig:testwmapI} shows that the two maps are nearly identical.
The rms of their difference is 2.7~mK, slightly larger that what one would expect from the instrumental noise, evidence that, in this case, the map-making procedure adds a non-negligible contribution.
\begin{figure}
	\includegraphics[angle=90, width=\columnwidth]{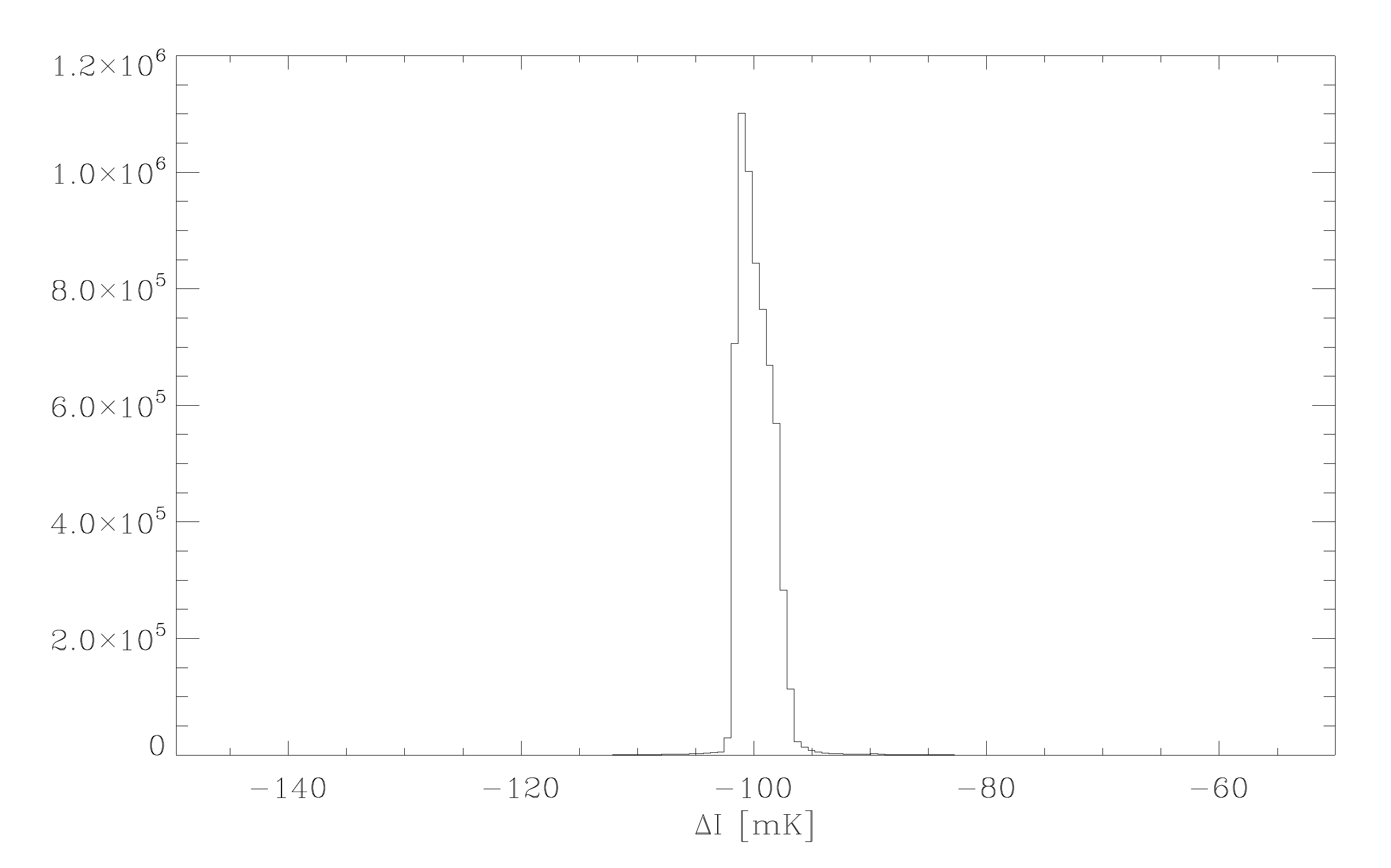}
       \caption{Distribution of the difference between Stokes~$I$ input and reconstructed map in the no instrumental noise case.
                 Note that there is an offset by about -100~mK, evidence of  mean signal loss.}
    \label{Fig:testwmapHistINoNoiseI}
\end{figure}

To verify this, the no noise case is shown in Figure~\ref{Fig:testwmapHistINoNoiseI}. 
The rms is 1.5~mK, with some asymmetry, with 16\% and 84\% cumulative distribution limits at ($-1.6, +1.4$)~mK. 
The map-making procedure error is thus smaller than, but comparable to, the expected statistical error (2.26~mK) and accounts for the larger total rms we obtain in the instrumental noise case.

This  however has a minor impact on our maps. 
First, the error budget is dominated by the confusion limit (9~mK) and the map-making procedure contribution to the total rms is negligible at $\sim$1\%. 
Then, compared to the sky signal, the map-making error ensures $S/N > 15$ everywhere in the sky and an ample $S/N > 100$ compared to the mean signal. 
The map-making procedure thus reconstructs the input map with negligible error compared to the sky signal.

\section{Ground Emission}\label{Sec:ground}

Ground emission is estimated and cleaned from low emission areas, after the data are calibrated and before the map-making procedure is applied.
For each set of azimuth scans - east or west - all points at the same declination share the same  ground emission contamination. 
Following \citet{Wolleben06} and \citet{Carretti10}, all data taken in low emission areas is averaged. 
Stokes $Q$ and $U$ change signs which makes the averaged sky component tend to zero and the final average value gives a reliable estimate of the ground emission. 

It is worth noting an important caveat that affects all surveys where the ground emission is estimated this way: 
     because the estimate is done in constant declination rings, the average 
     sky emission at same declination will be subtracted and 
     any average declination dependence will be subtracted with the ground. 
     This residual cannot be estimated precisely (it would require an a priori knowledge of the actual sky emission), however a few considerations suggest this is small. In addition to the points listed above, sky emission is mainly a function of Galactic coordinates, rather than Declination. That makes the residual subtracted term tend to zero.  A constant component of the signal in Galactic coordinates produces a full modulation in polarization angle along a declination ring, so its average would approach zero. The signal components on smaller angular scales also have an average which tends to zero. Overall the residual subtracted signal will be small compared to the typical signal even in low emission areas.

\begin{figure}
	\includegraphics[angle=90, width=\columnwidth]{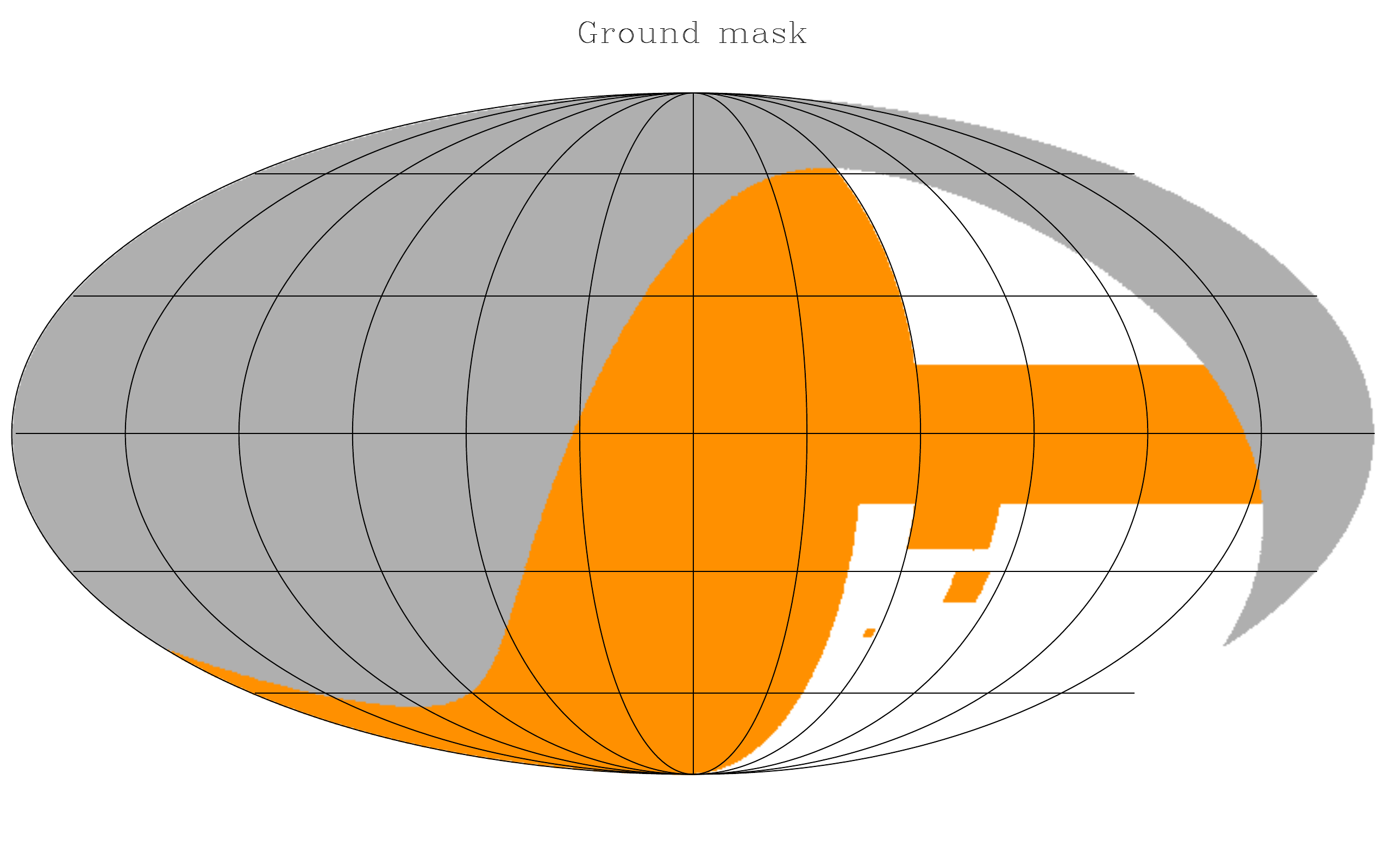}
    \caption{Area used to estimate the ground emission (white). The orange areas were masked out. 
                 The image is in Galactic coordinates centred at the Galactic Centre, 
	                       grid lines are  spaced by 30$^\circ$.}
    \label{Fig:grd_mask}
\end{figure}

We selected low-emission areas using WMAP maps as a guide. 
After cleaning, these were confirmed by the final S-PASS maps themselves with the only addition to the mask being the region centred approximately at $l=290^\circ$, $b=-20^\circ$, as shown in Figure~\ref{Fig:grd_mask}.

\begin{figure}
	\includegraphics[width=\columnwidth]{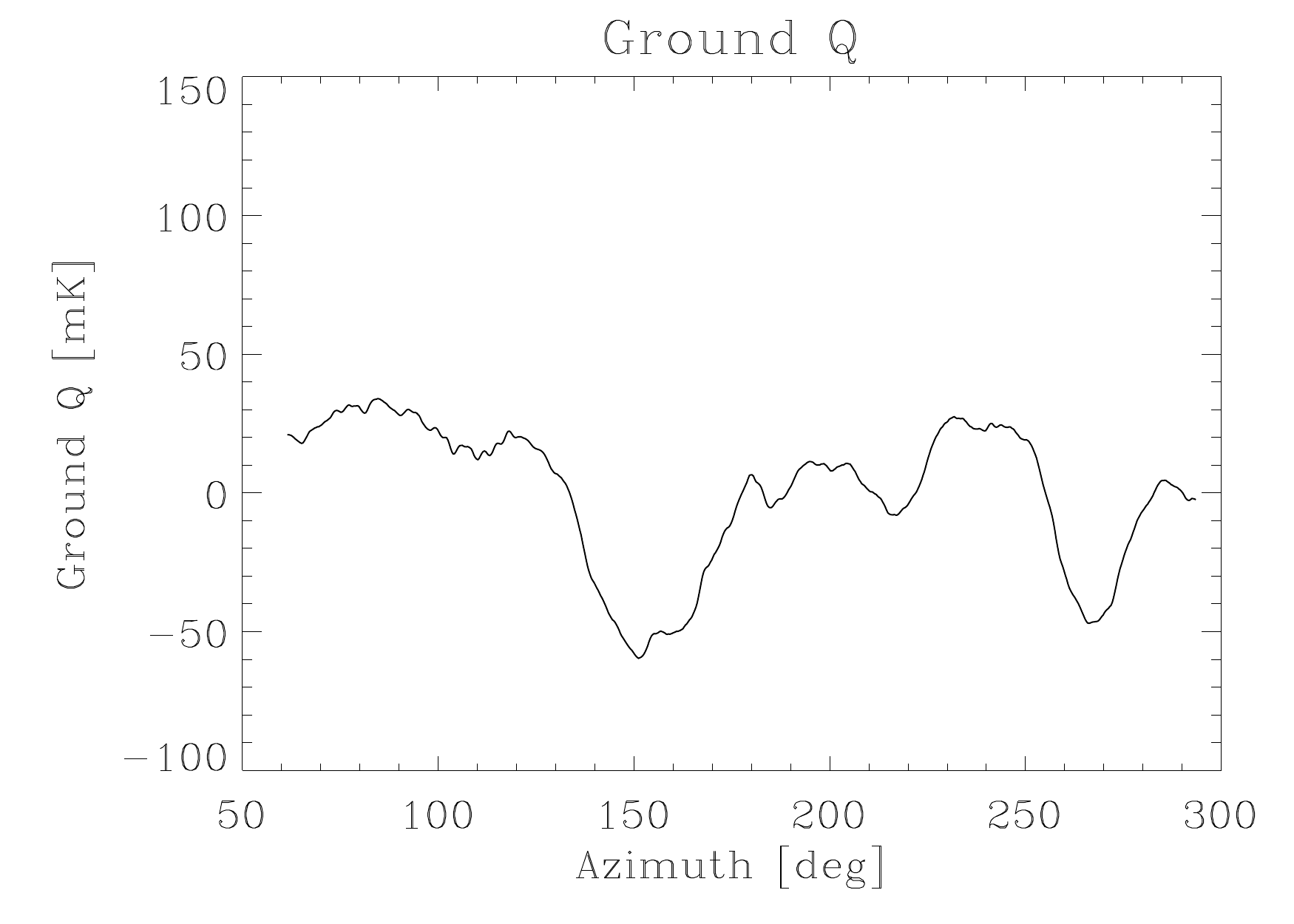}
	\includegraphics[width=\columnwidth]{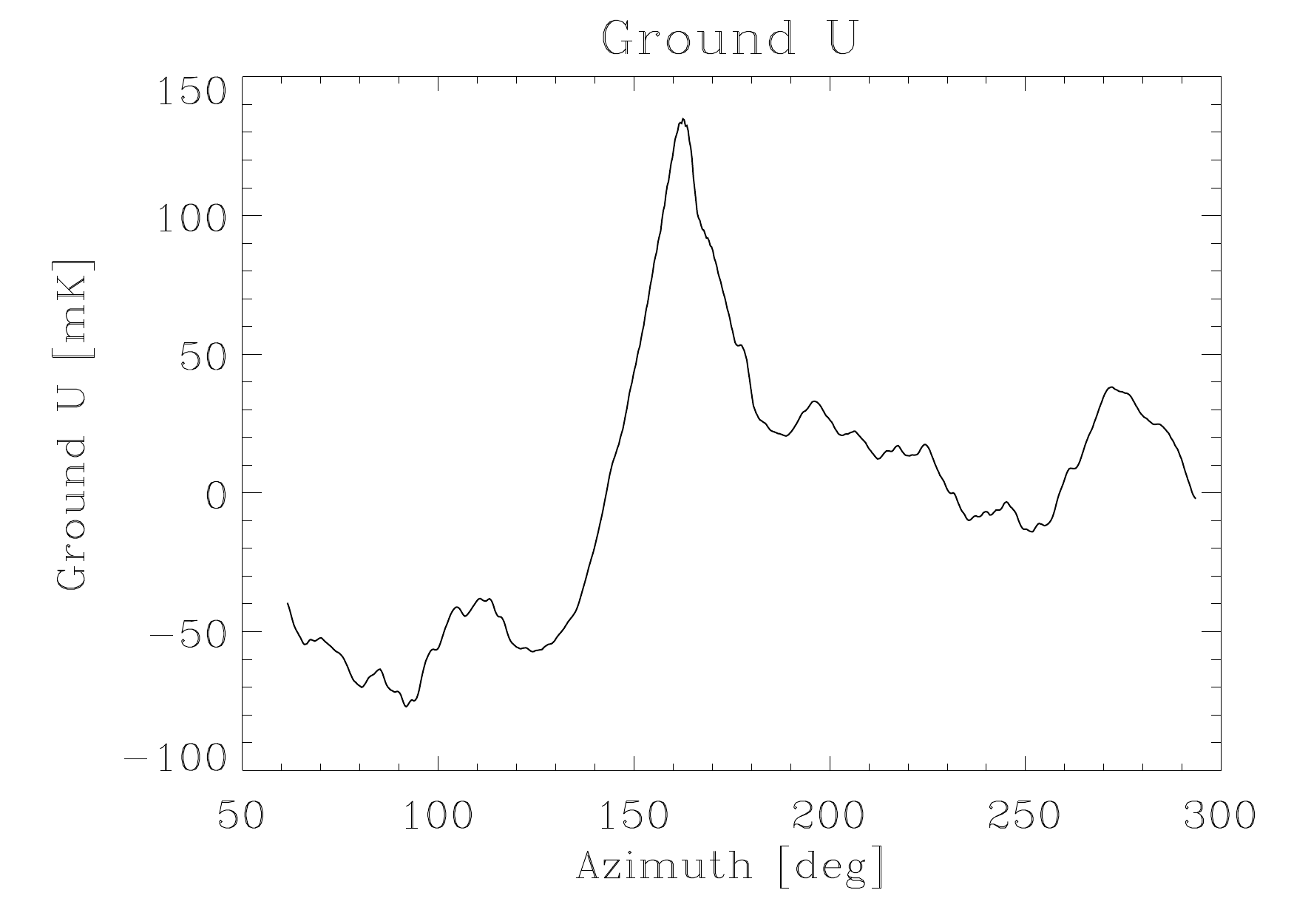}
    \caption{Ground emission profile for Stokes $Q$ (top) and $U$ (bottom) versus the azimuth range covered by the S-PASS scans.}
    \label{Fig:groundProf}
\end{figure}

Figure~\ref{Fig:groundProf} shows the ground emission profile for Stokes $Q$ and $U$. 
The ground component is about an order of magnitude lower than what was measured by \citet{Wolleben06}, proving the benefit of the AZ scan-based strategy.
These ground emission profiles are subsequently subtracted from the AZ scans.

Were there errors in such a procedure, they would leave residual ground emission that, because of the scanning strategy, would appear as rings concentric to the south Celestial pole.
As we will see in the next Section, the maps we obtain have no obvious signs of such structures, even where the signal is low, a sign that residual contamination is negligible compared to the sky signal.

To obtain a more quantitative estimate of the possible residual contamination, we have compared cleaned east and west scans at the same declination.
Ground emission for these two sets of scans is different and independent, while the average sky emission at the same declination is the same (but see later), so any difference between the two could only be due to residual ground emission. 
In particular, west scans at azimuth $(360^\circ - {\rm AZ})$ share the same Dec as east scans at azimuth AZ. 
Thus, an estimate of possible residual ground contamination can be obtained from
\begin{equation}
  \Delta X = \frac{X_w(360^\circ - {\rm AZ}) - X_e({\rm AZ})}{2}\;\;\;\; {\rm where}\; X = Q, U, L, 
  \label{EQ:groundRes}
\end{equation}
and $e$ and $w$ denote east and west scans.
In reality, even if the declination is the same, the parallactic angle differs in general between two points.
This means that the sky emission is not identical, but  mixed between $Q$ and $U$ depending on the difference in parallactic angle. 
The quantity in Equation~\eqref{EQ:groundRes}  thus accounts for any additional difference due to the sky for $Q$ and $U$ and represents a worst case of the ground residual leftover. 
To ameliorate any such effect, we  only use data for which the polarized emission is lower than 20~mK; this corresponds to about 50\% of the sky covered by S-PASS and allows all azimuths to be checked. 
Note that the linear polarization $L = \sqrt{Q^2 + U^2}$ is not affected by the PA rotation issue and represents a more accurate estimate of the residual ground contamination. 

\begin{figure}
	\includegraphics[width=\columnwidth]{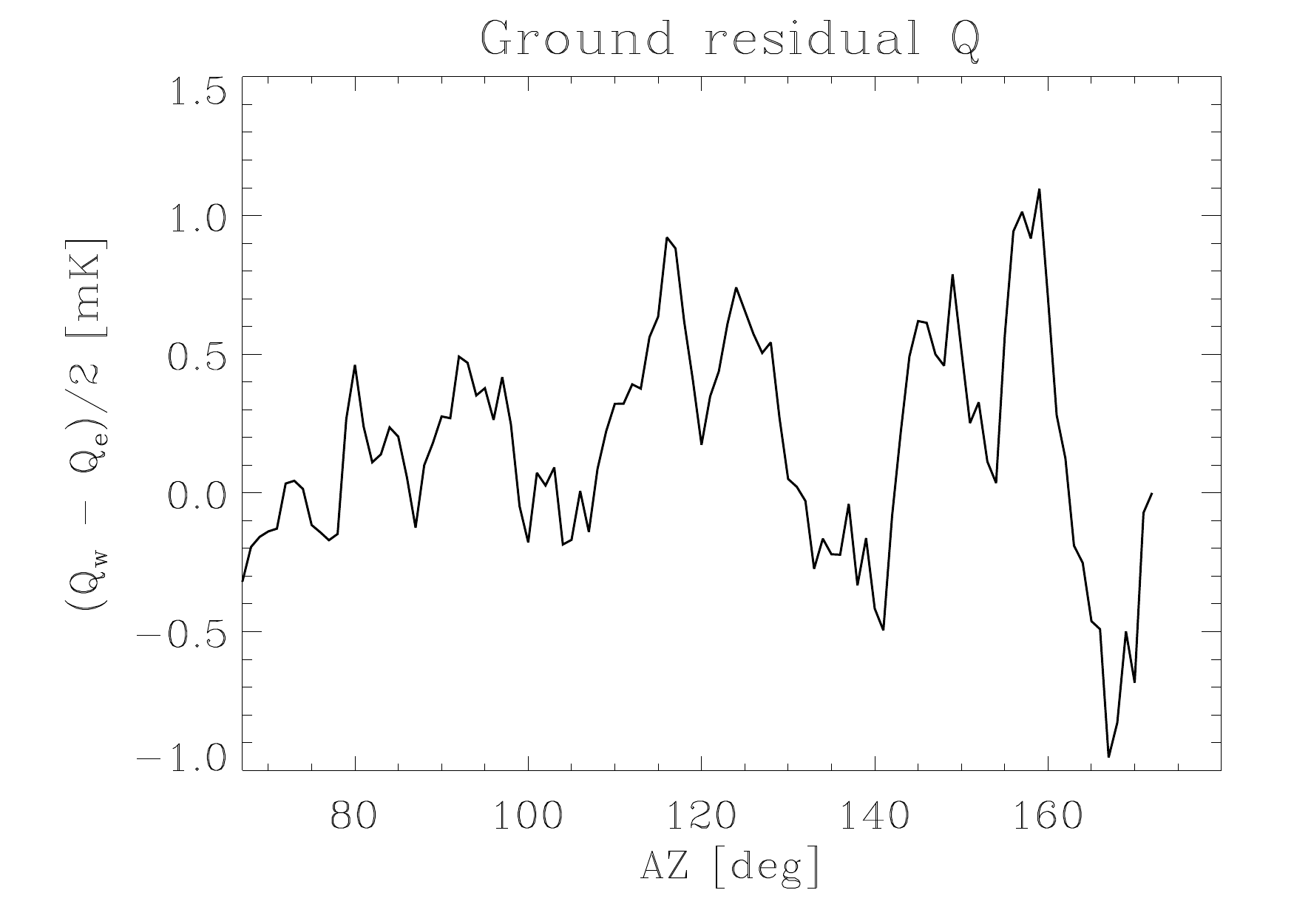}
	\includegraphics[width=\columnwidth]{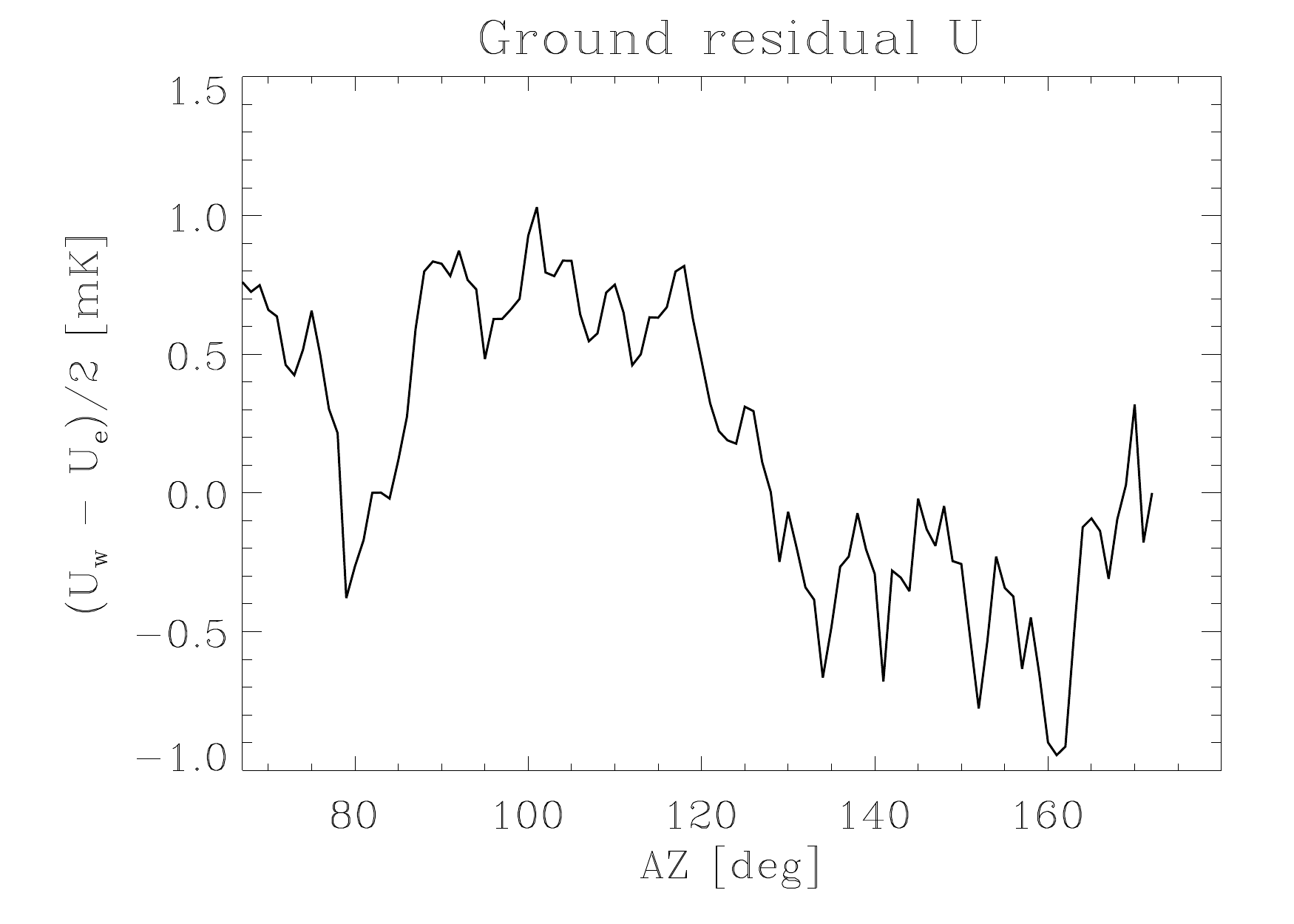}
	\includegraphics[width=\columnwidth]{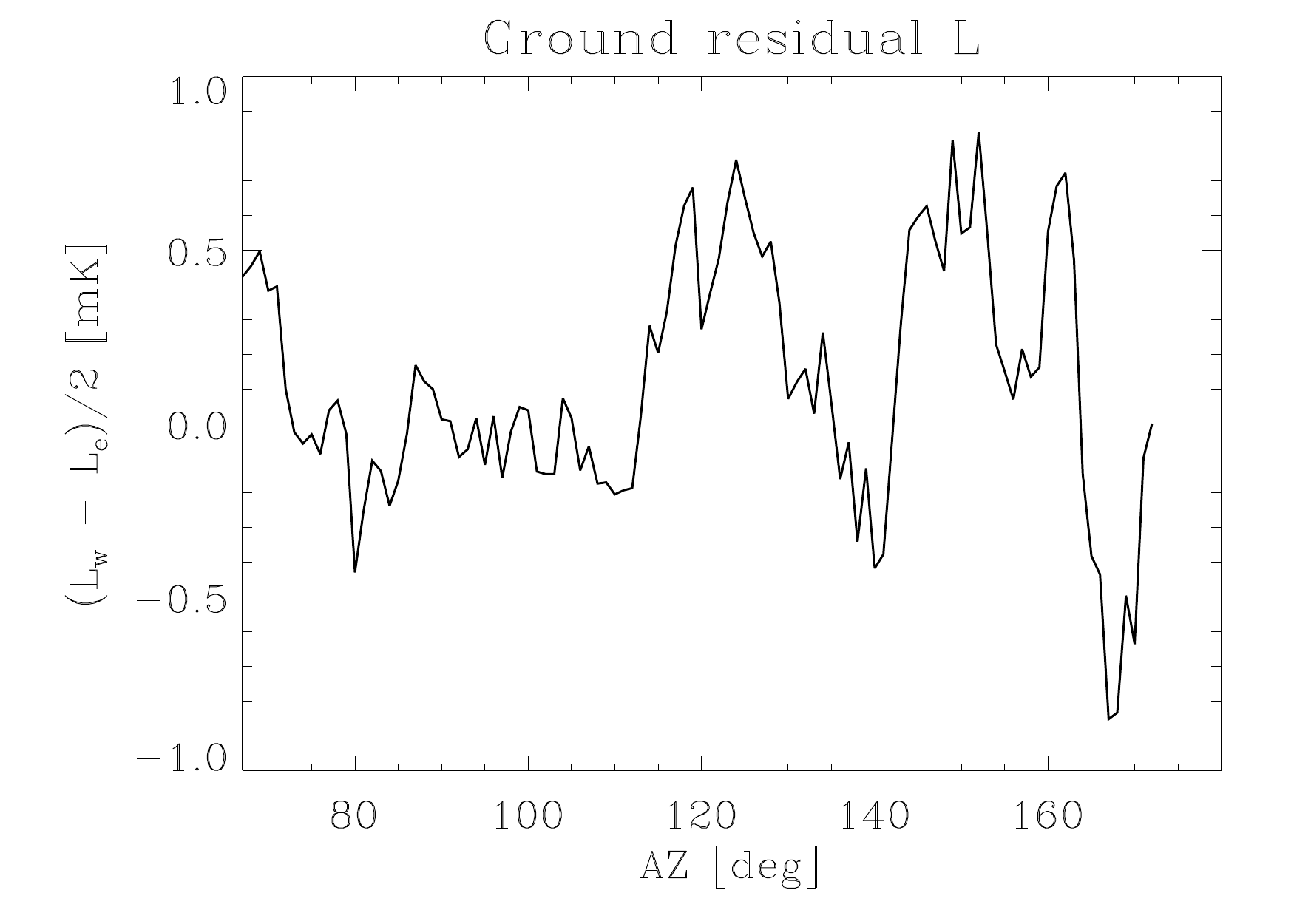}
    \caption{Ground emission residual estimates after cleaning for Stokes $Q$ (top), $U$ (middle), and polarized intensity $L$ (bottom).
	          The quantity plotted is $(X_w(360^\circ-{\rm AZ}) - X_e({\rm AZ})) / 2$ where $X$ = $Q$, $U$, and $L$, respectively -- see Equation~\eqref{EQ:groundRes}. 
	          West scans at azimuth (360-AZ) see the same sky seen by east scans at azimuth, so the difference measures
	          the residual contamination.}
    \label{Fig:groundRes}
\end{figure}

Figure~\ref{Fig:groundRes} shows the  residual contamination, and Table~\ref{Tab:groundRes}  reports the rms values.
The rms of the ground residual for linear polarization $L$ is $\sigma_{g, L} = 0.35$~mK, which we hereafter quote as the ground emission residual contribution to the error budget of our maps.

\begin{figure}
	\includegraphics[width=\columnwidth]{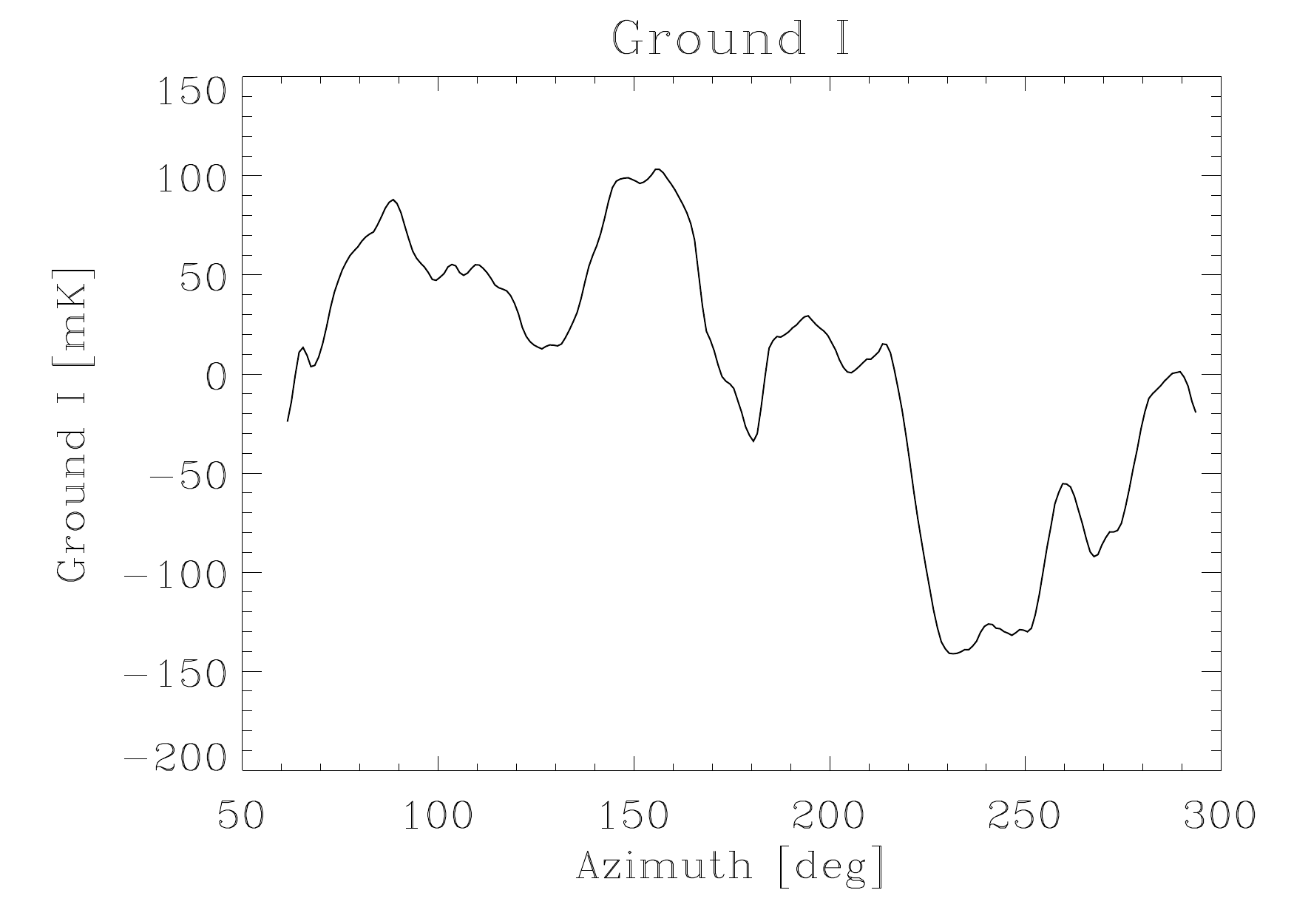}
    \caption{Ground emission profile of Stokes $I$. Details are as for Figure~\ref{Fig:groundProf}.}
    \label{Fig:groundProfI}
\end{figure}

\begin{figure}
	\includegraphics[width=\columnwidth]{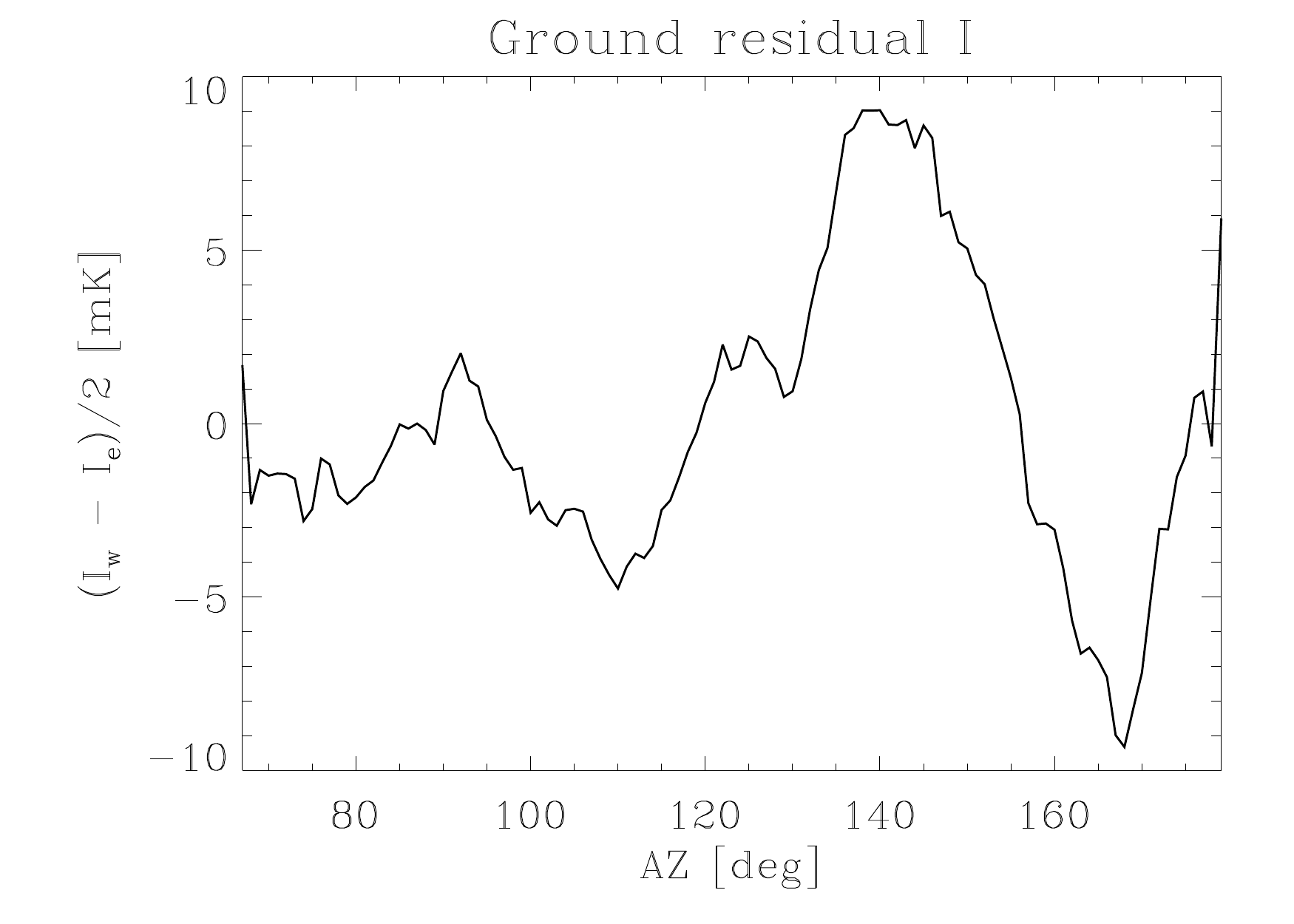}
    \caption{Ground emission residual estimate after cleaning for Stokes $I$. Details are as for Figure~\ref{Fig:groundRes}. }
    \label{Fig:groundResI}
\end{figure}

Stokes~$I$ does  not average to zero, but an approach similar to that for $Q$ and $U$ measurement was used in our analysis.
In particular, we averaged the data taken in low emission areas in azimuth bins that share the same ground emission. 
This makes the mean signal emission in low emission areas taken out, but this is not an issue because the mean level is lost in any case. 
Figure~\ref{Fig:groundProfI} shows the ground emission profile estimated this  way.
Residual emission is estimated as for $Q$ and $U$ as the semi-difference between the mean emission of east and west scans after the data are cleaned:
\begin{equation}
  \Delta I = \frac{I_w(360^\circ - {\rm AZ}) - I_e({\rm AZ})}{2} .
  \label{EQ:groundResI}
\end{equation}
Figure~\ref{Fig:groundResI} plots the residual error as a function of AZ; its rms value is $\sigma_{g, I} = 4.3$~mK (Table~\ref{Tab:groundRes}). 


\begin{table}
	\centering
	\caption{Standard deviation (rms) of the estimated ground emission residual of the polarized and total intensity emission shown in Figures~\ref{Fig:groundRes} and~\ref{Fig:groundResI} ($Q$, $U$, linear polarized intensity $L$, and total intensity $I$}
	\begin{tabular}{lccr} 
		\hline
		 & rms [mK] \\
		\hline
		$Q$ & 0.40 \\
		$U$ & 0.51 \\
		$L$ & 0.35 \\
		$I$ & 4.3\\
		\hline
	\end{tabular}
	\label{Tab:groundRes} 
\end{table}

\section{Maps}\label{Sec:maps}

\begin{figure*}
	\includegraphics[angle=90, width=2.0\columnwidth]{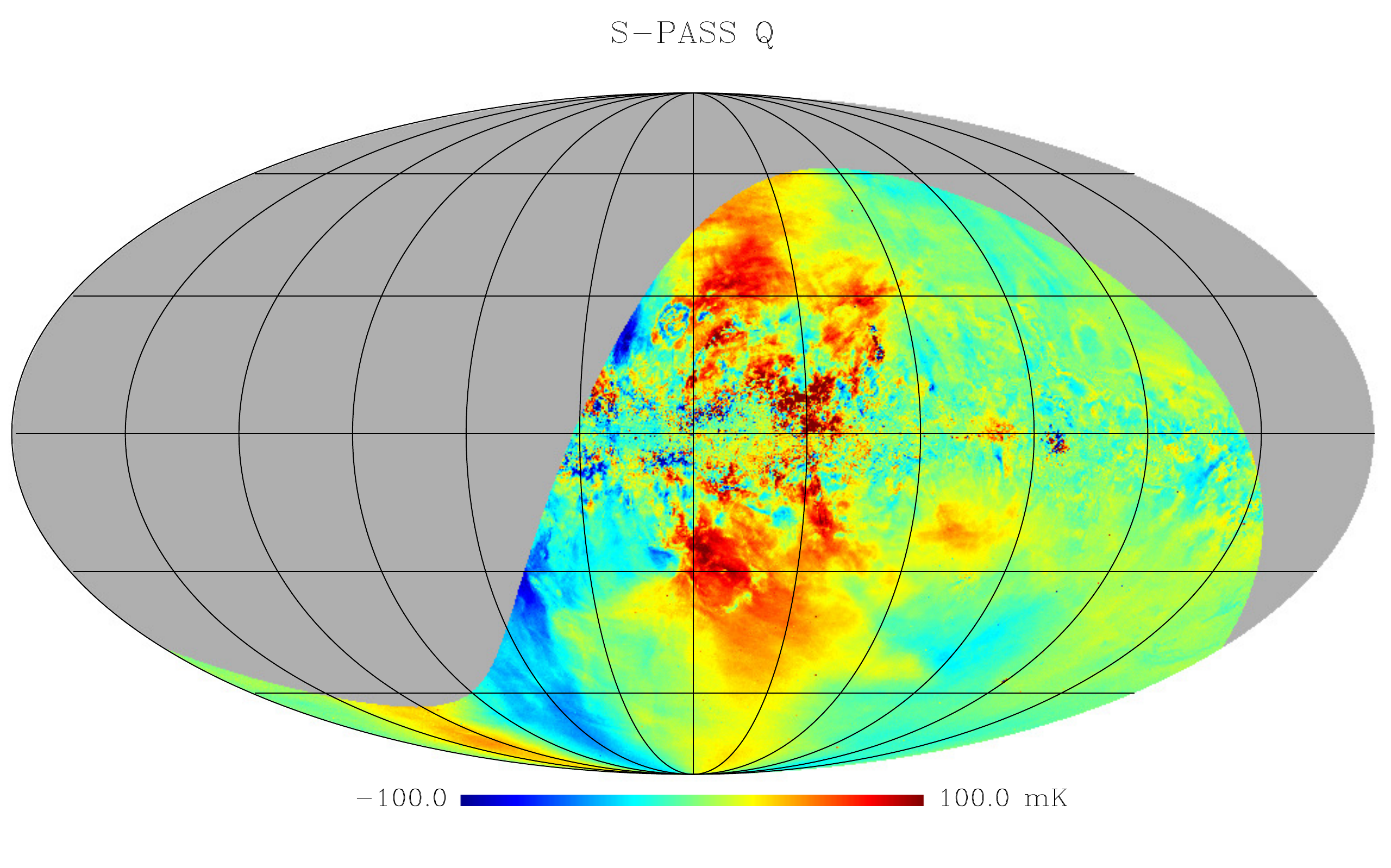}
	\includegraphics[angle=90, width=2.0\columnwidth]{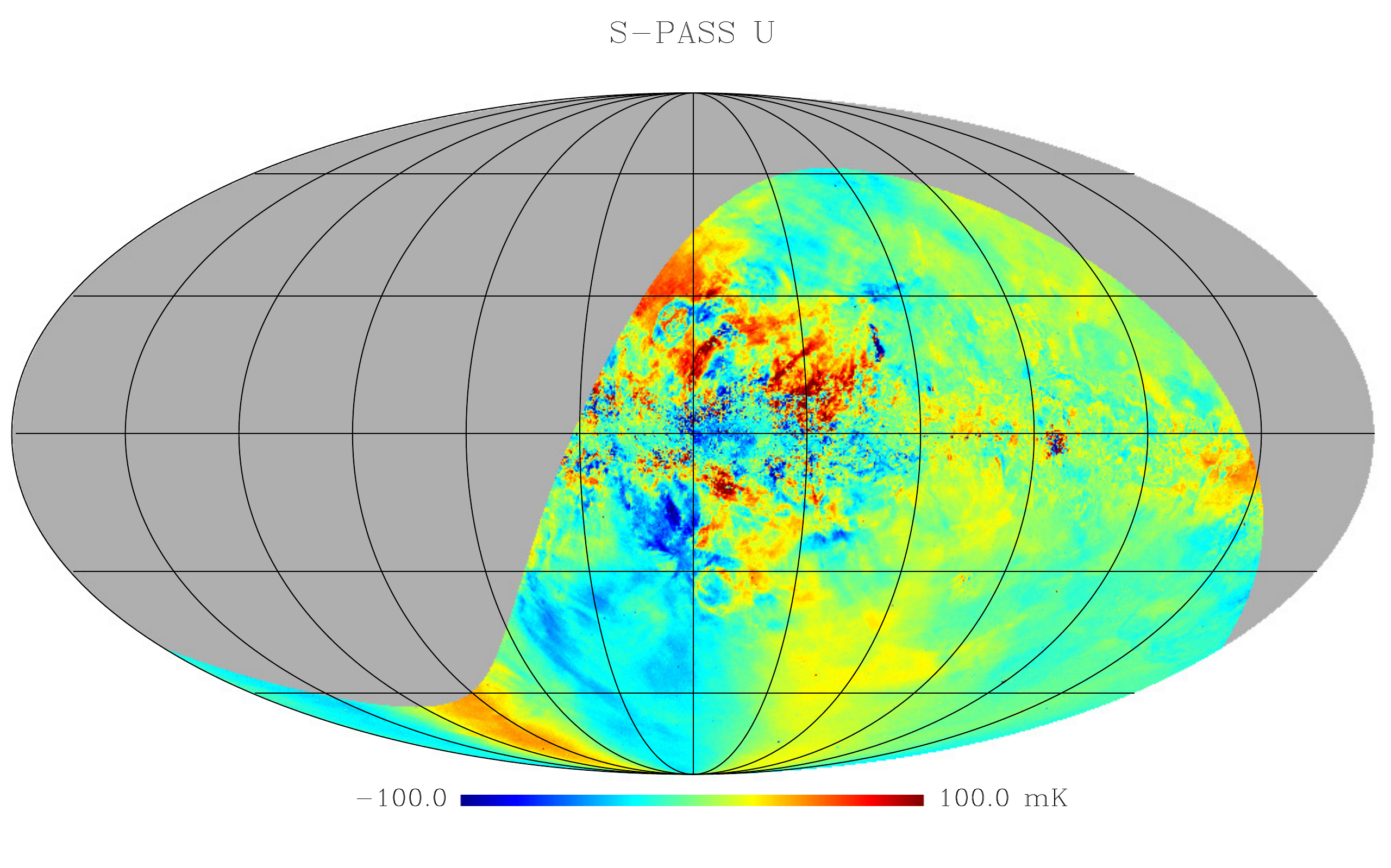}
    \caption{S-PASS maps of Stokes~$Q$ (top) and $U$ (bottom). Maps are in Mollweide projection, Galactic coordinates, with the Galactic centre at the centre and longitude increasing leftward.}
    \label{Fig:spassMaps}
\end{figure*}
\begin{figure*}
	\includegraphics[angle=90, width=2.0\columnwidth]{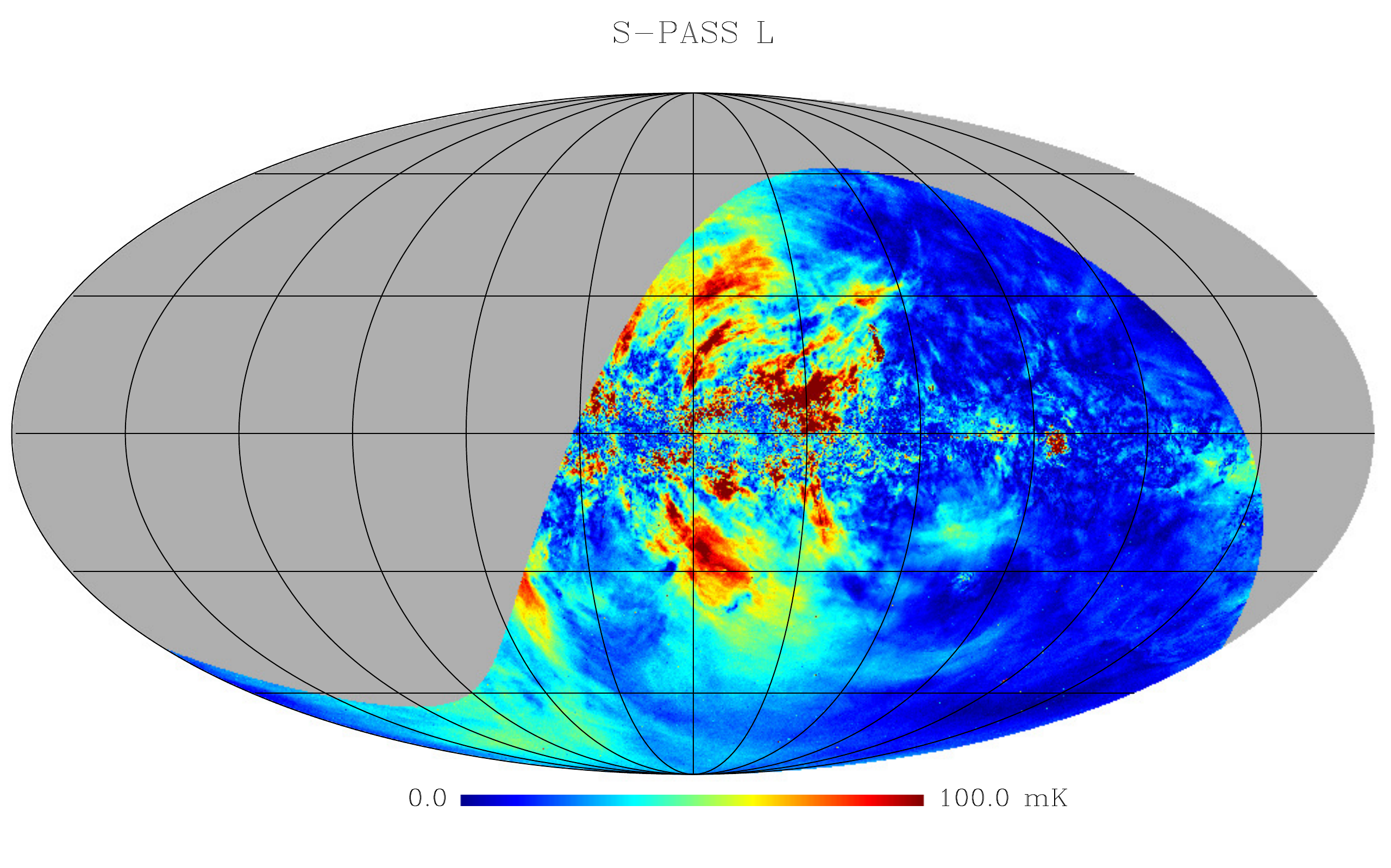}
    \caption{S-PASS map of linear polarized intensity $L$. The map is in Mollweide projection, Galactic coordinates, with the Galactic centre at the centre and longitude increasing leftward.}
    \label{Fig:spassMapsL}
\end{figure*}
\begin{figure*}
	\includegraphics[angle=90, width=2.0\columnwidth]{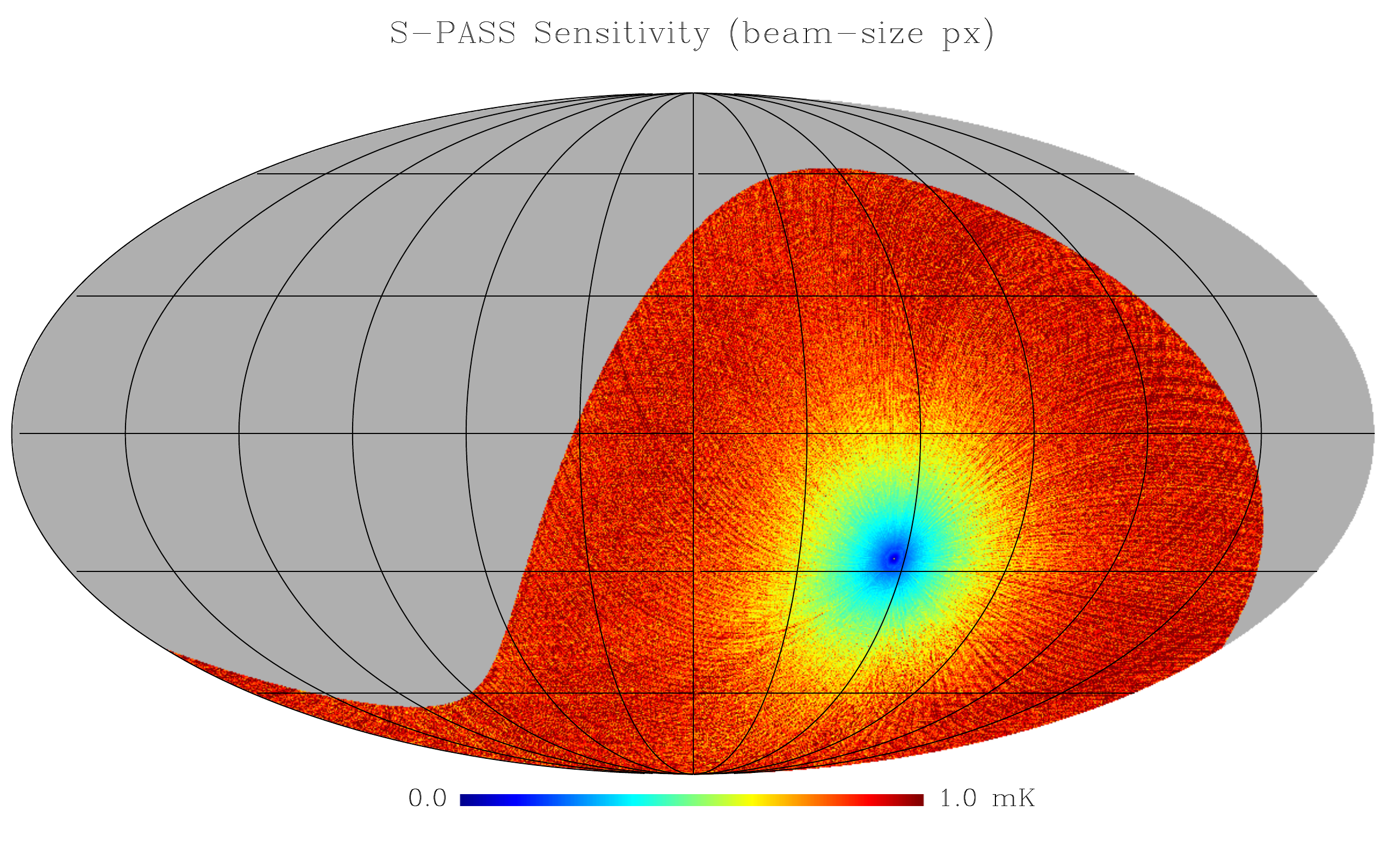}
    \caption{S-PASS sensitivity map on the beam--size scale. The map is in Mollweide projection, Galactic coordinates, with the Galactic centre at the centre and longitude increasing leftward.}
    \label{Fig:spassMapsS}
\end{figure*}

The map-making procedure explained above was applied to the observed data. 
Polarization maps are shown in Figure~\ref{Fig:spassMaps} (Stokes $Q$ and $U$) and Figure~\ref{Fig:spassMapsL} (Polarized Intensity $L$). 
All maps are in brightness temperature $T_b$ units.  
These images have been rebinned to $nside = 512$ (pixels of 6.8~arcmin) to give a better idea of the data quality and sensitivity on a beamsize scale. 
The polarized intensity has been debiased using~(e.g., \citealt{Wardle74}):
\begin{equation}
 L = \sqrt{Q^2 + U^2 - \sigma^2_{\rm px}} 
\end{equation}
where $\sigma_{\rm px}$ is the pixel sensitivity.

\begin{figure}
	\includegraphics[width=\columnwidth]{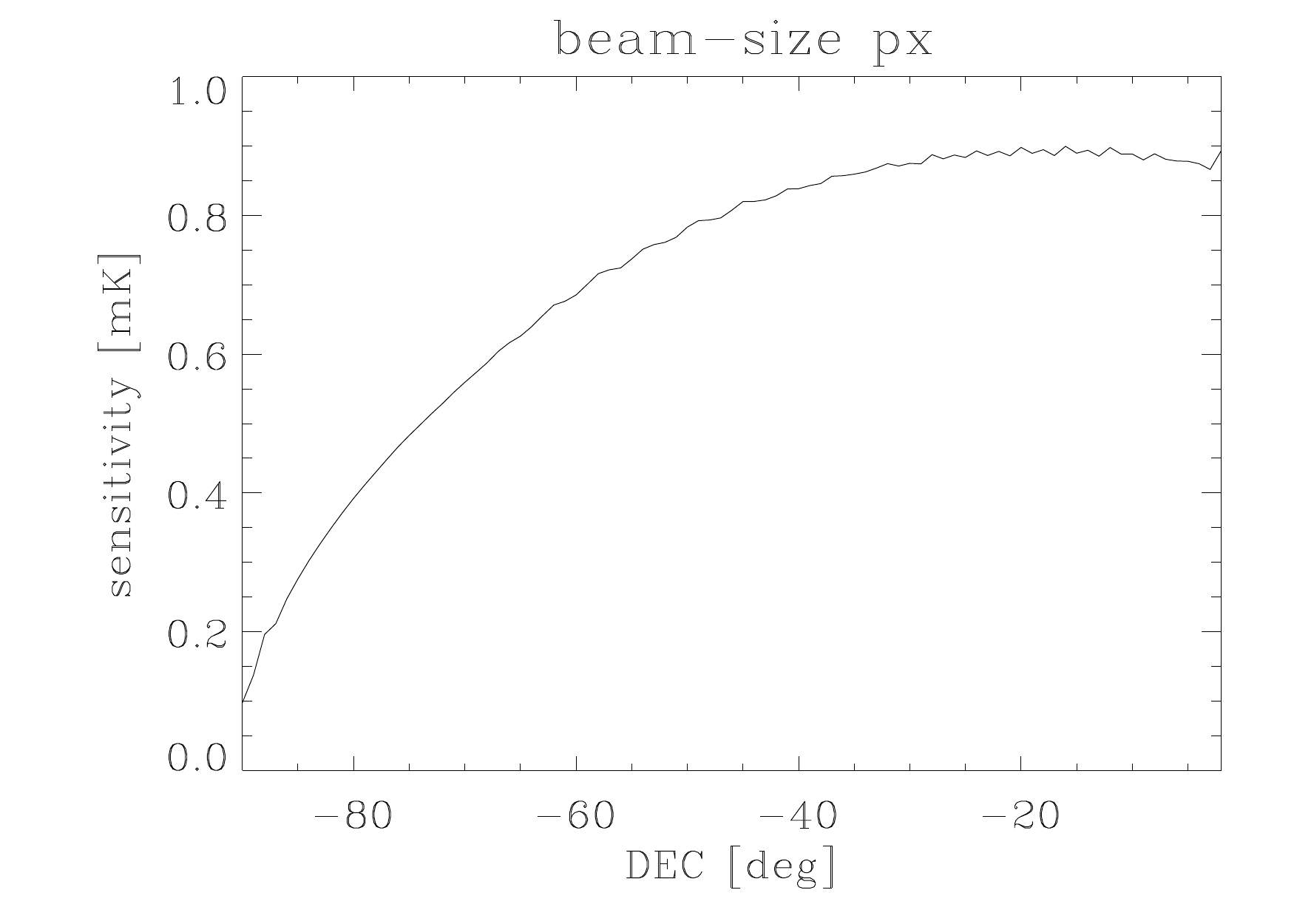}
    \caption{S-PASS sensitivity on a beamsize scale, averaged over all pixels at a given declination.}
    \label{Fig:sensProf}
\end{figure}

Figure~\ref{Fig:spassMapsS} shows the map of the sensitivity on beamsize pixels (1-$\sigma$), and Figure~\ref{Fig:sensProf} the sensitivity profile versus declination, where the sensitivity is averaged over all pixels at the same declination. The mean sensitivity is $\sigma_{b} = 0.81$~mK; it is worst at Dec~$=-18.4^\circ$ ($\sigma_{\rm max} = 0.89$~mK), where the scan spacing is largest, and is best ($\sigma_{\rm SCP} = 0.1$~mK) at the South Celestial Pole, where all scans converge.  

\begin{figure*}
	\includegraphics[angle=90, width=2.0\columnwidth]{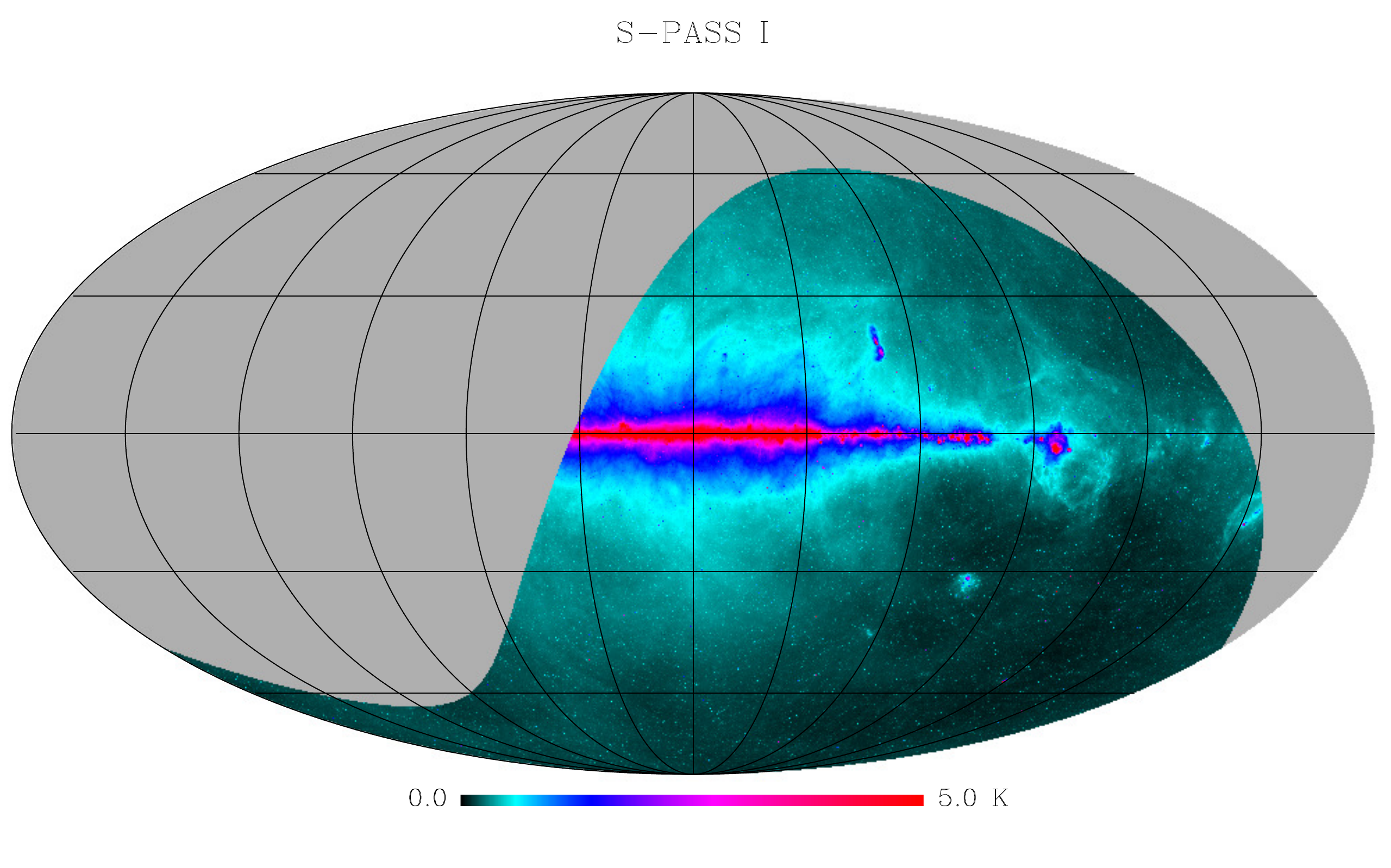}
    \caption{S-PASS Stokes $I$ image. The map is in Mollweide projection, Galactic coordinates, with the Galactic centre at the centre and longitude increasing leftward.}
    \label{Fig:spassMapsI}
\end{figure*}

The S-PASS Stokes~$I$ map shown in Figure~\ref{Fig:spassMapsI} incorporates the mean offset calibration described in Section~\ref{Sec:mapmaking}. 
The rms fluctuations are dominated by the confusion limit $\sigma_{I, (\rm CL)}  = 9$~mK. 
This dominates the error budget compared to all other terms -- i.e., map-making residual, ground emission residual, and instrument noise.

\subsection{Signal statistics and map description}

\begin{figure}
	\includegraphics[width=\columnwidth]{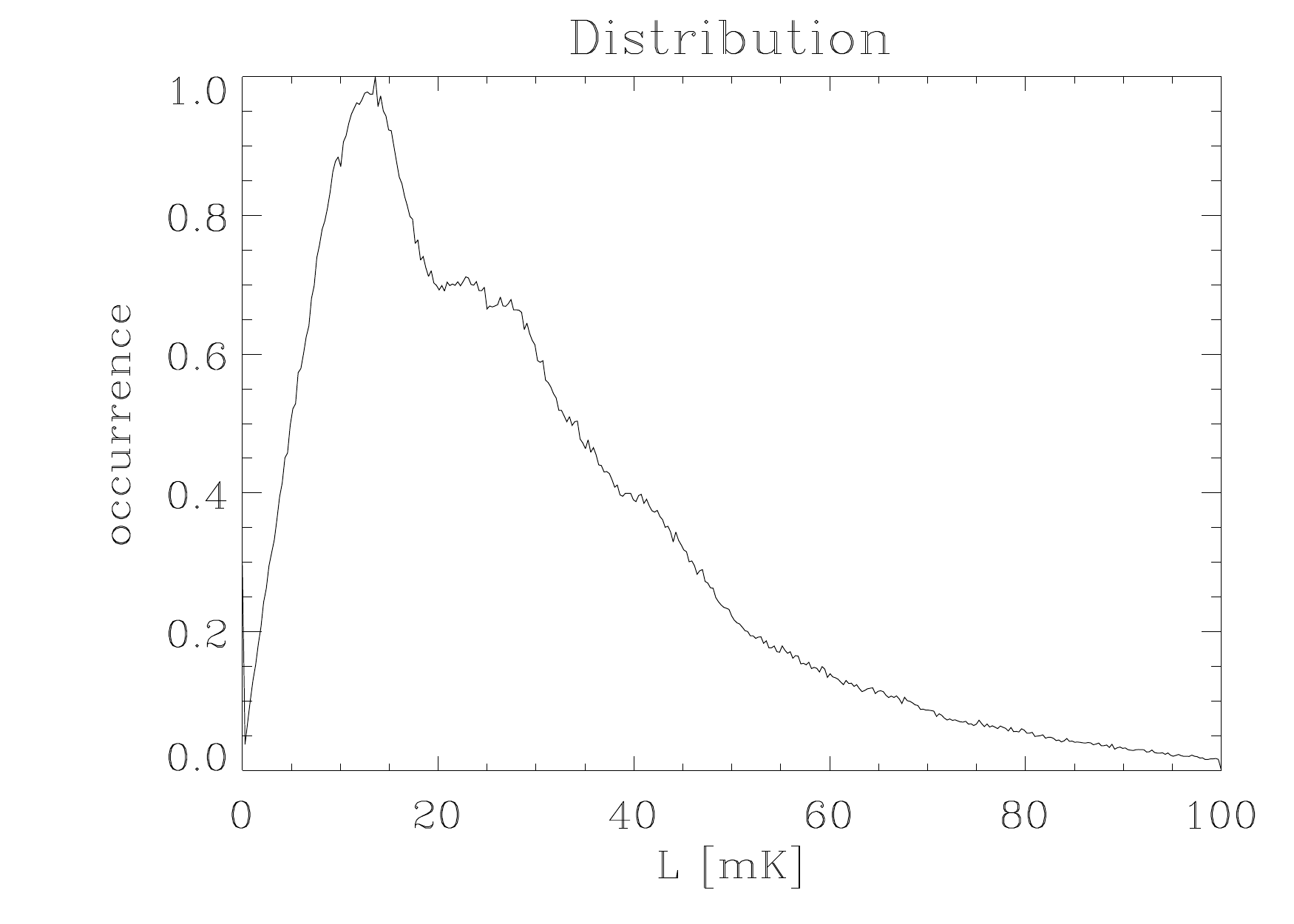}
	\includegraphics[width=\columnwidth]{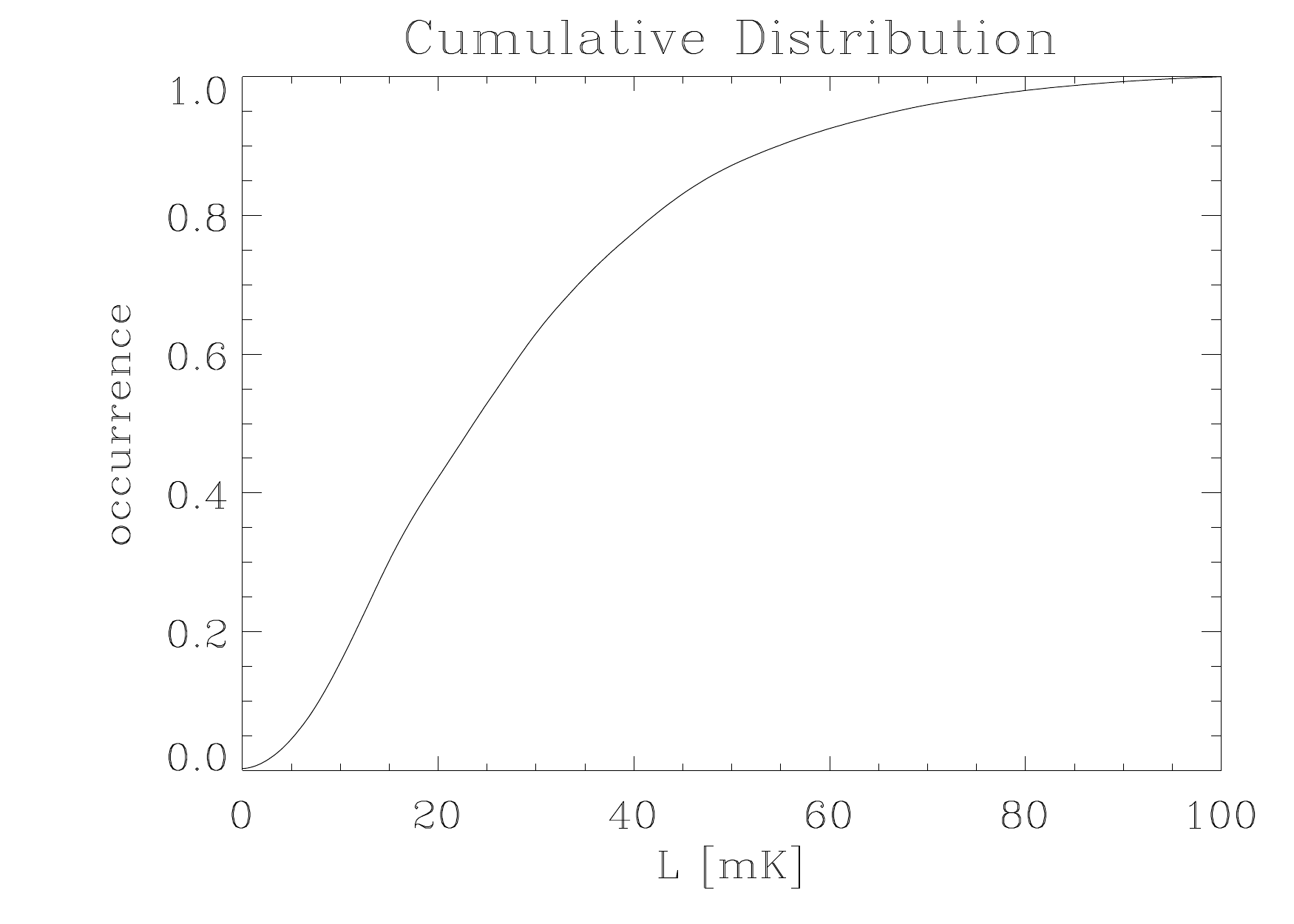}
    \caption{Distribution (top) and cumulative distribution (bottom) of the S-PASS polarized intensity $L$.}
    \label{Fig:polSignalDistr}
\end{figure}

The polarized signal distribution is shown by the histogram and  cumulative distribution in Figure~\ref{Fig:polSignalDistr}. The mean emission is $\bar{L} = 29$~mK.  

At the low emission end there is a peak at $L\sim 13$~mK, which  can be regarded as the typical emission in the low emission areas and a reference  to use by any experiment dealing with the Galactic Synchrotron emission as a foreground contaminant (e.g.\ CMB experiments).  

For stronger polarized emission regions, there is a flat plateau spanning 20--30~mK, followed by a slow roll--off to high emission. 

From the cumulative distribution we find that 98.6\% of the S-PASS area has a Signal-to-Noise ratio $S/N > 3$ and that 50\% of the S-PASS sky is fainter than 23.6~mK. 

The $Q$ and $U$ maps show that much of the depolarization screen seen at lower frequency of $|b| < 30^\circ$  is lifted at 2.3~GHz, and signal visible well below that edge.
Smooth and extended structures, evidence of little Faraday depolarization or modulation, stretch almost down to the Galactic Plane, with only the few degrees across the plane still modulated by Faraday Rotation or depolarized. 
The largest depolarization regions are in the Inner Galaxy on either side of the Galactic Plane, reaching up to $|b| \sim 10^\circ$. 
At all other longitudes signs of strong  depolarization or FR are limited to lower latitudes, except a few individual regions. e.g. the Gum Nebula or $\zeta$~Oph at ($l$, $b$)~$\sim$~($5^\circ$, $25^\circ$).

Extended and smooth structures are visible in the maps. 
The most striking feature stretches from the south Galactic cap  at around $l = 330^{\circ} - 360^{\circ}$, $b=-60^{\circ}$  to the northernmost end of the S-PASS area, a length of $100^\circ$ or more. 
These structures are the radio polarization counterparts of the $\gamma$-Ray Fermi Bubbles \citep{Carretti13a, Su10} emanating from the Galactic Centre.

\begin{figure}
	\includegraphics[angle=90, width=\columnwidth]{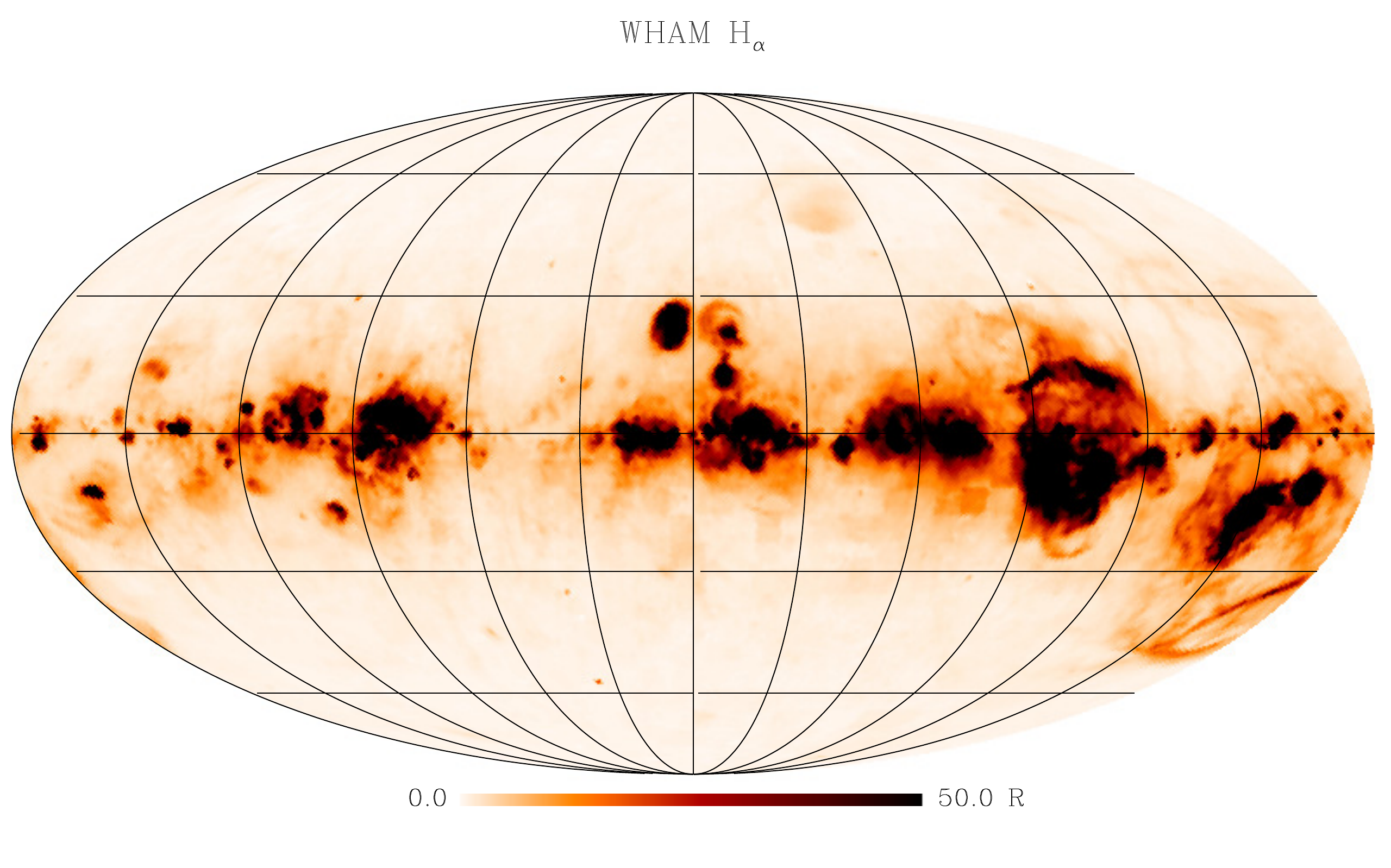}
    \caption{H$_\alpha$ emission as seen by the WHAM Sky Survey (\citealt{Haffner10}, Haffner et al. 2019 in prep.). 
                This map shows the total emission integrated over all spectral channels.   
                  }
    \label{Fig:whamMap}
\end{figure}

\begin{figure*}
	\includegraphics[angle=0, width=2.\columnwidth, viewport=50 570 520 820]{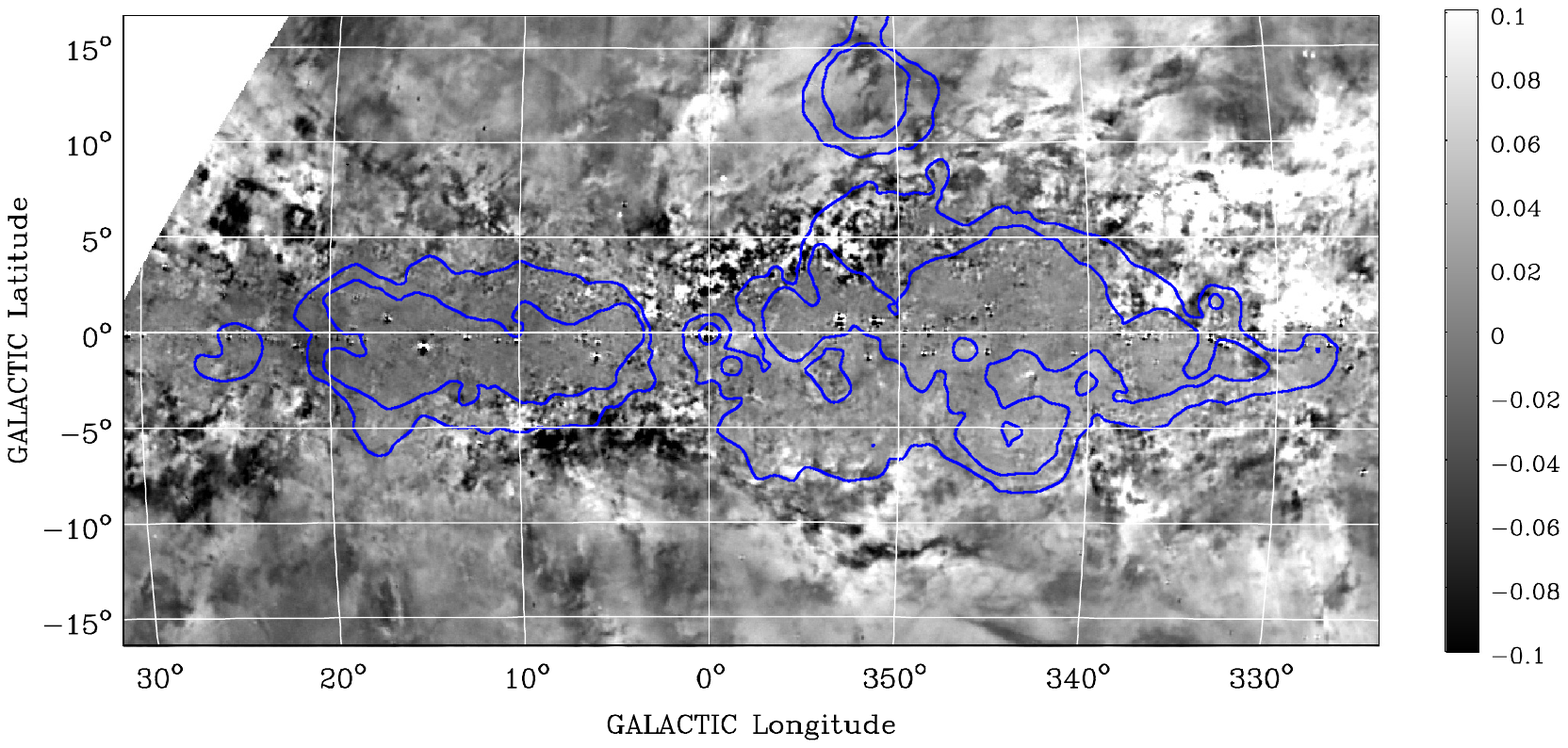}
    \caption{S-PASS Stokes $Q$ map (colour bar units are K) with H$_\alpha$ emission contours from WHAM map of Figure~\ref{Fig:whamMap}. 
                  Contours are at 22 and 44~R.
                 }
    \label{Fig:spassWhamCont}
\end{figure*}

\begin{figure}
	\includegraphics[angle=0, width=\columnwidth]{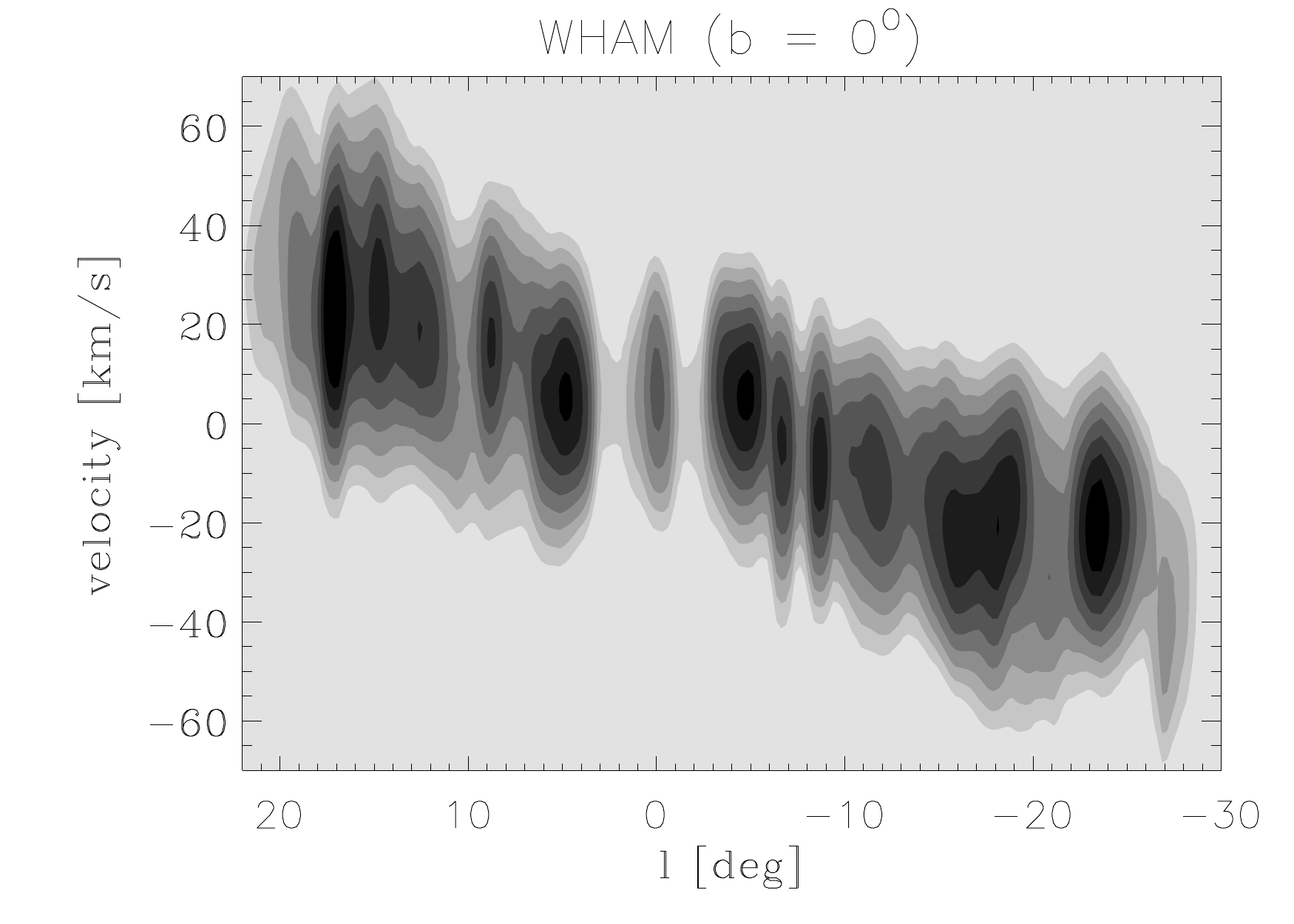}
	\includegraphics[angle=0, width=\columnwidth]{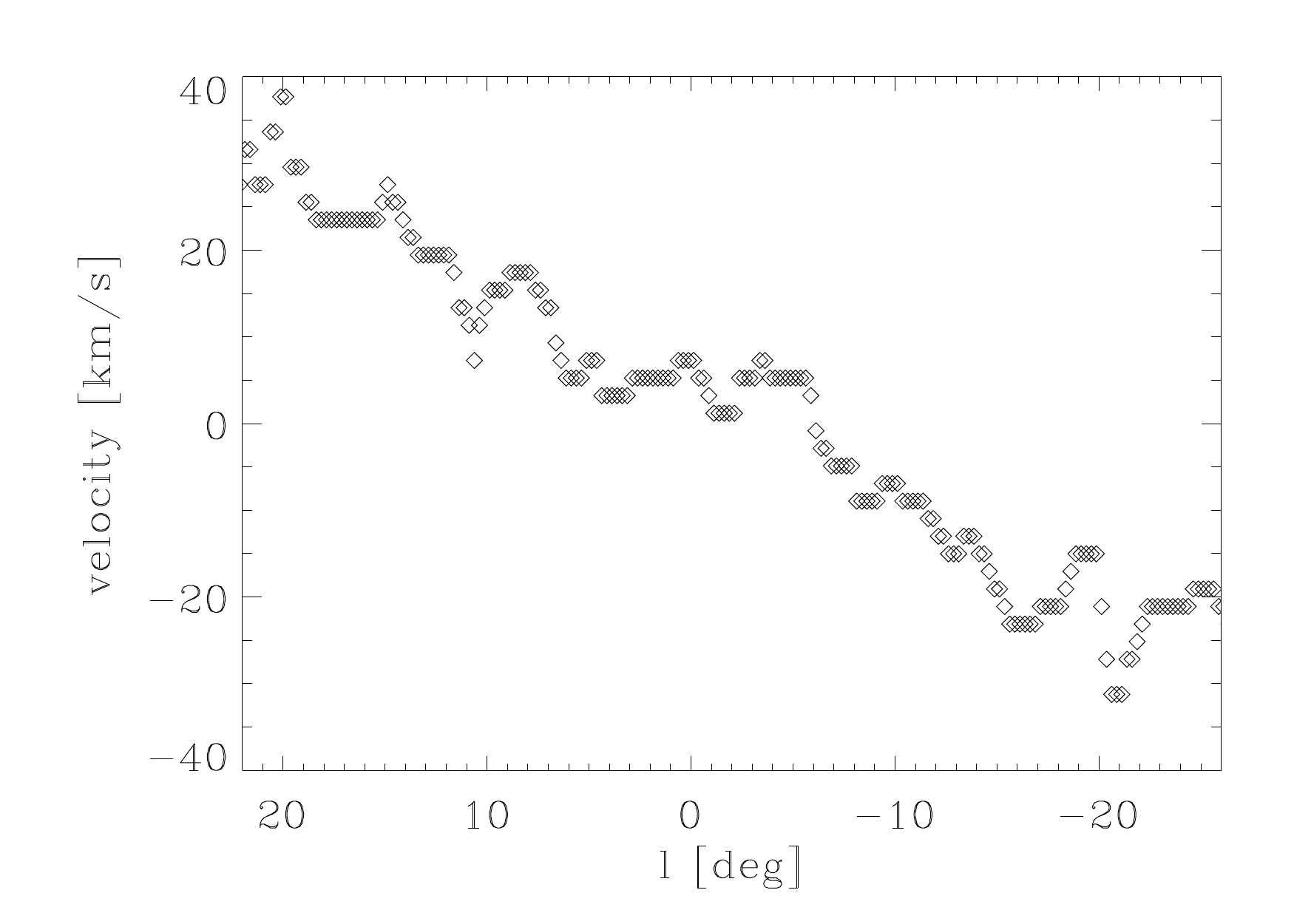}
    \caption{{\it Top:} Position-velocity (radial velocity in the Kinematic Local Standard of Rest -- LSRK -- system) map of the H$_\alpha$ emission of the two large regions in the central part of the Inner Galaxy 
                  (Galactic longitude $l \sim [20^\circ, -30^\circ]$) 
                  from the WHAM Sky Survey (\citealt{Haffner10}, Haffner et al. 2019 in prep.) for a latitude slice centred at $b = 0^\circ$ of width $\Delta b = 1^\circ$.
                  Contours starts at 1~R/(km/s) and scale by a factor $\sqrt{2}$ each.
                  {\it Bottom:} For the same slice, the velocity where the emission peaks at each longitude bin plotted versus Galactic longitude.  
                 }
    \label{Fig:whamPosVel}
\end{figure}

The two central depolarization regions at $l \sim [0^{\circ}$, $20^{\circ}]$ and $l \sim [335^{\circ}, 360^{\circ}]$, respectively, closely correspond to two large areas of H$_\alpha$ emission as seen in the WHAM Sky Survey\footnote{maps available at http://www.astro.wisc.edu/wham-site/wham-sky-survey/wham-ss/}  in Figure~\ref{Fig:whamMap} (\citealt{Haffner10}, Haffner et al. 2019 in prep.). 
H$_\alpha$ is a tracer of the ionised medium that, combined with the magnetic field, generates FR and related effects. 
Figure~\ref{Fig:spassWhamCont} shows contours of the two H$_\alpha$ ISM clouds overlaid with the Stokes~$Q$ emission and shows there is an excellent correlation between the cloud edges and the S-PASS areas where the signal is depolarized or heavily FR modulated.
 We show Stokes $Q$ because we deem it better suited than the polarized intensity $L$ for showing areas of heavy FR modulation which 
generate variations of the polarization angle that $L$ cannot capture.
Stokes $U$, not shown here, gives similar results. 
The position-velocity plot of the two clouds (Figure~\ref{Fig:whamPosVel}) reveals that the velocity changes linearly from 30~km~s$^{-1}$ at $l\sim20^\circ$ to  $-30$~km~s$^{-1}$ at $l\sim340^\circ$, 
consistent with the kinematics of a spiral arm at a distance from the Sun of 2~kpc (e.g. see \citealt{Dickey13}) that matches the Sagittarius arm,  midway between our location and the Galactic bulge. 

Frequency spectral index behaviour is studied by~\citet{Krachmalnicoff18} who, comparing  S-PASS, WMAP, and Planck maps, find the distribution peaks at $\hat\beta = -3.2$ with an rms spread 
$\sigma_\beta \sim 0.2$. The same authors also studied the spatial behaviour via  polarized angular power spectra (APS). Besides the global spectra at different Galactic latitude cuts they also computed it in $\sim400$~deg$^2$ areas (1\% of the sky each), showing the best sky spots for CMB investigations.

At mid and high Galactic latitudes, where Faraday Rotation is negligible, 
S-PASS maps are an excellent match to higher frequency data, with no decorrelation within noise limits \citep{Krachmalnicoff18}. This shows that, not only is depolarization negligible, but that Faraday Rotation is small, making maps at this frequency  an excellent data set to study foreground contamination in CMB experiments.
\citet{Krachmalnicoff18} also use S-PASS  to characterise polarized Galactic synchrotron emission 
with unprecedented sensitivity.

\begin{figure}
	\includegraphics[width=\columnwidth]{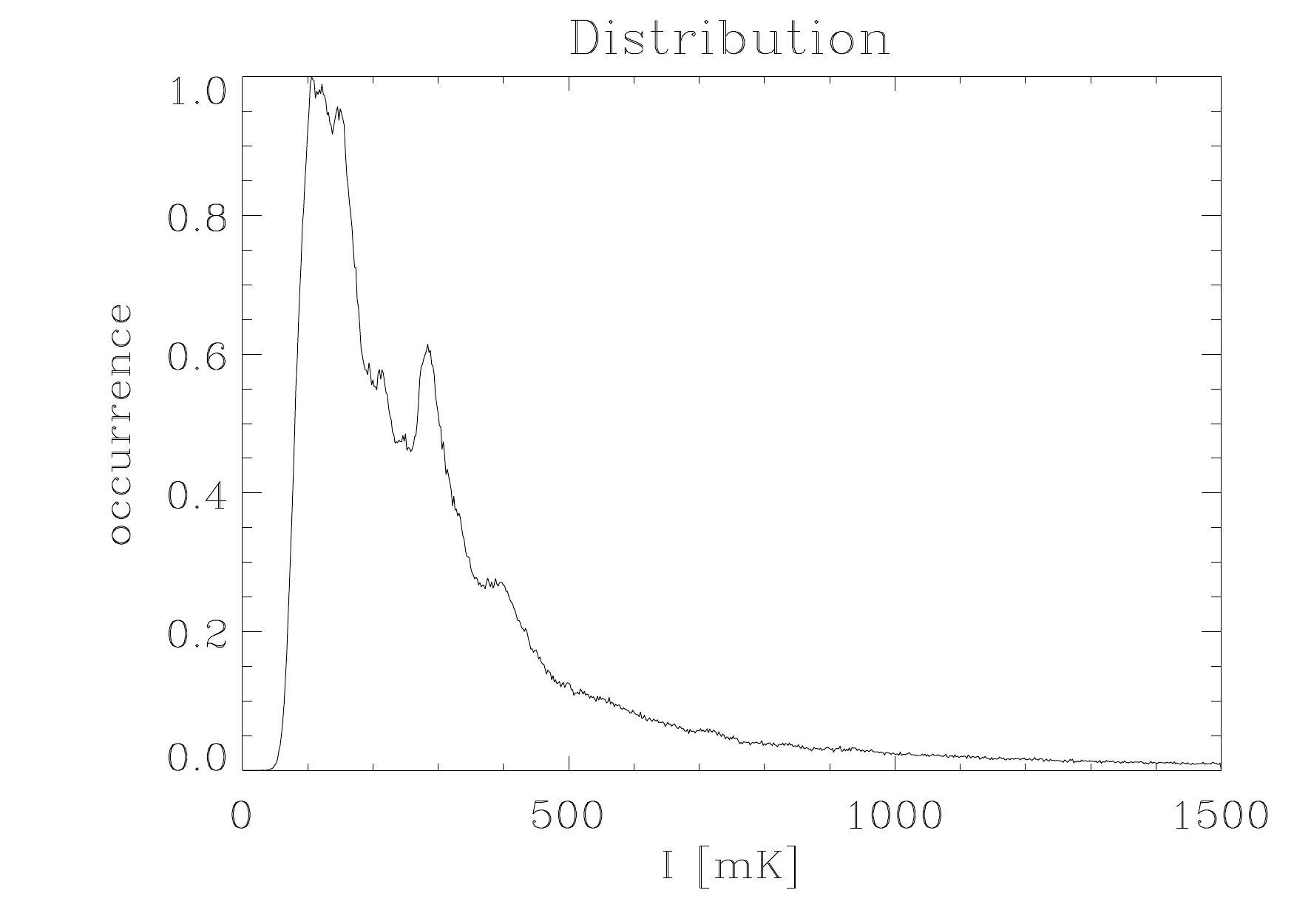}
	\includegraphics[width=\columnwidth]{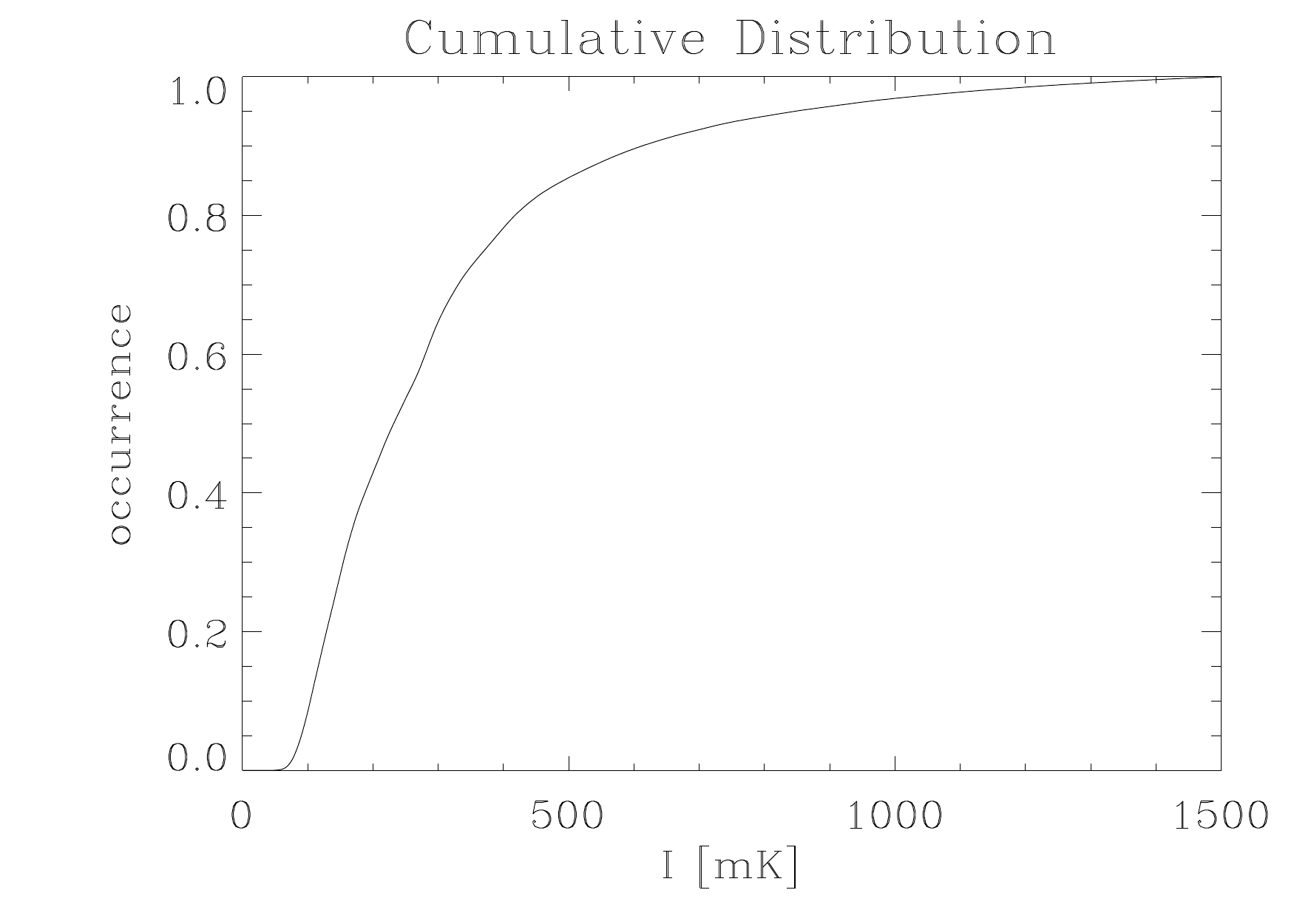}
    \caption{Distribution (top) and cumulative distribution (bottom) of the S-PASS total intensity $I$.}
    \label{Fig:ISignalDistr}
\end{figure}

Figure~\ref{Fig:ISignalDistr} shows the distribution and  cumulative distribution of Stokes $I$. The mean emission is $\overline{I} = 420$~mK, with 50\% of the pixels with emission lower than 230~mK. 
At the low emission end there is a broad peak, broadly spanning the range 90--180~mK, that can be regarded as the typical emission range in the low emission areas of the map. 

\begin{figure*}
	\includegraphics[angle=90, width=1.3\columnwidth]{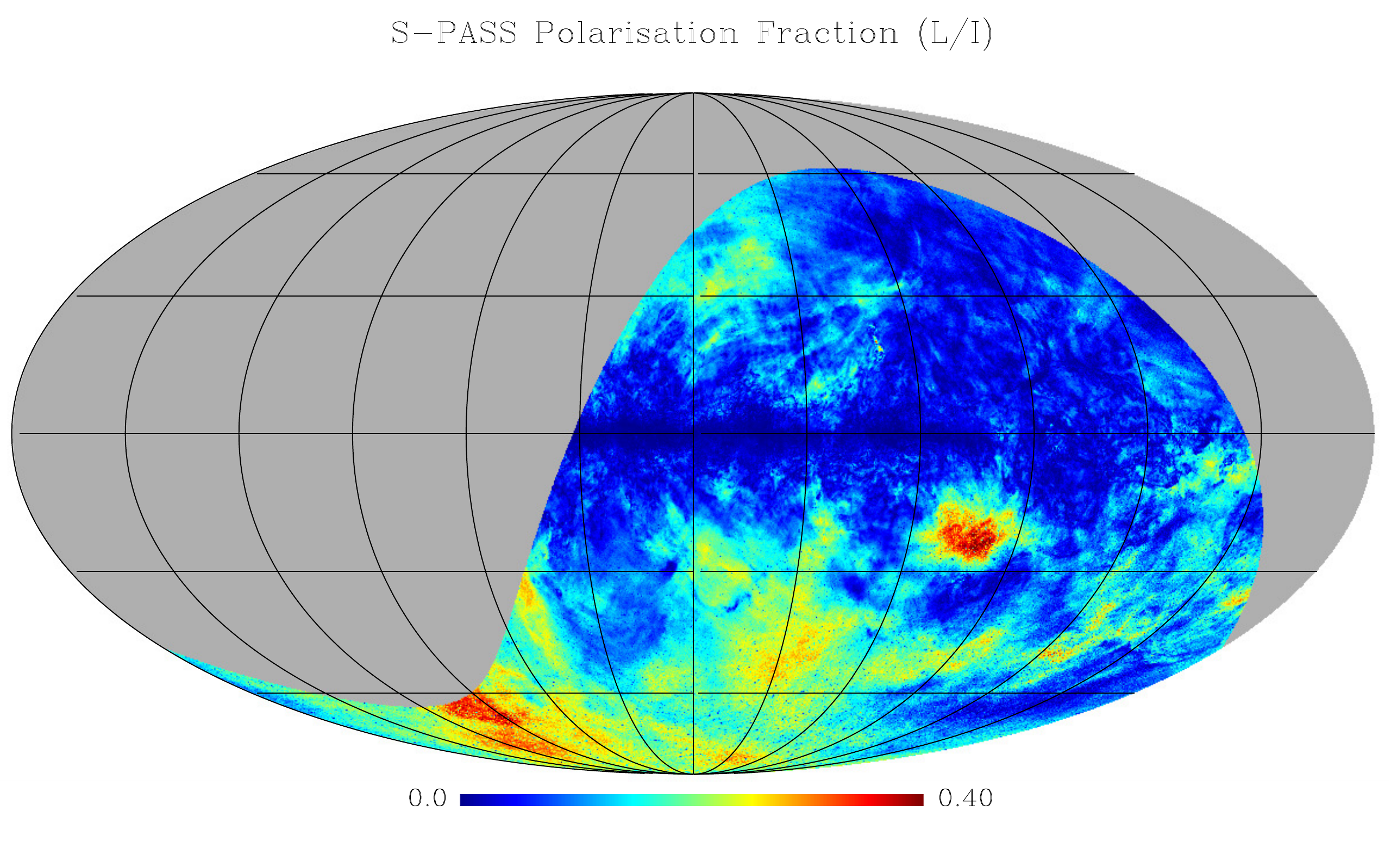}
    \caption{Fractional polarization $L/I$ from S-PASS.} 
    \label{Fig:spassMapsFrac}
\end{figure*}

\begin{figure}
	\includegraphics[width=\columnwidth]{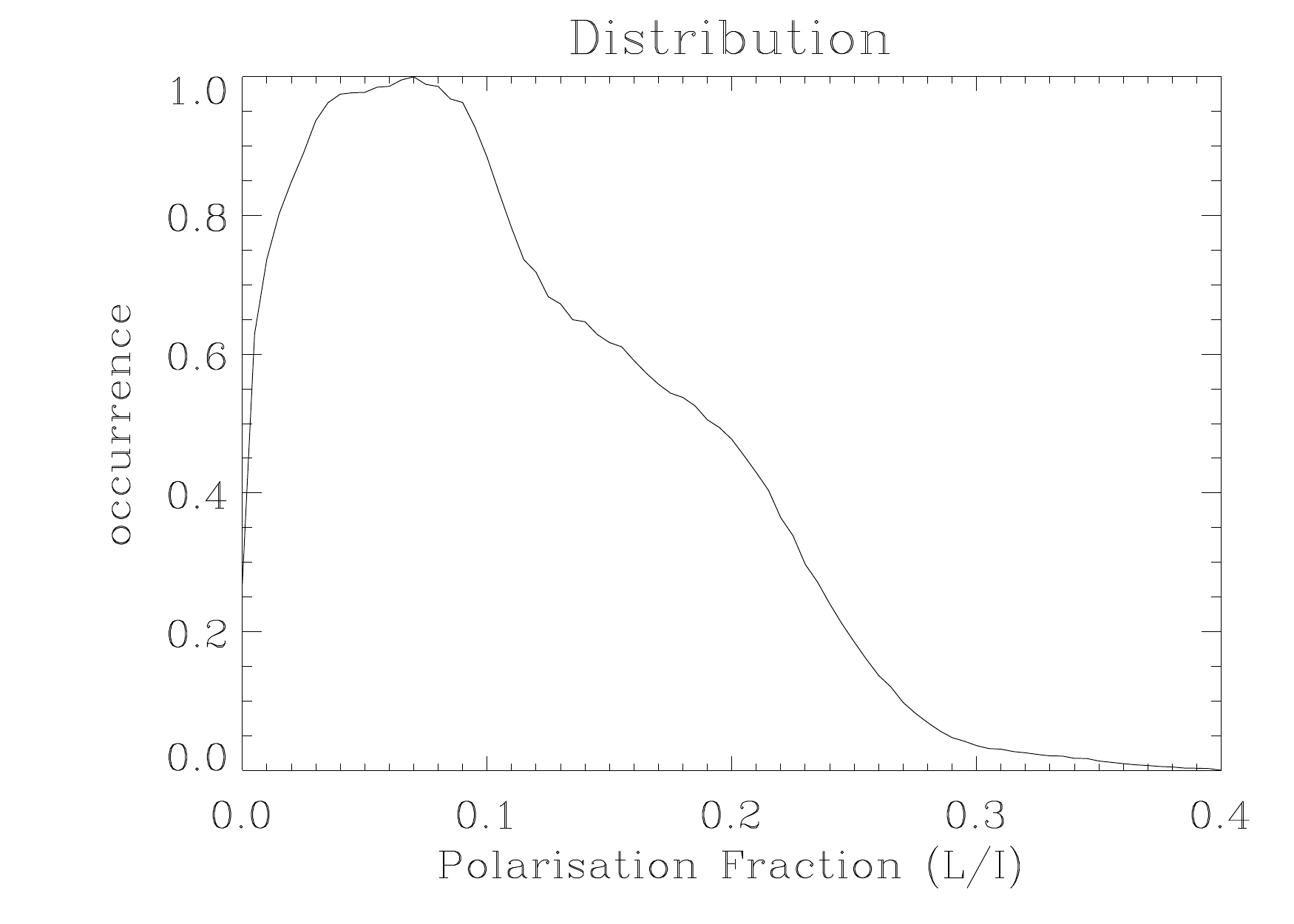}
    \caption{Distribution of the fractional polarization  $L/I$.}
    \label{Fig:FracPolDistr}
\end{figure}

Figure~\ref{Fig:spassMapsFrac} shows the map of polarization fraction $L/I$,  with the distribution of pixel values shown in Figure~\ref{Fig:FracPolDistr}.
%
%
\begin{figure}
	\includegraphics[width=\columnwidth]{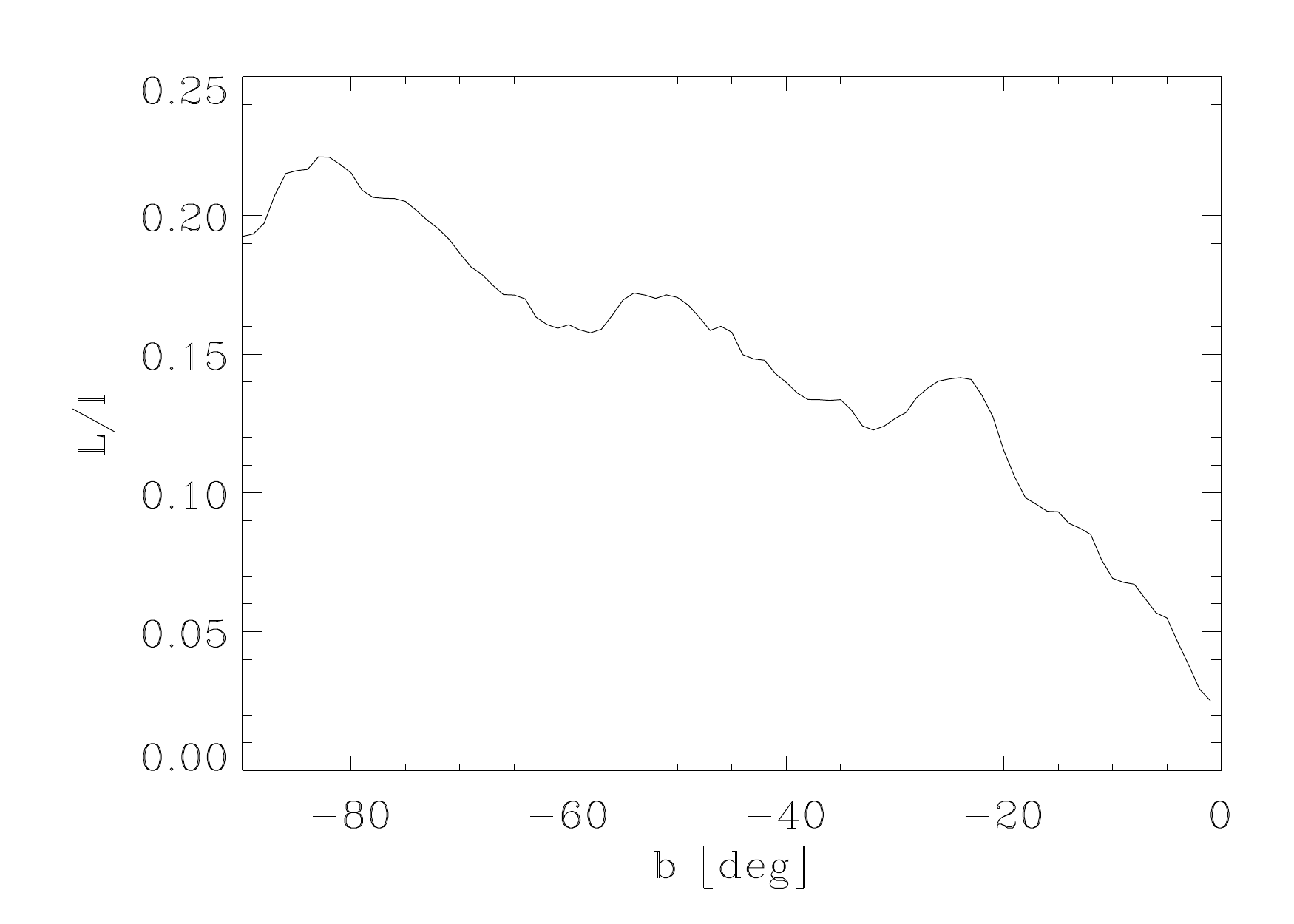}
	\includegraphics[width=\columnwidth]{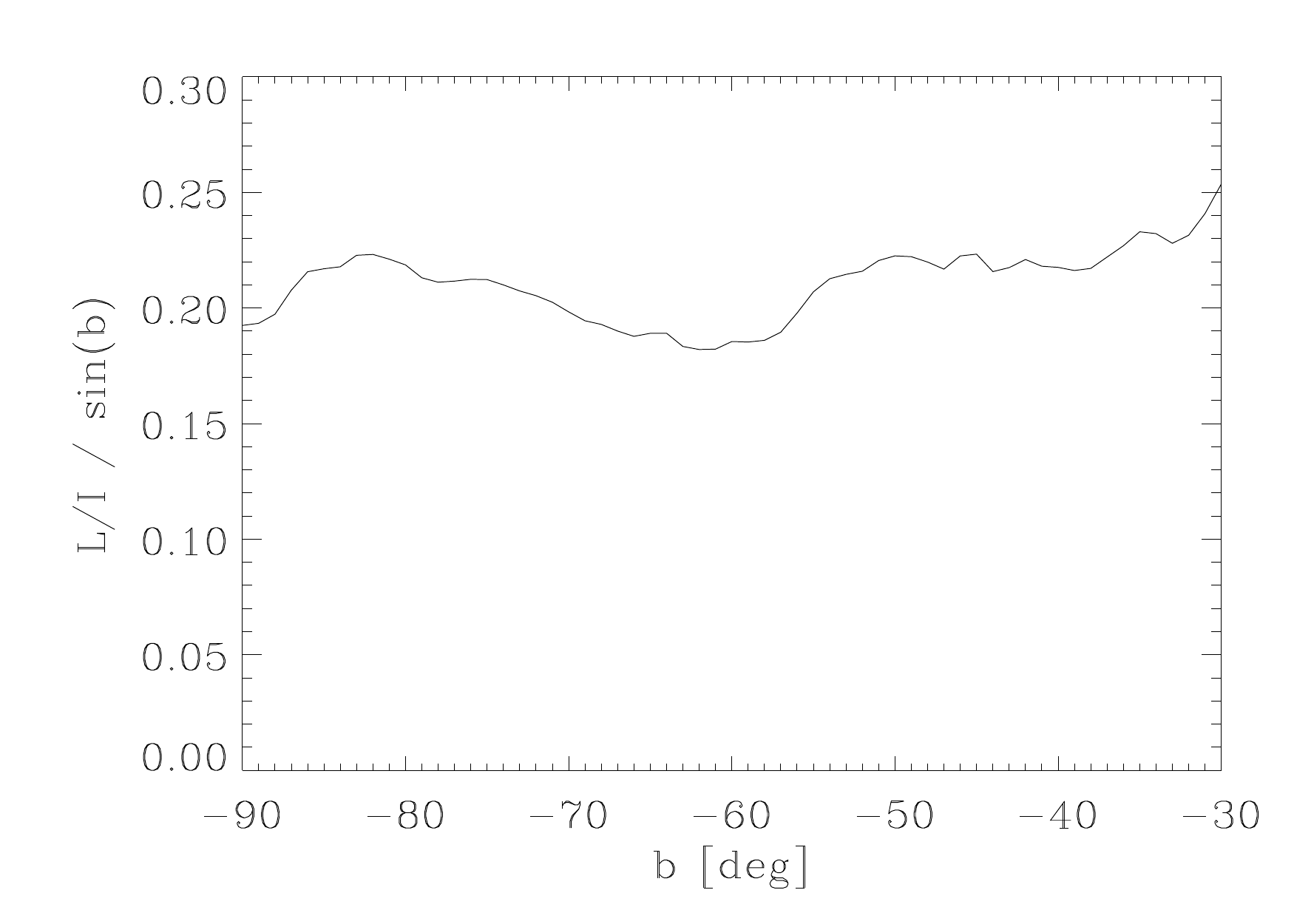}
    \caption{Average fractional polarization $L/I$ (top) and corrected for $\sin b$ (bottom) versus Galactic latitude $b$. 
    The latter is expected to be constant in the case that the local emission below the Galactic Plane has a polarization angle parallel to the Galactic Plane.}
    \label{Fig:FracProf}
\end{figure}
An asymmetry is observed, with the southern hemisphere generally showing higher polarization fraction.
The polarization fraction in the Galactic Plane is close to zero, with the lowest spots approximately corresponding to Milky Way spiral arm locations. 
This is due to three effects: (1) the synchrotron emission in the Galactic plane being mixed with unpolarized free-free emission, leading to a lower total polarization fraction; (2) the tangled magnetic field through the thick spiral ams leads to line-of-sight depolarization, a pure geometrical effect independent of the frequency; and (3)  higher free electron column density that leads to higher Faraday Rotation, and in turn depolarization, in the Plane and at spiral arm locations. 
The north-south asymmetry  might be related to Stokes $I$ emission from the Local Arm and the Gould Belt that is offset to northern Galactic latitudes in most of the area covered by S-PASS, in particular in the inner Galaxy, as also appears clear from the S-PASS Stokes~$I$ map.

The South Galactic Cap has typically high polarization fractions, from 25 to 40\%. This is not that far from the maximum for synchrotron emission, and might be evidence that the local, off-the-plane magnetic field (below the Galactic plane at the Solar location) is mostly parallel to the Galactic disc with little vertical component. In such a case $L/I$ should behave as $(L/I)  \sin(|b|)$  at high latitudes, on average. This is supported by Figure~\ref{Fig:FracProf} where $(L/I)$ corrected for $\sin(|b|)$ is approximately constant versus $b$ from the south Galactic pole down to $b\sim -40^\circ$. This is consistent with the small vertical component of the local, off-the-plane magnetic field found with RM analysis of the South Galactic Cap \citep{Mao10}.

The polarization fraction distribution peaks in the range 2-10\%, but values up to 20-30\% are quite common, and polarization fractions up to 40\% can be observed.

\subsection{Rotation Measure}

\begin{figure}
	\includegraphics[angle=90, width=\columnwidth]{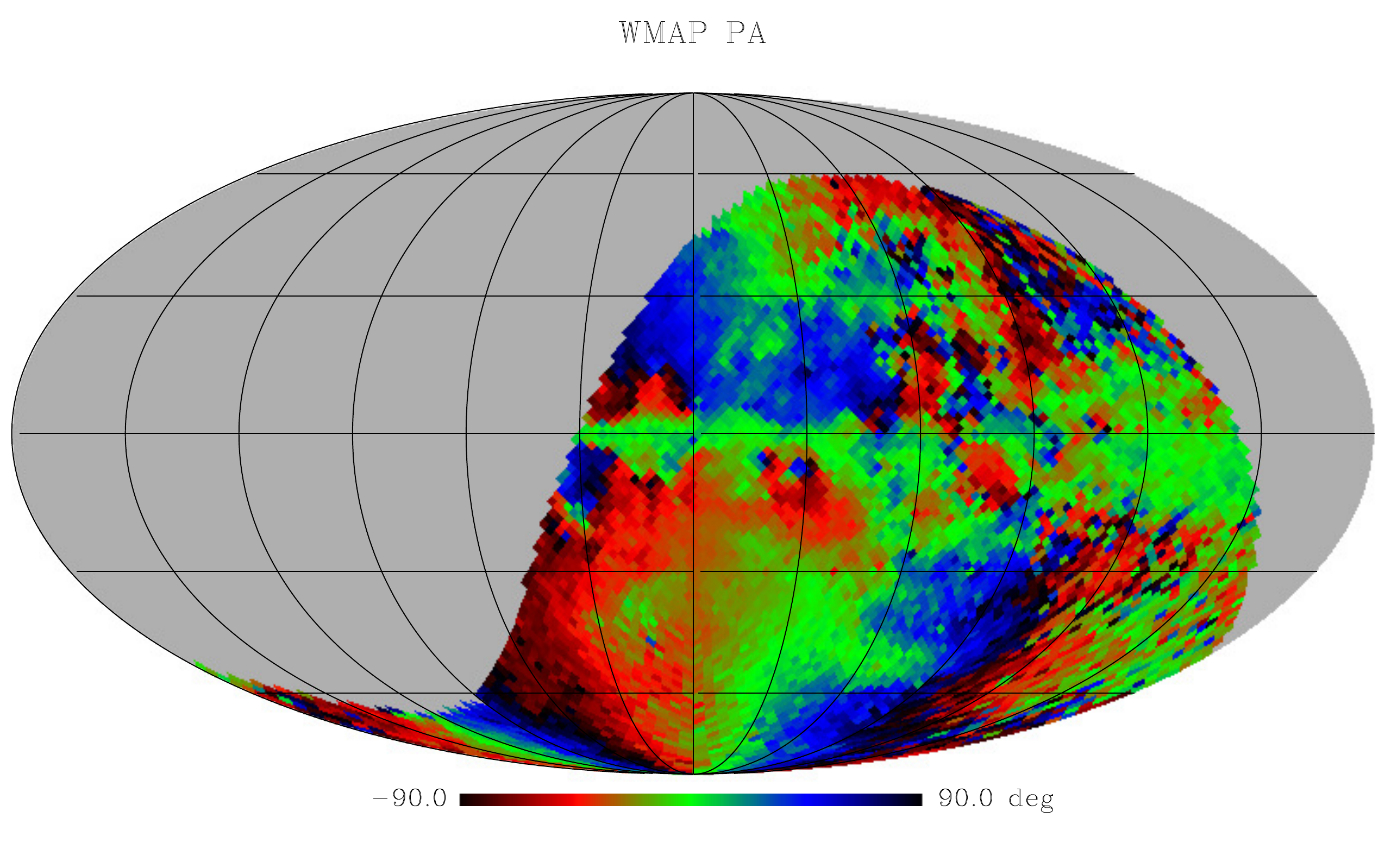}
	\includegraphics[angle=90, width=\columnwidth]{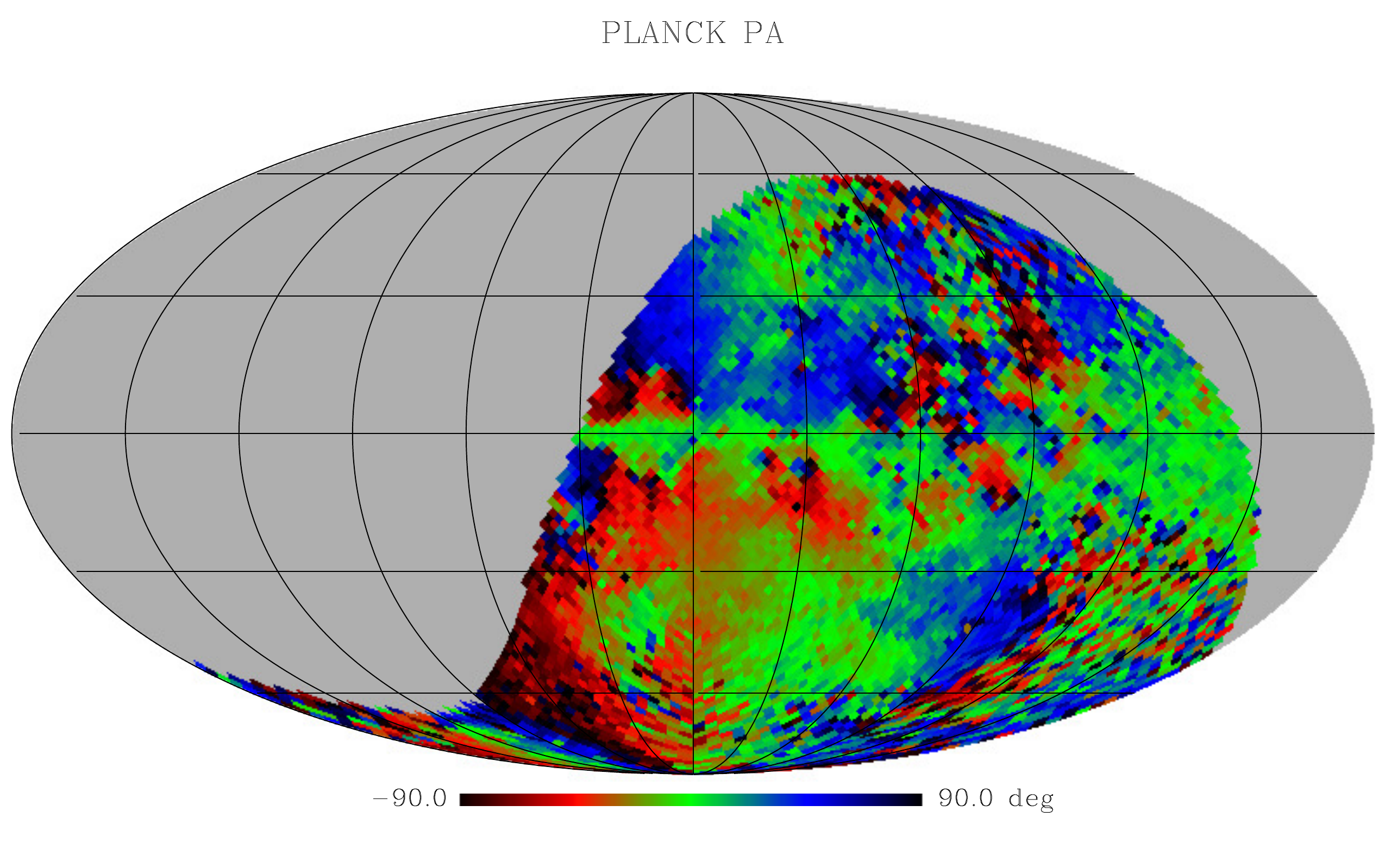}
	\includegraphics[angle=90, width=\columnwidth]{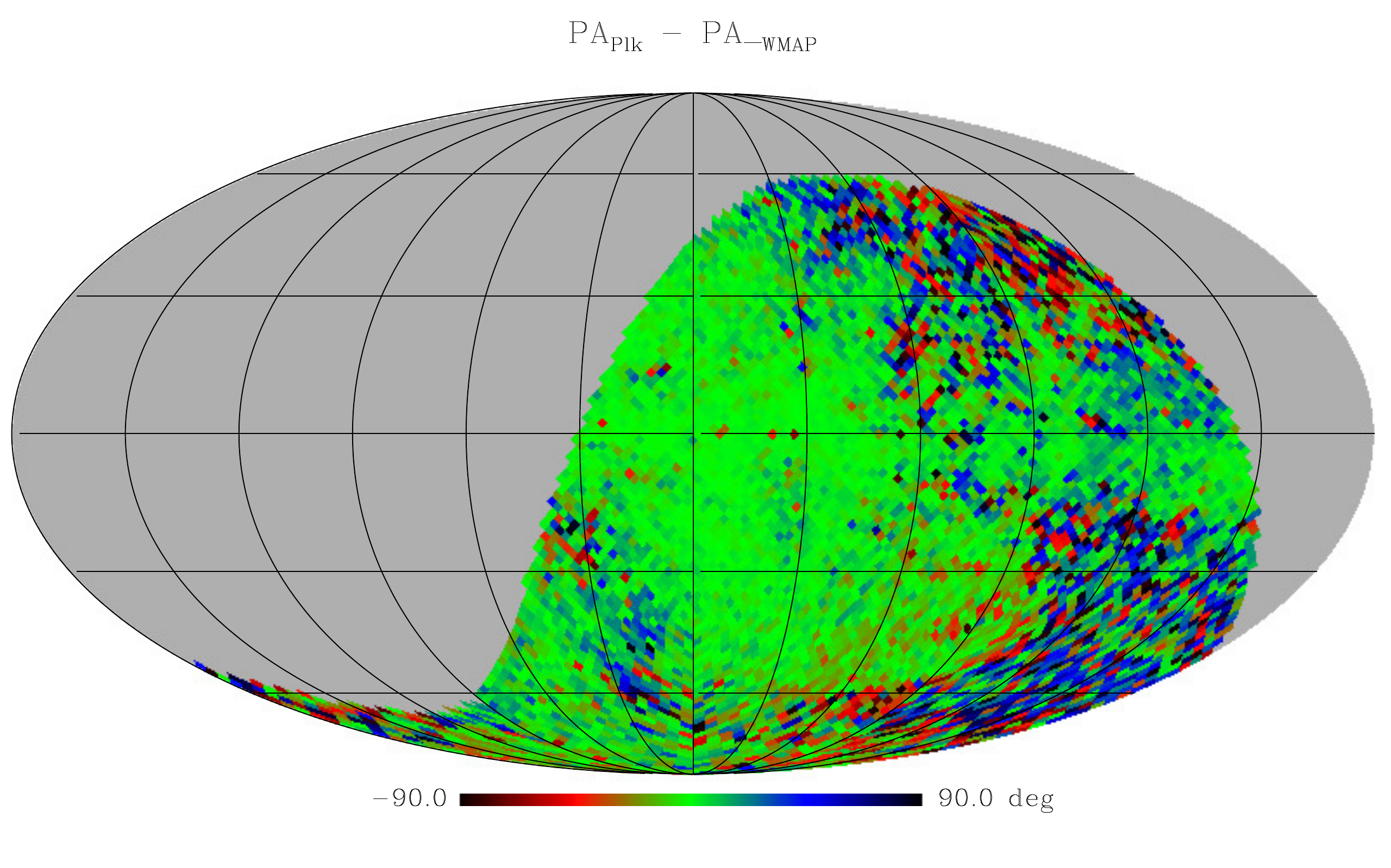}
    \caption{WMAP polarization angle map at 23~GHz (top), Planck's at 30~GHz (middle), and their difference (bottom).}
    \label{Fig:wmap_planck_PA}
\end{figure}

We combined S-PASS polarization maps with  WMAP~\citep{Bennett13} and Planck~\citep{Planck18} archival polarization data at 23~GHz and 30~GHz, respectively  -- we assumed their nominal frequencies,  22.8 and 28.4~GHz, respectively.
However, because of the  low signal at high frequencies there are regions where meaningful RM measurements cannot be obtained.
To select the most reliable areas for RM measurement, we compare the polarization angles in 
Figure~\ref{Fig:wmap_planck_PA}, which shows WMAP and Planck polarization angle maps and their difference in the area covered by S-PASS. 
To increase sensitivity the data were binned in pixels of $2^\circ$ ($nside = 32$).
While the two maps generally match well, there are areas with large differences that can exceed 30-40$^\circ$,  up to 90$^\circ$. 
All these areas have low polarized emission.

These differences cannot be attributed to Faraday Rotation. 
A difference of  $45^\circ$ between  22.8 and 28.4~GHz would mean a RM of 12760~rad~m$^{-2}$,  far too high for those latitudes where values of  the order of 10-20~rad m$^{-2}$ or less are measured by other tracers at high latitudes (e.g. extragalactic sources at high latitudes, \citealt{Oppermann15}, or in the south Galactic cap~\citealt{Mao10}). 

Instead, the differences are possibly residual errors in either or both maps,  perhaps 
due to residual zero-offset calibration errors that, in low emission regions, turn into large errors in polarization angle.
Regardless of their origin and given we cannot discriminate which of the two maps is most affected, we exclude from our analysis all pixels 
where the WMAP-Planck angle difference exceeds 15$^\circ$.
 
For all other pixels, RM is computed by a best fit procedure to the S-PASS, WMAP, and Planck data using the linear relation:
\begin{equation}
 \phi_\lambda  = {\rm RM} \, \lambda^2 + \phi_0 ,
\end{equation}
where $\phi_\lambda$ is the polarization angle at the wavelength $\lambda$ and $\phi_0$ is the intrinsic polarization angle at $\lambda = 0$.
 The data are not sufficient to resolve the $n$--$\pi$ ambiguity, so $n=0$ is assumed.

\begin{figure}
	\includegraphics[angle=90, width=\columnwidth]{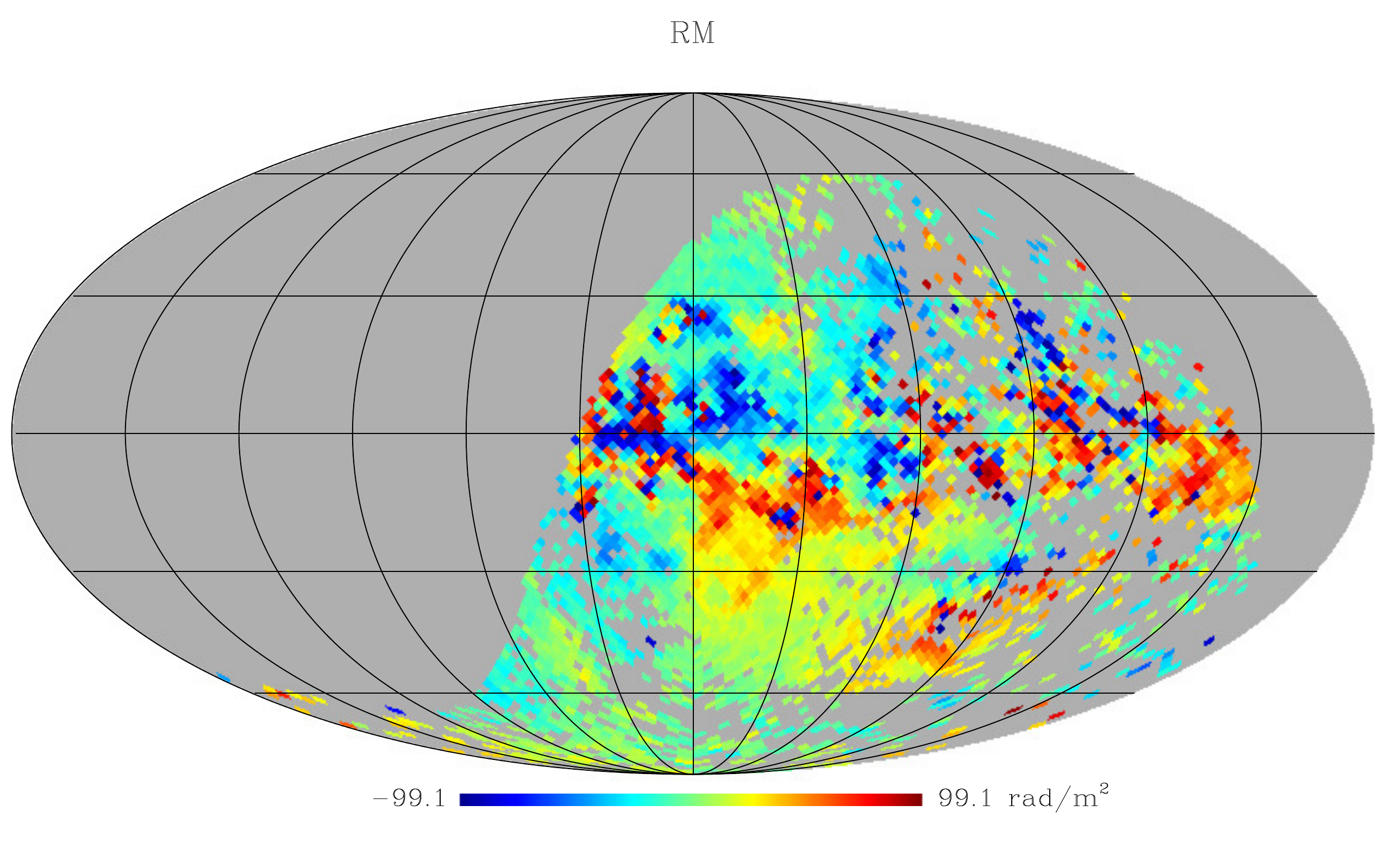}
	\includegraphics[angle=90, width=\columnwidth]{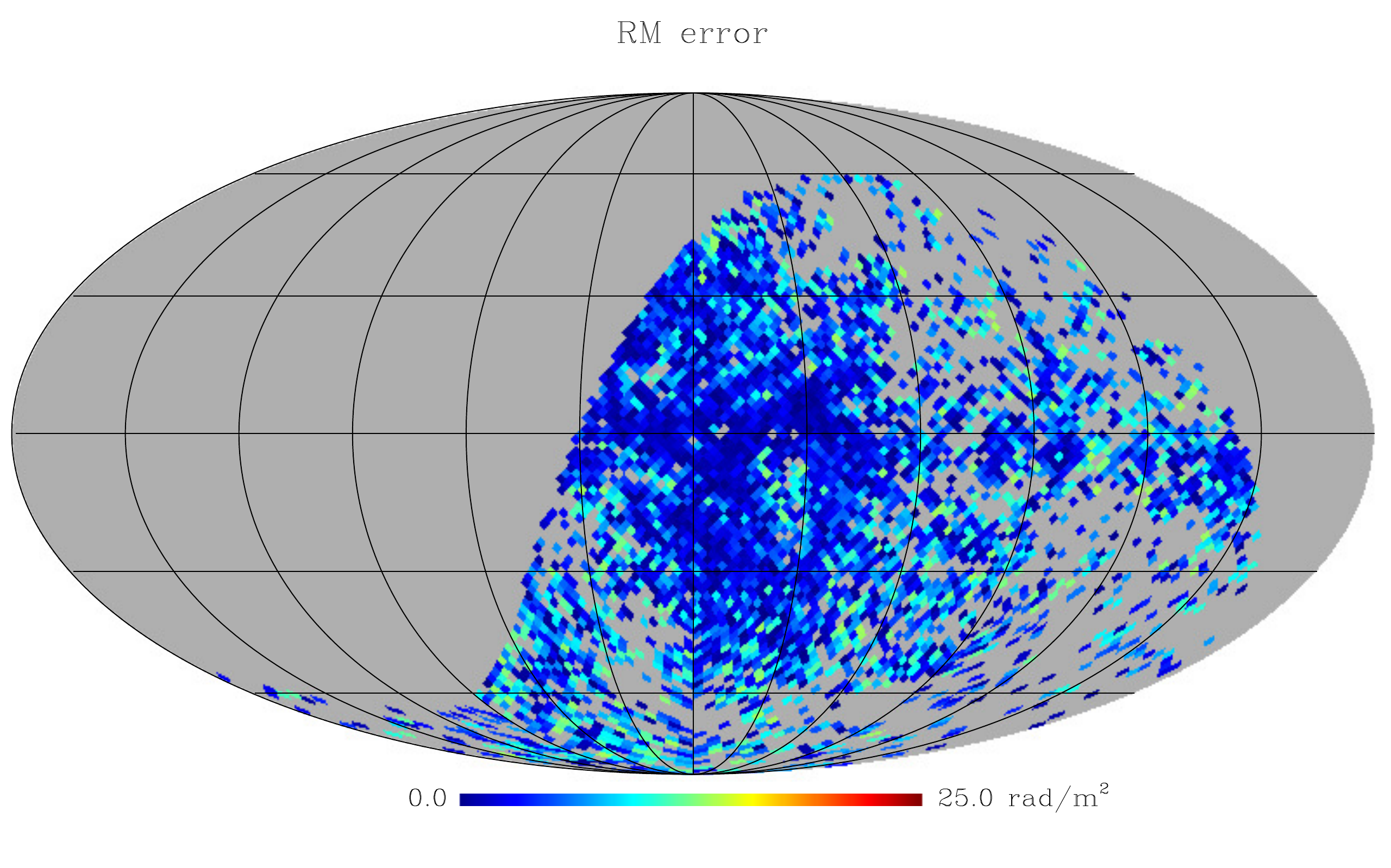}
    \caption{Top: Map of RM obtained combining best fitting S-PASS (2.3 GHz), WMAP (22.8 GHz) and Planck (28.4 GHz) polarization angle maps. Bottom: RM errors (1-$\sigma$) from the fit procedure. }
    \label{Fig:RM}
\end{figure}

The RM map is shown in Figure~\ref{Fig:RM} along with the error computed from the linear fit procedure.  
The area of the Fermi Bubbles looks to have low RMs, negative in the north lobe, positive in the south one. The Bubbles are overlapped by higher RMs regions generated by structures in the foreground. The high RMs generated by the HII region of the nearby $\zeta$~Oph is obvious, as the local ring-like structure G353-34 and the arc of stronger RMs at the S and W edge of the HI Supershell GSH 006-15+7 analysed by~\citet{Thomson18}. The latter protrudes from the Sagittarius arm some 1.5~kpc from the Sun. The Orion area and the Gum Nebula also stand out.

It is worth mentioning that in areas of strong depolarization the lambda-squared law fails, these RMs are underestimated.

\section{Scientific results from S-PASS}\label{Sec:science}
A very diverse range of 
astrophysics research has been conducted with S-PASS data so far, from the study of the local ISM, to CMB foregrounds and Cosmic Web, via the structure of our Galaxy (to mention only a few highlights).
In Table~\ref{Tab:science} we give a list of the papers published to date that employ S-PASS data. Further details can be found in each individual paper. More work, not reported here, is in progress and to be published soon.
\begin{table*}
	\centering
  \caption{Scientific journal articles based on S-PASS data published to date.}
  \begin{tabular}{rlll}
    \hline
  & Title & Topic & reference \\
    \hline
1 & Giant magnetized outflows from the centre &  Discovery of the  Milky Way's giant,  & \citet{Carretti13a} \\
  &   of the Milky Way                                          & radio lobes  the ped radio counter-  & \\
  &                                                                       &  parts of the  $\gamma$--ray Fermi Bubbles  & \\
   & & & \\
2 & Detection of a radio bridge in Abell 3667     &  Discovery of ICM extended emission  & \citet{Carretti13b}\\
   &                                                                      &  connecting the cluster centre to outer  &    \\
   &                                                                      &  relics  &    \\
   & & & \\
3   &  Absolutely calibrated radio polarimetry of the    &  Analysis of the magnetic properties   & \citet{Sun14}\\
   &  inner Galaxy at 2.3 and 4.8~GHz                         &  of Inner Galaxy spiral arms                                           &    \\
   & & & \\
 4  &  Galactic interstellar turbulence across the   &  First ISM turbulence map on an all-sky  & \citet{Iacobelli14}\\
   &  southern sky seen through spatial gradients        &     size scale                                                   &    \\
   &  of the polarization vector                             &   & \\
   & & & \\
5   &  A Radio-Polarisation and Rotation Measure             &  Study of the nature of Gum Nebula  & \citet{Purcell15}\\
   &   Study of the Gum Nebula and Its Environment         &                                                                                 &    \\
   & & & \\
6   & Magnetic Field Disorder and Faraday Effects   &  Fractional polarization behaviour with   & \citet{Lamee16}\\
   & on the Polarization of Extragalactic Radio          &  frequency and other parameters of the                                                                                  &    \\
   &                Sources                                                & S-PASS compact sources bright sample & \\
    & & & \\
7   & A Southern-Sky Total Intensity Source      &    Compact source catalogue of S-PASS & \citet{Meyers17}\\
   & Catalogue at 2.3 GHz from S-Band         &        (Stokes $I$ only)                                                     &    \\
   &  Polarisation All-Sky Survey Data           &  & \\
    & & & \\
8   &  A new perspective on turbulent Galactic     &  ISM turbulence analysis tools with & \citet{Robitaille17}\\
   &  magnetic fields through comparison of linear         &   multi--scale wavelets and       &    \\
   &  polarization decomposition techniques         &       E/B--Modes decompositions                              &    \\
     & & & \\
9   &   Limiting magnetic fields in the cosmic web     &  New upper limits on the Cosmic Web    & \citet{Brown17}\\
   &   with diffuse radio emission                                &  synchrotron emission and magnetic       &    \\
   &                                                                             &   field intensity                                         &    \\
    & & & \\
10   & Advanced Diagnostics for the Study of Linearly   &  New diagnostic tools to study diffuse linear   & \citet{Herron18}\\
   & Polarized Emission. II. Application to Diffuse          &      polarized emission                   &    \\
   & Interstellar Radio Synchrotron Emission         &                                                                                 &    \\
    & & & \\
11   & The jet/wind outflow in Centaurus A:     &  New findings in Cen A lobes/outflows & \citet{McKinley18}\\
   &  a local laboratory for AGN feedback        &                                                                                 &    \\
    & & & \\
12 & Ghost of a shell: magnetic fields of Galactic     & Magnetic field analysis of a Galactic   & \citet{Thomson18}\\
   & supershell GSH~006-15+7                          &      supershell                                                                           &    \\
    & & & \\
13  &  Interstellar magnetic cannon targeting the    &    Discovery of a new, large, local cavity & \citet{Robitaille18}\\
   &  Galactic halo : A young bubble at the origin of   &  in the inner Galaxy and its dynamic           &    \\
   &  the Ophiuchus and Lupus molecular complexes          &  analysis                                                                               &    \\
    & & & \\
14   & The S-PASS view of polarized Galactic     &  High precision analysis of Galactic   & \citet{Krachmalnicoff18}\\
   &  Synchrotron at 2.3~GHz as a contaminant   &   polarized synchrotron emission  as  a CMB  &    \\
   &  to CMB observations                                     &   foreground over the entire southern sky                              &    \\
    & & & \\
15   & S-PASS/ATCA: a window on the magnetic     &  Compact sources RM catalogue of    & \citet{Schnitzeler19}\\
   &    universe in the southern hemisphere        &   the S-PASS sample followed-up with  &    \\
   &                                                                      &    broad--band ATCA observations &    \\
    & & & \\
16   & S-band Polarization All Sky Survey (S-PASS):     &   S-PASS survey description paper  & This paper\\
   &   survey description and maps        &    &    \\
 \hline
    \end{tabular}
    \label{Tab:science}
\end{table*}

\section{Data release}\label{Sec:data}
With this paper we make the  S-PASS maps publicly available.
More specifically we will release the following maps, in both HEALPix format and Aitoff projection:
\begin{description}
 \item{-} Stokes $Q$, 
 \item{-} Stokes $U$, 
 \item{-} Stokes $I$, 
 \item{-} Stokes $Q$, $U$ pixel sensitivity 
 \item{-} RM map
 \item{-} RM error map 
\end{description}

Data will be made available on the web site {\it https://sites.google.com/inaf.it/spass} 
and on 
the  Legacy Archive for Microwave Background Data Analysis (LAMBDA\footnote{https://lambda.gsfc.nasa.gov/}) web site.

\section{Summary and conclusions}\label{Sec:conc}
We have observed the entire southern sky at Dec~$< -1^\circ$ at 2.3~GHz with the Parkes Radio Telescope, realising a detailed polarization map with resolution of 8.9~arcmin that preserves information on all angular scales, including the  overall mean emission of Stokes~$Q$ and $U$. 
Given the high resolution (for a single--dish telescope) of our measurements, our data encompass a spatial dynamic range of some 1200, one of the largest ever achieved in similar studies.
The mean sensitivity of S-PASS on a beamsize scale is 0.81~mK. 
We account for systematic errors, including ground contamination (the largest source of uncertainty), to an accuracy of 0.35~mK.

One of the major goals of the survey was to preserve the global offset of Stokes $Q$ and $U$. 
This has been achieved with a novel scanning strategy based on long azimuth scans at the EL of the South Celestial Pole at the telescope location taken at both east and west azimuths, combined with a  map-making procedure which makes use  of a basket-weaving technique as well as parallactic angle modulation. 
This procedure has been tested with realistic simulations and shown to be able to recover the global mean emission with an accuracy better than 0.5\%. 

Simulations demonstrate that the reconstruction error of our technique adds a negligible error even in low emission areas ($L < 20$~mK) and achieves a S/N~$> 50$ elsewhere. 
This ensures high quality data with the error budget led by instrumental noise and flux calibration accuracy with negligible additional errors.

The survey has been successful in unveiling the polarized emission from the Galactic disc and the disc-halo transition region, previously 
been hidden by Faraday depolarization at lower frequencies. This allows improved studies of Galactic magnetism with diffuse emission, 
including large-scale Galactic structure, the Galactic magnetic field, and ISM turbulence. 
A number of scientific analyses have already been conducted and more are possible in a number of diverse science areas.

We find that the mean polarized emission at 2.3 GHz is 29~mK and the typical emission in low emission areas is $\sim 13$~mK. 
50\% of the S-PASS area has polarized emission fainter than 23.6~mK, and 98.6\% of the pixels have S/N~$>3$.  

We also computed the RM map of the diffuse emission combining S-PASS with archival higher frequency maps from WMAP and Planck at the locations where WMAP and Planck are consistent. 
The resolution of this map is $2^\circ$, with sensitivity limited by the poor S/N ratio of the high frequency data. We expect to obtain a higher resolution and higher sensitivity RM map when S-PASS is combined with the upcoming GMIMS southern surveys.

Although not the primary goal of S-PASS,   the data have also been used to generate Stokes~$I$ images whose rms error is set by the confusion limit (9~mK).  
The offset calibration, impossible to constrain in total intensity with our observing technique, has been obtained using archival data of absolutely calibrated observations of the south Celestial pole.

\section*{Acknowledgements}
This work has been carried out in the framework of the S-band Polarization All Sky Survey collaboration (S-PASS).
We thank John Reynolds and Andrew Hunt for modifying the Parkes drive software and turning an idea in our mind into reality, 
Warwick Wilson for creating the backend configuration required by S-PASS, the whole Parkes Operations team who made observing with such a complicated setup smooth,
and The Dish for having behaved seamlessly despite being nearly 50 years old. 
We thank team member Stefano Cortiglioni for his contribution. We thank Tom Landecker for fruitful discussions on Brightness Temperature calibration and Nicoletta Krachmalnicoff for generating the PSM Stokes $I$ map.
We thank the referee for constructive comments.
The Parkes radio telescope is part of the Australia Telescope National Facility which is funded by the Commonwealth of Australia for operation as a National Facility managed by CSIRO. 
This work was partly funded by ASI under the project ASI I/016/07/0.
The Dunlap Institute is funded through an endowment established by the David Dunlap family and the University of Toronto. 
R.M.C. was the recipient of an Australian Research Council Future Fellowship (FT110100108).
B.M.G. acknowledges the support of the Natural Sciences and Engineering Research Council of Canada (NSERC) through grant RGPIN-2015-05948, and of the Canada Research Chairs program.
X.H.S. is supported by the National Natural Science Foundation of China under grant no. 11763008.
M.H. acknowledges funding from the European Research Council (ERC) under the European Union Horizon 2020 research and innovation programme (grant agreement No 772663).
We acknowledge the use of the Miriad package \citep{Sault95}.
We acknowledge the use of the PSM, developed by the Component Separation Working Group (WG2) of the Planck Collaboration.
The Wisconsin H--Alpha Mapper and its Sky Survey have been funded  primarily through awards from the U.S. National Science Foundation.
This research made use of Montage. It is funded by the National Science Foundation under Grant Number ACI-1440620, and was previously funded by the National Aeronautics and Space Administration's Earth Science Technology Office, Computation Technologies Project, under Cooperative Agreement Number NCC5-626 between NASA and the California Institute of Technology.







\appendix
\onecolumn
\section{ Stokes $Q$ and $U$ scans offset calibration}
\label{Sec:appA}

This Section presents the equations to find the best offset set for Stokes $Q$ and $U$ scans.

Let ${\bf Y}_i$ be the polarization vector (Q, U) of the $i^{\rm th}$ piece of data measured in the instrument reference frame:
\begin{equation}
{\bf Y}_i 
=
\begin{bmatrix}
    Q_{m,i} \\
    U_{m,i}
\end{bmatrix},
\end{equation}
${\bf X}_i$ the polarization vector of sky emission in the sky reference frame:
\[ 
{\bf X}_i 
=
\begin{bmatrix}
    Q_i \\
    U_i
\end{bmatrix}, \addtag
\]
${\bf A}_{is}$ the offsets of $Q$ and $U$ of the scan $is$:
\[ 
{\bf A}_{is} 
=
\begin{bmatrix}
    A^Q_{is} \\
                  \\
    A^U_{is}
\end{bmatrix}, \addtag \label{eq:a}
\]
and ${\bf R}(\theta)$ the 2D rotation matrix between two local 2D reference frames centred at the pointing direction rotated by the angle $\theta$, so that ${\bf R}(-2\phi)$ is the rotation matrix to convert $Q$ and $U$ from the instrument to the sky reference frame 
 ($\phi$ is the parallactic angle).

For the sample $i$ taken in the scan ${is}$
\[ 
{\bf X}_i  = {\bf R}(-2\phi_i) ({\bf Y}_i + {\bf A}_{is} ). \addtag
\]

The best offset set is obtained minimising the square differences between the sky emission observed in the same pixel with different scans:
\begin{align}
S^2  & = \sum_{is=2}^{n_s} \sum_{ik=1}^{is-1} \sum_{p=1}^{n_{p}^{is,ik}} w_p \left|{\bf X}_{p,is} - {\bf X}_{p,ik}\right|^2 \nonumber \\
        & = \sum_{is=2}^{n_s} \sum_{ik=1}^{is-1} \sum_{p=1}^{n_{p}^{is,ik}}  w_p \left|\left({\bf R}_{p,is} {\bf Y}_{p,is} - {\bf R}_{p,ik} {\bf Y}_{p,ik}\right) + \left({\bf R}_{p,is} {\bf A}_{is} - {\bf R}_{p,ik} {\bf A}_{ik}\right)\right|^2 \label{eq:s2}
\end{align}
where $n_s$ is the number of scans, ${n_{p}^{is,ik}}$ the number of pixels where the two scans $is$ and $ik$ cross, ${\bf Y}_{p,ix}$ the polarized emission vector at the pixel $p$ observed with the scan $ix$, ${\bf R}_{p,ix} = {\bf R}(-2\phi_{p,ix})$, ${\bf A}_{ix}$ the offset of the scan $ix$, and $w_p$ a weight for pixel $p$. In this work we have used $w_p = 1$, but it is included here to give the most general formulation.

The minimisation is done compared to all the $2 n_s$ free parameters ${\bf A}_{i}$,  
\begin{align}
\frac{\partial S^2}{\partial A^Q_i} & = 0,  \,\,\,\,\,{\rm for}\,\,\, i=1, n_s \\
\frac{\partial S^2}{\partial A^U_i} & = 0
\end{align}
\\
or in a more compact format
\\ 
\begin{equation}
\frac{\partial S^2}{\partial {\bf A}_i} = 
        \begin{bmatrix}
        0\\
        0
        \end{bmatrix} \,\,\,\,\,{\rm for}\,\,\, i=1, n_s \\
\end{equation}
\\
where we define 
\\
\begin{equation}
\frac{\partial}{\partial {\bf A}_i}  = 
   \begin{bmatrix}
     \frac{\partial}{\partial A_i^Q} \\     
     \frac{\partial}{\partial A_i^U}      
   \end{bmatrix} \label{eq:vecder}
\end{equation}
\\
Let us denote 
\\
\begin{equation}
{\bf D}_{p,is,ik} = \left({\bf R}_{p,is} {\bf Y}_{p,is} - {\bf R}_{p,ik} {\bf Y}_{p,ik}\right) + \left({\bf R}_{p,is} {\bf A}_{is} - {\bf R}_{p,ik} {\bf A}_{ik}\right)
\end{equation}
\\
and $D_Q$, $D_U$ its two components. 
\\
Considering that
\\
\begin{align}
\frac{\partial }{\partial A^Q_i}\left({\bf R}_{p,j} {\bf A}_{j}\right) & = 
 \begin{bmatrix}
    R^{11}_{p,j}\delta_{i,j} \\
    R^{21}_{p,j}\delta_{i,j} \\
  \end{bmatrix}
   = 
 \begin{bmatrix}
    \cos(-2\phi_{p,j})\,\delta_{i,j} \\
    -\sin(-2\phi_{p,j})\,\delta_{i,j} \\
  \end{bmatrix},
 \\
\frac{\partial }{\partial A^U_i}\left({\bf R}_{p,j} {\bf A}_{j}\right) & = 
 \begin{bmatrix}
    R^{12}_{p,j}\delta_{i,j} \\
    R^{22}_{p,j}\delta_{i,j} \\
  \end{bmatrix}
=
 \begin{bmatrix}
    \sin(-2\phi_{p,j})\,\delta_{i,j} \\
    \cos(-2\phi_{p,j})\,\delta_{i,j} \\
  \end{bmatrix},
\end{align}
\\
the partial derivates of each term of Equation~\eqref{eq:s2} are
\\
\begin{align}
\frac{\partial }{\partial A^Q_i}\left|{\bf D}_{p,is,ik}\right|^2  & = 
    2 \left[\left({\bf R}^T_{p,is} {\bf D}_{p,is,ik}\right)_Q  \delta_{i,is} -  \left({\bf R}^T_{p,ik} {\bf D}_{p,is,ik}\right)_Q  \delta_{i,ik} \right]   \\
\frac{\partial }{\partial A^U_i}\left|{\bf D}_{p,is,ik}\right|^2  & = 
    2 \left[\left({\bf R}^T_{p,is} {\bf D}_{p,is,ik}\right)_U  \delta_{i,is} -  \left({\bf R}^T_{p,ik} {\bf D}_{p,is,ik}\right)_U  \delta_{i,ik} \right]   
\end{align}
\\
where the subscripts $_Q$ and $_U$ denotes the Q and U components and $T$ the transpose matrix, or using the compact format of Equations~\eqref{eq:vecder}
\\
\begin{equation}
  \frac{\partial }{\partial {\bf A}_i}\left|{\bf D}_{p,is,ik}\right|^2  = 
    2 \left[\left({\bf R}^T_{p,is} {\bf D}_{p,is,ik}\right)  \delta_{i,is} -  \left({\bf R}^T_{p,ik} {\bf D}_{p,is,ik}\right) \delta_{i,ik} \right].
    \label{eq:oneder}
\end{equation}
\\
From Equations~\eqref{eq:s2} and~\eqref{eq:oneder}, the solving Equation~\eqref{eq:vecder}  can be written as
\\
\begin{equation}
        \sum_{is\neq i} \sum_{p=1}^{n_{p}^{is,i}}  w_p \left[{\bf A}_i - {\bf R}^T_{p,i}{\bf R}_{p,is} {\bf A}_{is} + {\bf Y}_{p,i} - {\bf R}^T_{p,i}{\bf R}_{p,is} {\bf Y}_{p,is}\right) = 
        \begin{bmatrix}
        0\\
        0
        \end{bmatrix}   \,\,\,\,\,{\rm for}\,\,\, i=1, n_s \\
\end{equation}
Rearranged to focus on the variables ${\bf A}_i$ to solve for, it turns into the $n_s$ pairs of equations:
\begin{equation}
         \left(\sum_{is\neq i} \sum_{p=1}^{n_{p}^{is,i}}  w_p \right) {\bf A}_i +  \sum_{is\neq i} \left( -\sum_{p=1}^{n_{p}^{i,is}}  w_p {\bf R}^T_{p,i}{\bf R}_{p,is}\right) {\bf A}_{is} 
          = \sum_{is\neq i} \sum_{p=1}^{n_{p}^{is,i}}  w_p  \left({\bf R}^T_{p,i}{\bf R}_{p,is} {\bf Y}_{p,is}  - {\bf Y}_{p,i} \right)    \,\,\,\,\,{\rm for}\,\,\, i=1, n_s \\
\end{equation}
 \\
 that is the system of $2 n_s$ linear equations to solve to find the best set of offset values ${\bf A}_i$.
\\
It can also be written in matrix form as
\\
\begin{equation}
{\bf M}\cdot{\bf A} = {\bf B}
\end{equation}
\\
where {\bf M} is the $n_s\times n_s$ matrix  whose elements are the $2 \times 2$ matrixes
\\
\begin{align}
&{\bf M}_{i,i}  = \sum_{is\neq i} \sum_{p=1}^{n_{p}^{is,i}}  w_p {\bf I} \\
&{\bf M}_{i,is}  = -\sum_{p=1}^{n_{p}^{i,is}}  w_p {\bf R}^T_{p,i}{\bf R}_{p,is}, \,\,\,\,\,{\rm for}\,\,\, i \neq is ,
\end{align}
\\
${\bf A}_i$ is defined in Equation~\eqref{eq:a}, and 
\begin{equation}
    {\bf B}_i = \sum_{is\neq i} \sum_{p=1}^{n_{p}^{is,i}}  w_p  \left({\bf R}^T_{p,i}{\bf R}_{p,is} {\bf Y}_{p,is}  - {\bf Y}_{p,i} \right).
\end{equation}

\section{ Stokes $I$ scans offset calibration}
\label{Sec:appB}

The equations to estimate the best set of offset for Stokes $I$ scans are obtained as for Stokes $Q$ and $U$, except that the 2-component vectors ${\bf Y}_i$, 
${\bf X}_i$, and ${\bf A}_{is}$ are replaced the scalars $y_i$, $x_i$, and $A_{is}$, and the rotation matrix ${\bf R}$ with the scalar unity $1$.

Following the same steps one gets the system of $n_s$ equations to solve:
\begin{equation}
         \left(\sum_{is\neq i} \sum_{p=1}^{n_{p}^{is,i}}  w_p \right)  A_i +  \sum_{is\neq i} \left( -\sum_{p=1}^{n_{p}^{i,is}}  w_p \right) A_{is} 
          = \sum_{is\neq i} \sum_{p=1}^{n_{p}^{is,i}}  w_p  \left(  y_{p,is}  -  y_{p,i} \right)    \,\,\,\,\,{\rm for}\,\,\, i=1, n_s \\
\end{equation}
\\
In matrix form:
\\
\begin{equation}
{\bf M}\cdot{\bf A} = {\bf B}
\end{equation}
\\
where {\bf M} is the $n_s\times n_s$ matrix  of elements
\\
\begin{align}
&{\bf M}_{i,i}  = \sum_{is\neq i} \sum_{p=1}^{n_{p}^{is,i}}  w_p  \\
&{\bf M}_{i,is}  = -\sum_{p=1}^{n_{p}^{i,is}}  w_p , \,\,\,\,\,{\rm for}\,\,\, i \neq is ,
\end{align}
\\
and $n_s$-element vector:
\begin{equation}
    {\bf B}_i = \sum_{is\neq i} \sum_{p=1}^{n_{p}^{is,i}}  w_p  \left( y_{p,is}  - y_{p,i} \right).
\end{equation}
\\
It is worth noticing this system is degenerate because of the lack of parallactic angle modulation that $Q$ and $U$ benefit from (see main text).
It is solved through the Singular Value Decomposition (SVD) method, a powerful tool to solve ill-conditioned systems.


\bsp	
\label{lastpage}
\end{document}